\documentclass[11pt,oneside,reqno]{article}
\usepackage{amstext}
\usepackage{amssymb}
\usepackage[a4paper]{geometry}
\usepackage{color}
\usepackage{float}
\usepackage{booktabs}
\usepackage{mathrsfs}
\usepackage{multirow}
\usepackage{graphicx}
\usepackage{setspace}
\makeatletter

\providecommand{\tabularnewline}{\\}
\newcommand{\lyxdot}{.}

\usepackage{amsfonts}
\usepackage{amstext}
\usepackage{amsthm}
\usepackage{breqn}

\newtheorem{thm}{Theorem}
\newtheorem{lem}{Lemma}
\newtheorem{cor}{Corollary}
\newtheorem{prop}{Proposition}
\newtheorem{asm}{Assumption}
\newtheorem*{asm1'}{Assumption 1'}
\newtheorem*{asm2'}{Assumption 2'}
\newtheorem*{asm3'}{Assumption 3'}
\newtheorem*{asm4'}{Assumption 4'}

\theoremstyle{definition}
\newtheorem{rem}{Remark}

\usepackage{xr}
\usepackage{hyperref}
\makeatletter
\newcommand*{\addFileDependency}[1]{
	\typeout{(#1)}
	\@addtofilelist{#1}
	\IfFileExists{#1}{}{\typeout{No file #1.}}
}
\makeatother

\newcommand*{\myexternaldocument}[1]{%
	\externaldocument{#1}%
	\addFileDependency{#1.tex}%
	\addFileDependency{#1.aux}%
}
\myexternaldocument{Mono_matching10}
\myexternaldocument{Supp_mono_matching10}

\makeatother

\begin{document}
\title{Isotonic propensity score matching\thanks{We are grateful to Markus Frölich, Daniel Gutknecht, Phillip Heiler, Lihua Lei, Yoshi Rai, Christoph Rothe, Carsten Trenkler, and participants at the econometrics seminar at Mannheim 2022, NASMES 2023, and IAAE 2023, for helpful comments and discussions. We also would like to thank a co-editor and anonymous referees for helpful comments to revise the paper.}}
\author{Mengshan Xu\thanks{Department of Economics, University of Mannheim, L7 3-5, 68161, Mannheim, Germany. Email: mengshan.xu@uni-mannheim.de} and Taisuke Otsu\thanks{Department of Economics, London School of Economics, Houghton Street, London, WC2A 2AE, UK. Email: t.otsu@lse.ac.uk}}
\maketitle
\begin{abstract}
We propose a one-to-many matching estimator of the average treatment
effect based on propensity scores estimated by isotonic regression.
This approach is predicated on the assumption of monotonicity in the
propensity score function, a condition that can be justified in many
economic applications. We show that the nature of the isotonic estimator
can help us to fix many problems of existing matching methods, including
efficiency, choice of the number of matches, choice of tuning parameters,
robustness to propensity score misspecification, and bootstrap validity.
As a by-product, a uniformly consistent isotonic estimator is developed
for our proposed matching method.
\end{abstract}

\section{Introduction}

In both randomized experiments and observational studies, matching
estimators are widely used to estimate treatment effects. This paper
proposes a novel one-to-many propensity score matching method of the
average treatment effect (ATE), where the propensity score is assumed
to be monotone increasing in the exogenous covariate and is estimated
by the isotonic regression. Our matching scheme is exact, i.e., for
the outcome $Y$, the binary treatment $W$, the covariate $X$, and
a sample of size $N$, the matched set for the $i$-th unit is defined
as
\[
\mathcal{J}(i)=\left\{ j=1,\dots,N:W_{j}=1-W_{i}\text{ and }\tilde{p}(X_{j})=\tilde{p}(X_{i})\right\} ,
\]
where $\tilde{p}(\cdot)$ is a uniformly consistent isotonic estimator
of the propensity score developed in Section \ref{subsec:Uniformly-consistent-isotonic}.
For multi-dimensional covariates $X$, we employ a monotone index
model and consider the matched set:
\[
\mathcal{J}(i)=\left\{ j=1,\dots,N:W_{j}=1-W_{i}\text{ and }\tilde{p}_{\tilde{\alpha}}(X_{j}^{\prime}\tilde{\alpha})=\tilde{p}_{\tilde{\alpha}}(X_{i}^{\prime}\tilde{\alpha})\right\} ,
\]
where $\tilde{p}_{\tilde{\alpha}}(\cdot)$ is a uniformly consistent
monotone single-index estimator of the propensity score developed
in Section \ref{sec:Multi}.

Remarkably, the isotonic estimator proves to be especially well-suited
as the initial nonparametric estimator in a two-stage semiparametric
approach to estimating the ATE. It incorporates features of both matching
and weighting estimators into the second-stage ATE estimator, addressing
at least five issues commonly encountered by existing matching methods
in the causal inference literature.

First, it is well known that the existing matching estimators of the
ATE with a fixed number of matches are inefficient (Abadie and Imbens,
2006) since they do not balance bias and variance in the second-stage
estimation. In comparison, our isotonic matching estimator is more
efficient. In the univariate case, our method attains the semiparametric
efficiency bound; in the multivariate case, where the efficiency bound
becomes more complicated, we show that our proposed estimator performs
better than those based on a fixed number of matches with propensity
scores derived from widely used parametric models such as probit and
logit, which are prevalent in applied research.

Second, although the performance of fixed-number matching estimators
can be improved by increasing the number of matches with the sample
size, the efficiency gain is somewhat artificial (Imbens, 2004) since
the optimal number of matches and data-dependent ways of choosing
it have been open questions. However, these issues are addressed by
recent papers by Armstrong and Kolesár (2021) and Lin \emph{et al.}
(2023). By specifying a large enough Lipschitz constant, Armstrong
and Kolesár (2021) showed that the matching estimator with the number
of matches set to one is minimax optimal if the conditional mean is
restricted to be Lipschitz; by adding an estimated correction term.
Lin \emph{et al.} (2023) gave the optimal number of matches for a
bias-corrected matching estimator. In this paper, we argue that the
isotonic estimator can provide an alternative solution: It gives a
piece-wise monotone increasing estimator, which partitions observations
into different groups. Within these groups, the treated and untreated
observations have the same estimated propensity scores, so they can
be naturally matched to each other without the need of choosing the
number of matches, weights, and relevant distance measures. (For our
method, the distance is zero under any measure.) In contrast, these
choice problems are unavoidable in traditional methods for both covariates
matching and propensity score matching, no matter whether they are
based on the inverse variance matrix (e.g., Abadie and Imbens, 2006)
or the (empirical) density function (e.g., Imbens, 2004) of covariates.
Surprisingly, the set of the matching counterparts adaptively selected
by the isotonic estimator automatically becomes the optimal choice
in the second stage, in that it achieves the semiparametric efficiency
bound of ATE (Hahn, 1998) for the univariate case.

Third, compared to other semiparametric matching methods, where the
first stage propensity score is estimated with kernel or series-based
techniques, our method is more practical in a twofold sense. It is
not only free from the choice of the optimal number of matches, as
mentioned in the second point above, but also does not involve smoothing
parameters of conventional nonparametric methods, such as series length
or bandwidth. In general, choosing the tuning parameters of a first-stage
nonparametric estimator remains a difficult open question in the semiparametric
estimation literature. The MSE optimal tuning parameter is usually
not a good choice since the optimal first-stage estimator of the nuisance
function does not imply the optimality of the second-stage semiparametric
estimation (Bickel and Ritov, 2003). To ensure the $\sqrt{N}-$consistency
of a semiparametric estimator, ``undersmoothed'' tuning parameters
should be applied (Newey, 1994). But it is difficult to find a clear
standard for shrinking tuning parameters below their MSE optimal values.
The non-smooth nature of the isotonic estimator, on the other hand,
turns out to automatically render an adequate amount of undersmoothing.
At the cost of a monotonicity assumption imposed on the nuisance function,
our proposed estimator avoids this choice problem
while still maintaining other desirable properties of a decent semiparametric
estimator, such as $\sqrt{N}-$consistency or efficiency.

Fourth, compared to popular parametric models of propensity scores,
such as probit and logit, our proposed method contains a nonparametric
first stage, so it is more robust to model misspecification. We acknowledge
that combined with a single index structure, the probit and logit
models can also approximate many different data-generating processes.
But our method will always be more robust than them since both probit
and logistic functions are monotone increasing themselves. In other
words, the isotonic regression can well estimate all the data generating
processes that can be well approximated by probit or logit model,
but not vice versa. In addition, this robustness is achieved without
costing the efficiency (compared to parametric methods) of the second-stage
matching estimator.

Fifth, it is well known that the nonparametric bootstrap of the fixed-number
matching estimator is invalid in the presence of continuous covariates
(Abadie and Imbens, 2008). In the past decade, much work has been
done to solve this problem by proposing cleverly structured wild bootstrap
procedures. Otsu and Rai (2017) proposed a consistent wild bootstrap
for covariates matching, and their approach was extended by Bodory
\emph{et al.} (2016) and Adusumilli (2020) to propensity score matching
estimators. In our paper, we show that all these intricate bootstraps
are no longer necessary in the case of monotone increasing propensity
scores since the nonparametric bootstrap inference is asymptotically
valid for our isotonic matching estimator.

Our method relies on the monotonicity assumption on propensity scores.
Monotonicity is a natural shape restriction that can be justified
in many applications in social science, economic studies, and medical
research. Well-known examples in economics include the demand function,
which is usually monotone decreasing in prices, and the supply or
the utility functions, which are often monotone increasing in quantities.
Furthermore, many functions derived from cumulative distribution functions
(CDF) inherit the monotonicity from the latter. For example, in a
threshold-crossing binary choice model 
\begin{equation}
Y=\begin{cases}
1 & \text{if }X^{\prime}\beta_{0}>\varepsilon\\
0 & \text{if }X^{\prime}\beta_{0}\leq\varepsilon
\end{cases},\label{eq:binary}
\end{equation}
the conditional expectation of $Y$ on $X$ can be written as $\mathbb{E}[Y|X]=\mathbb{P}(Y=1|X)=F_{\varepsilon}(X^{\prime}\beta_{0})$,
where $F_{\varepsilon}(\cdot)$ is the CDF of an independent noise
$\varepsilon$. If we assume $\varepsilon\sim N(0,1)$, \eqref{eq:binary}
becomes a probit model; if we assume $\varepsilon\sim\mathrm{Logistic}(0,\frac{\pi^{2}}{3})$,
it becomes a logit model. Although both parametric models are widely
applied in estimating the probability of treatments, we can relax
the distributional assumptions on $\varepsilon$ and express \eqref{eq:binary}
with a semiparametric model $Y=F_{\varepsilon}(X^{\prime}\beta_{0})+\nu$,
where $F_{\varepsilon}(\cdot)$ is a nonparametric link function.
We emphasize that the link function is monotone increasing by construction.
See Cosslett (1983, 1987, 2007), Matzkin (1992), and Klein and Spady
(1993) for more discussions of the model \eqref{eq:binary}.\footnote{Although this paper explores monotonicity of the propensity score
function, our isotonic regression approach may be extended to the
regression-based estimators with monotonicity constraints on the expected
outcome functions $\mathbb{E}[Y(1)|X]$ and $\mathbb{E}[Y(0)|X]$.
However, it should be noted that if monotonicity is imposed on the
link functions of index models, the regression-based approach is clearly
more restrictive than the propensity-score-based approach (because
monotonicity on $F_{\varepsilon}(\cdot)$ is not substantive).}

One of the main challenges of developing the asymptotic properties
of the proposed estimator is the inconsistency of the isotonic estimator
at its boundaries, sometimes called the ``spiking'' problem in the
literature. If the dependent variable is binary, there is a non-trivial
probability for a non-shrinking group of left-end estimates to be
exactly zero even under the strict overlap condition, regardless of
the sample size; the right-end estimates have a similar issue. As
a result, the matched sets for observations at two ends are empty,
and we cannot construct a valid sample analog of ATE. Furthermore,
observations near two ends are matched according to inconsistently
estimated propensity scores, which are biased towards zero or one,
resulting in a detrimental effect on the ATE estimator similar to
the one caused by limited overlaps (Khan and Tamer, 2010; Rothe, 2017; among others). Although truncating those observations,
whose propensity scores (either estimated parametrically or nonparametrically)
are closer to 0 and 1, is widely implemented in applied work, this
strategy has two caveats if one works with the isotonic estimator.
The first problem is the size of truncation: If too little was truncated,
it might be insufficient to correct the boundary problem. A safe choice
of truncation in the literature for different problems involving isotonic
estimators is to truncate the first and last $\alpha_{N}$-th quantile,
with $\alpha_{N}\sim N^{-1/3}$ (or up to a logarithmic factor, see
Wright, 1981; Durot, Kulikov and Lopuhaä, 2013;
and Babii and Kumar, 2021). However, this truncation scheme is too
much for our purpose. In fact, for any $\alpha_{N}$ such that $\alpha_{N}N^{1/2}\to\infty$,
the truncated ATE estimator might be no longer $\sqrt{N}$-consistent.\footnote{This problem is not universal for every semiparametric estimator.
For example, for a partially linear model $Y=X\beta+\psi(Z)+\varepsilon$,
we can truncate more than its $N^{-1/2}$-th quantile, and the estimator
of $\beta$ maintains $\sqrt{N}$-consistency. In fact, one can get
$\sqrt{N}$-rates even if $\beta$ is estimated from an arbitrary
sub-sample with a size proportional to $N$ since different $X$'s
are linked to the same $\beta$. However, for ATE, in general, the
truncated parts directly constitute estimation bias.} Second, as discussed in Appendix \ref{subsec:Uni-rate-of=000020II},
one of the key conditions for $\sqrt{N}$-consistency and efficient
estimation of ATE is \eqref{eq:uni_rate_II} below, but whether this
condition still holds after truncation is unclear. To solve these
two problems, we extend the everywhere-consistent isotonic estimator
of Meyer (2006) to a uniformly consistent isotonic (hereafter, UC-isotonic)
estimator, which is by design to suit our two-stage semiparametric
matching estimator. The proposed estimation procedure does not involve
any truncation, the above-mentioned favorable properties of the isotonic
estimator remain intact, and the full set of data is utilized in both
the first stage estimation of the propensity score and the second
stage estimation of ATE. 

Our proposed method builds on the large literature of causal inference
for covariate and propensity score matching estimators, e.g., Rosenbaum
and Rubin (1983, 1984), Rosenbaum (1989), Heckman, Ichimura and Todd
(1997, 1998), Heckman, Ichimura, Smith and Todd (1998), Dehejia and
Wahba (1999), Abadie and Imbens (2006, 2008, 2011, 2016), Imbens (2004),
Frölich (2004), Frölich, Huber and Wiesenfarth (2017), Otsu and Rai
(2017), Bodory, Camponovo, Huber and Lechner (2016), Adusumilli (2020),
among others. The propensity score matching estimators studied in
the literature mainly use parametrically estimated propensity scores,
such as probit and logit. Our proposed method, in contrast, uses a
special type of nonparametric estimator, the isotonic estimator, to
estimate the propensity score. 

The isotonic estimator has a long history. The earlier work includes
Ayer \emph{et al.} (1955), Grenander (1956), Rao (1969, 1970), and
Barlow and Brunk (1972), among others. The isotonic estimator of a
regression function can be formulated as a least square estimation
with a monotonicity constraint. Suppose that the conditional expectation
$\mathbb{E}[Y|X]=p_{0}(X)$ is monotone increasing. Then, for an iid
random sample $\{Y_{i},X_{i}\}_{i=1}^{N}$, the isotonic estimator
is the minimizer of the sum of squared errors, $\min_{p\in\mathcal{M}}\sum_{i=1}^{N}\{Y_{i}-p(X_{i})\}^{2},$
where $\mathcal{M}$ is the class of monotone increasing functions.
The minimizer can be calculated with the pool adjacent violators algorithm
(Barlow and Brunk, 1972), or equivalently by solving the greatest
convex minorant of the cumulative sum diagram $\{(0,0),(i,\sum_{j=1}^{i}Y_{j}),i=1,\ldots,N\}$,
where the corresponding $\{X_{i}\}_{i=1}^{N}$ are ordered sequence.
See Groeneboom and Jongbloed (2014) for a comprehensive discussion
of different aspects of isotonic regression. 

Our work is linked to the vast literature on semiparametric estimation,
e.g., Chamberlain (1987), Robinson (1988), Newey (1990, 1994), van
der Vaart (1991), Andrews (1994), Hahn (1998), Ai and Chen (2003),
Bickel and Ritov (2003), Chen, Linton and Van Keilegom (2003), Chen
and Santos (2018), among others. In most of the works cited above,
nonparametric methods involving smoothing parameters were applied
at the initial stage, while our work uses the isotonic estimation
that is non-smooth and does not involve smoothing parameters. On the
other hand, the double machine learning estimators (hereafter, DML;
see, e.g., Robins, Rotnitzky and Zhao, 1995; Chernozhukov \emph{et
al}., 2017, 2018; among others) provide efficient estimators of the
ATE that do not rely on subjective choices of smoothing parameters,
thereby, to some extent, sharing many advantages of our approach.
We provide a detailed comparison between the isotonic propensity score
matching estimator and the DML for the ATE in Section \ref{subsec:DML}. 

There are some authors working on concrete semiparametric models with
plug-in isotonic estimators. Huang (2002) studied the properties of
the monotone partially linear model, and his work was extended by
Cheng (2009) and Yu (2014) to the monotone additive model. Balabdaoui,
Durot and Jankowski (2019) studied the monotone single index model
with the monotone least square method, and Groeneboom and Hendrickx
(2018), Balabdaoui, Groeneboom and Hendrickx (2019), and Balabdaoui
and Groeneboom (2021) (the last two papers are called BGH hereafter)
developed a score-type approach for the monotone single index model
and show the single index parameter can be estimated at $\sqrt{N}$-rate.
Building on previous works, Xu (2021) studied a general framework
of semiparametric Z-estimation with plug-in isotonic estimators, monotone
single-index estimators, or monotone additive estimators, and applied
the generic result to inverse probability weighting (IPW) estimators
of ATE. For the augmented IPW (AIPW) model, Qin \emph{et al.} (2019)
and Yuan, Yin and Tan (2021) applied the monotone single index model
to estimate the propensity score, then plugged the estimated propensity
scores with other estimates of potential outcomes into a doubly-robust
moment function. Their asymptotic results rely on the consistent estimations
of both propensity scores and potential outcomes, and thus differ
from our approach.

In terms of applying isotonic regression to estimate the ATE, the
primary difference between this paper and Chapter 3 of Xu (2021) is
that we address the boundary issue inherent in the isotonic estimator,
while Xu (2021) relies on a stronger assumption adapted from Assumption
5.1 in Newey (1994). In the process of writing this paper, we have
gradually realized that this assumption does not automatically apply
to the IPW estimator, although it straightforwardly holds for some
other semiparametric models, such as the monotone partially linear
model and the monotone single index model, wherein the plugged-in
isotonic estimator is not in the denominator. Compared to Chapter
3 of Xu (2021), the main contributions of this paper are: (i) proposing
a UC-isotonic estimator that is suitable as the first-stage estimator
in a propensity score matching estimator of the ATE; (ii) revealing
the equivalence between the matching estimator and the IPW estimator
when the first-stage propensity score is estimated via UC-isotonic
regression; and (iii) based on this equivalence, enriching the literature
on propensity score matching by introducing a new approach that addresses
several problems of the existing matching methods, as detailed at
the beginning of this introduction.\footnote{At almost the same time, an independent work by Liu and Qin (2022)
derived a similar equivalence result for the average treatment effect
on treated (ATT). Recently, a revised version of Liu and Qin (2022)
is published as Liu and Qin (2024). There are two main differences
between our paper and their papers. First, we formally address the
boundary problem of the isotonic estimator and achieve the $\sqrt{N}$-normality
of the ATE estimator by proposing a uniformly consistent isotonic
estimator. Second, our asymptotic analysis of the model with multivariate
covariates in Section \ref{sec:Multi} focuses on a more general case,
where the influence of the estimation errors from the parametric component
of the first-stage monotone single index model is maintained.}

The rest of the paper is organized as follows. After introducing the
setting and notations, Section \ref{sec:main} shows the implementation
and asymptotic properties of the proposed isotonic matching estimator
with a univariate covariate. Section \ref{sec:Comparison} compares
our approach with existing matching estimators as well as the double
machine learning estimator for the ATE. The univariate results are
extended to the case of multivariate covariates in Section \ref{sec:Multi},
where the propensity score is modeled by a semiparametric single-index
model with an unknown monotone increasing link function. In Section
\ref{sec:Boot}, we establish the validity of the nonparametric bootstrap.
Monte-Carlo simulation studies are presented in Section \ref{sec:Monte-Carlo}.
All proofs are presented in Appendix, while additional theoretical
details and simulation comparisons are provided in Supplementary Material.

\section{Main results\protect\label{sec:main}}

\subsection{Setup and isotonic propensity score\protect\label{subsec:setup}}

Suppose we observe the triple $(Y,W,X)$ drawn randomly from the product
space $\mathcal{Z}=\mathbb{R}\times\{0,1\}\times\mathcal{\mathcal{X}}$.
Within the triple, $W\in\{0,1\}$ is a binary treatment variable,
$Y=W\cdot Y(1)+(1-W)\cdot Y(0)$ is an outcome variable with potential
outcomes $Y(1)$ and $Y(0)$ for $W=1$ and $0$, respectively, and
$X$ is a scalar covariate with continuous domain $\mathcal{X}=[x_{L},x_{U}]\subset\mathbb{R}$.
In this section, we tentatively assume $X$ is scalar, and discuss
extensions for multivariate $X$ in Section \ref{sec:Multi}. Without
loss of generality, $\{Y_{i},W_{i},X_{i}\}_{i=1}^{N}$ is an iid sample
of $(Y,W,X)$ and is ordered by $X$. If $X$ is continuously distributed,
we should have $X_{1}<X_{2}<\cdots<X_{N}$ with probability one (i.e.,
no ties).

In this section, we consider the estimation of the ATE, $\tau=\mathbb{E}[Y(1)-Y(0)]$,
by matching the propensity score $p(x)=\mathbb{P}(W=1|X=x)=\mathbb{E}[W|X=x]$,
where $p(\cdot)$ is an unknown monotone increasing function. In particular,
we estimate $p(\cdot)$ by the isotonic estimator 
\begin{equation}
\hat{p}(\cdot)=\arg\min_{p\in\mathcal{M}}\sum_{i=1}^{N}\{W_{i}-p(X_{i})\}^{2},\label{eq:iso}
\end{equation}
where $\mathcal{M}$ is the class of all monotone increasing functions
defined on $\mathcal{X}$. Since Brunk (1958), this isotonic regression
estimator has been extensively studied in the statistics literature
(see, e.g., Barlow \emph{et al.}, 1972, and Groeneboom and Jongbloed,
2014, for an overview). One of the well-known features of isotonic
regression is that the estimator $\hat{p}(\cdot)$ is a monotone increasing
piecewise constant function with jump points at $\{X_{n_{k}}\}_{k=1}^{K}$
for some integer $K$ with $1\leq K\le N$. By these jump points,
the sample is divided into $K$ disjoint groups, with $\{n_{k}\}_{k=1}^{K}$
denoting the first indices of these $K$ groups. Further, we let $N_{k}$
denote the number of observations belonging to the $k$-th group.
Based on these definitions, it holds that $n_{k}+N_{k}=n_{k+1}$ for
each $k=1,\ldots,K-1$, and $\sum_{k=1}^{K}N_{k}=N$. Note that the
integer $K$ and the corresponding disjoint groups are automatically
determined by the isotonic estimation algorithm (see the formula (\ref{eq:uc_isoton})
below), rather than being chosen by the user. 

To avoid ambiguity caused by splitting a flat piece into several sub-pieces
with the same estimated value, we impose
\begin{equation}
\hat{p}(X_{n_{1}})<\hat{p}(X_{n_{2}})<\cdots<\hat{p}(X_{n_{K}}),\label{eq:unique}
\end{equation}
to ensure uniqueness of this partition (i.e., if $\hat{p}(X_{n_{k}})=\hat{p}(X_{n_{k+1}})$,
we simply combine the groups $k$ and $k+1$). Then, the isotonic
estimator $\hat{p}(\cdot)$ is characterized as follows.

\begin{asm} \label{asm:1} {[}Sampling{]} $\{Y_{i},W_{i},X_{i}\}_{i=1}^{N}$
is an independent and identically distributed (iid) sample of $(Y,W,X)\in\mathbb{R}\times\{0,1\}\times\mathcal{X}$,
where $\mathcal{X}=[x_{L},x_{U}]\in\mathbb{R}$. $X$ is continuously
distributed, and the sample $\{Y_{i},W_{i},X_{i}\}_{i=1}^{N}$ is
indexed according to $X_{1}<X_{2}<\cdots<X_{N}$. \end{asm}

\begin{prop} \label{prop:disjoint-grouping} Under Assumption \ref{asm:1},
the isotonic estimator $\hat{p}(\cdot)$ satisfying \eqref{eq:unique}
partitions the sample into $K$ disjoint groups in the sense that
for each $k=1,\ldots,K$ and $i=n_{k},\ldots,n_{k}+N_{k}-1$,
\begin{equation}
\hat{p}(X_{i})=\frac{1}{N_{k}}\sum_{j=n_{k}}^{n_{k}+N_{k}-1}W_{j}.\label{eq:iso_estimate_k}
\end{equation}
\end{prop}

To define our propensity score matching estimator based on $\hat{p}(\cdot)$,
let $N_{k,1}$ and $N_{k,0}$ denote the numbers of treated and controlled
observations within group $k$, i.e., $N_{k,1}=\sum_{i=n_{k}}^{n_{k}+N_{k}-1}W_{i}$
and $N_{k,0}=N_{k}-N_{k,1}$. Our one-to-many matching method is implemented
within each of these $K$ groups, and each treated (controlled) observation
in group $k$ will be matched with its $N_{k,0}$ ($N_{k,1}$) counterparts,
which belong to the same group and have the same value of the estimated
propensity score. The following results directly follow from Proposition
\ref{prop:disjoint-grouping}.

\begin{prop} \label{prop:matching-score}\textbf{ }Suppose Assumption
\ref{asm:1} holds.
\begin{description}
\item [{(i)}] {[}Isotonic estimator{]} For each integer $k=1,\ldots,K$,
the isotonic estimator $\hat{p}(\cdot)$ for $p(\cdot)$ is represented
as 
\begin{equation}
\hat{p}(x)=\frac{N_{k,1}}{N_{k}},\label{eq:PS_score}
\end{equation}
 for each $x\in\{X_{i}\}_{i=n_{k}}^{n_{k}+N_{k}-1}$.
\item [{(ii)}] {[}Existence of matching counterparts{]} For any $i\in\{1,\ldots,N\}$
with $0<\hat{p}(X_{i})<1$, the set of its matching counterparts $\{j:W_{j}=1-W_{i},\hat{p}(X_{j})=\hat{p}(X_{i})\}$
is non-empty.
\end{description}
\end{prop}

Before we proceed, we need to solve the problem of the potential lack
of matching counterparts for those $i$'s with $\hat{p}(X_{i})=0$
or $1$. Under the strict overlaps (in Assumption \ref{asm:3} below),
the problem is essentially associated with the inconsistency of the
isotonic estimator at the boundary. In the next subsection, we propose
a modified isotonic estimator that is uniformly consistent on $\mathcal{X}$. 

\subsection{Uniformly consistent isotonic estimator\protect\label{subsec:Uniformly-consistent-isotonic}}

Like other nonparametric estimators, the isotonic estimator is imprecise
at the boundary. If we apply the isotonic estimator to the binary
dependent variable $W$, there is a non-trivial probability of $\hat{p}(X_{i})=0$
or $1$ even if the true propensity score $p(x)$ is bounded away
from zero and one for all $x\in\mathcal{X}$. For example, if $\hat{p}(X_{1})=0$,
\eqref{eq:PS_score} implies $N_{1,1}=0$, i.e., there are no matching
counterparts for the treated units.

To fix this problem, we propose a modified isotonic estimator that
is uniformly consistent in its domain at a $\left(\log N\right)^{1/3}N^{-1/3}$
rate and is easy to implement. For the sample $\{W_{i},X_{i}\}_{i=1}^{N}$
with $X_{1}<\cdots<X_{N}$, we transform $\{W_{i}\}_{i=1}^{N}$ into
$\{\tilde{W}_{i}\}_{i=1}^{N}$ by averaging its first and last $\lfloor N^{2/3}\rfloor$
observations:
\begin{equation}
\tilde{W}_{i}=\begin{cases}
\frac{1}{\lfloor N^{2/3}\rfloor}\sum_{i=1}^{\lfloor N^{2/3}\rfloor}W_{i} & \text{for }i\leq\lfloor N^{2/3}\rfloor\\
W_{i} & \text{for }\lfloor N^{2/3}\rfloor<i\leq N-\lfloor N^{2/3}\rfloor\\
\frac{1}{\lfloor N^{2/3}\rfloor}\sum_{i=N-\lfloor N^{2/3}\rfloor+1}^{N}W_{i} & \text{for }i>N-\lfloor N^{2/3}\rfloor.
\end{cases}\label{eq:data_transform}
\end{equation}
Our proposed UC-isotonic estimator is obtained by implementing the
standard isotonic regression of $\tilde{W}$ on $X$:
\begin{equation}
\tilde{p}(x)=\begin{cases}
\max_{s\le i}\min_{t\ge i}\sum_{j=s}^{t}\tilde{W}_{j}/(t-s+1) & \text{for }x=X_{i}\\
\tilde{p}(X_{i}) & \text{for }X_{i-1}<x\leq X_{i}\\
\tilde{p}(X_{N}) & \text{for }x>X_{N}.
\end{cases}\label{eq:uc_isoton}
\end{equation}

A similarly modified estimator was proposed by Meyer (2006), where
she averaged the first and last $\lceil\log(N)\rceil$ dependent variables
instead of the first and last $\lfloor N^{2/3}\rfloor$ ones. The
choices are different because she focuses on the consistency of the
isotonic estimator itself, while we are interested in the performance
of the second-stage matching estimator. To achieve an $N^{-1/2}$
rate at the second stage, we need the isotonic estimator to be uniformly
consistent at a rate faster than $N^{-1/4}$, which won't be achieved
under Meyer's choice. Meyer (2006) presented a theorem
regarding the consistency of the modified estimator at the boundary;
however, a proof of consistency was not provided, nor was the rate
of convergence discussed.

In this paper, we formally establish the uniform convergence rate
of the modified isotonic estimator $\tilde{p}(\cdot)$. To this end,
we impose the following assumption. 

\begin{asm} \label{asm:2} {[}Monotonicity and continuity{]} (i)
$p(x)=\mathbb{E}[W|X=x]$ is a monotone increasing function of $x\in\mathcal{X}$,
(ii) $p(x)$ is continuously differentiable with its first derivative
$p^{(1)}(x)>0$ for all $x\in\mathcal{X}$, and (iii) $X$ has a continuous
density $f(x)$ satisfying that for some positive constants $\overline{f}$
and $\underline{f}$, it holds $\underline{f}<f(x)<\overline{f}$
all $x\in\mathcal{X}$. \end{asm}

Assumption \ref{asm:2} (i) is our main assumption, the monotonicity
of $p(\cdot)$. Assumption \ref{asm:2} (ii) is required for the $\sqrt{N}-$consistency
of the second-stage matching estimator. The same assumption has been
adopted by Groeneboom and Hendrickx (2018; Assumption A2) and by BGH
(Assumption A3 and its accompanying remark; see also Lemma 22 in the
supplementary material of BGH) in the context of the monotone single
index model. If we believe that the underlying propensity score function
has some flat parts where $p^{(1)}(x)=0$, we could first run an isotonic
estimation of $\tilde{W}_{i}+c\cdot X_{i}$ on $X_{i}$, where $c$
is a positive constant, to obtain $\tilde{p}_{c}(x)$. Then, by subtracting
the linear trend $c\cdot x$ from $\tilde{p}_{c}(x)$, we obtain a
consistent estimator of $p(x)$.\footnote{Technically, if $p(\cdot)$ has some flat parts where
$p^{(1)}(x)=0$, then the original estimator $\tilde{p}(\cdot)$ may
not satisfy the requirement in \eqref{eq:ebar-C} in Appendix \ref{appsub:Proof-of-T_univariate},
$|\delta(x)-\bar{\delta}_{N}(x)|\leq C_{0}|p(x)-\tilde{p}(x)|$. Flat
parts in $p(\cdot)$ imply that $p(x)-\tilde{p}(x)=0$ might hold
within an entire interval, potentially leading to a violation of \eqref{eq:ebar-C}.
In contrast, if $p(\cdot)$ is strictly monotone increasing, then
$p(\cdot)$ and $\tilde{p}(\cdot)$ will cross at most once within
each partition given by the isotonic estimator, since $\tilde{p}(\cdot)$
is a piecewise flat function. We refer to Sections 10.2-10.3 and Figure
10.1 of Groeneboom and Jongbloed (2014) for more details.} Assumption \ref{asm:2} (iii) imposes an upper and lower bound for
the density of $X$.

To avoid unnecessarily repeatedly defined notations, we let the same
set of notations, $K,$ $N_{k,1}$, $N_{k}$, and $n_{k}$, denote
the number of groups, the number of treated observations in group
$k$, the number of members in group $k$, and the index of the first
element of group $k$, under the grouping scheme given\emph{ }by the
UC-isotonic estimator $\tilde{p}(\cdot)$ ($N_{k,1}$ is calculated
with the original treatment variable $\{W_{i}\}_{i=1}^{N}$). We obtain
an analogous result to Proposition \ref{prop:matching-score} for
the UC-isotonic estimator.

\begin{prop} \label{prop:matching-score-modified} Under Assumptions
\ref{asm:1} and \ref{asm:2}, it holds
\begin{description}
\item [{(i)}] $N_{1}\geq\lfloor N^{2/3}\rfloor$ and $N_{K}\geq\lfloor N^{2/3}\rfloor$.
\item [{(ii)}] $\tilde{p}(x)=\frac{N_{k,1}}{N_{k}}$ for each $k=1,\dots,K$
and $x\in\{X_{i}\}_{i=n_{k}}^{n_{k}+N_{k}-1}.$
\item [{(iii)}] $N_{1}=O_{p}(N^{2/3})$ and $N_{K}=O_{p}(N^{2/3}).$
\end{description}
\end{prop}

Part (i) of this proposition says that all the averaged $W_{i}$'s
at the beginning and end of the data are absorbed in the first and
the last group. Part (ii) provides an analogous representation of
the UC-isotonic estimator $\tilde{p}(\cdot)$ as $\hat{p}(\cdot)$.
While Part (i) gives a lower bound of the sizes of the first and the
last group, Part (iii) gives (stochastic) upper bounds of them. Based
on this proposition, the uniform convergence rate of the UC-isotonic
estimator is obtained as follows.

\begin{thm} \label{thm:uc_isoton} Under Assumptions \ref{asm:1}
and \ref{asm:2}, it holds
\[
\sup_{x\in\mathcal{X}}|\tilde{p}(x)-p(x)|=O_{p}\left(\frac{\log N}{N}\right)^{1/3}.
\]
\end{thm}

Finally, to guarantee the existence of matching counterparts by $\tilde{p}(\cdot)$,
we impose the strict overlap condition.

\begin{asm} \label{asm:3} {[}Strict overlaps{]} There exist positive
constants $\underline{p}$ and $\bar{p}$ such that $0<\underline{p}\leq p(x)\leq\bar{p}<1$
for all $x\in\mathcal{X}$. \end{asm}

Assumption \ref{asm:3} is standard in the treatment effect literature.
It is necessary for the identification and $\sqrt{N}$-consistent
estimation of the ATE. Combining Proposition \ref{prop:matching-score-modified}
and Theorem \ref{thm:uc_isoton} with Assumption \ref{asm:3}, the
existence of the matching counterparts by $\tilde{p}(\cdot)$ is obtained
as follows.

\begin{cor} \label{cor:non-empty-matched-set} Suppose Assumptions
\ref{asm:1}-\ref{asm:3} hold. For each $i=1,\dots,N$, the set of
its matching counterparts $\{j=1,\dots,N:W_{j}=1-W_{i},\tilde{p}(x_{j})=\tilde{p}(x_{i})\}$
is non-empty with probability approaching one. \end{cor}

\subsection{Isotonic propensity score matching\protect\label{subsec:Isotonic-matching}}

Based on the UC-isotonic estimator $\tilde{p}(\cdot)$, the isotonic
propensity score matching estimator for the ATE $\tau$ can be implemented
as follows.
\begin{enumerate}
\item Transform the sample $\{Y_{i},W_{i},X_{i}\}_{i=1}^{N}$ indexed by
$X_{1}<\cdots<X_{N}$ into $\{Y_{i},\tilde{W}_{i},X_{i}\}_{i=1}^{N}$
using \eqref{eq:data_transform}.
\item Compute the UC-isotonic estimator $\tilde{p}(\cdot)$ using \eqref{eq:uc_isoton}.
\item For each $i=1,\ldots,N$, compute the matching counterparts
\begin{equation}
\mathcal{J}(i)=\left\{ j=1,\dots,N:W_{j}=1-W_{i}\text{ and }\tilde{p}(X_{j})=\tilde{p}(X_{i})\right\} .\label{eq:matching}
\end{equation}
\item Calculate the matching estimator for the ATE $\tau$ by
\begin{equation}
\hat{\tau}=\frac{1}{N}\sum_{i=1}^{N}(2W_{i}-1)\left(Y_{i}-\frac{1}{M_{i}}\sum_{j\in\mathcal{J}(i)}Y_{j}\right),\label{eq:ATE=000020sample}
\end{equation}
where $M_{i}=|\mathcal{J}(i)|$ is the number of matches for $i$.
\end{enumerate}
We proceed with the following assumptions.

\begin{asm} \label{asm:4} {[}Data generating process{]} (i) $\mathbb{E}[Y(0)^{2}]<\infty$
and $\mathbb{E}[Y(1)^{2}]<\infty$, (ii) $\mathbb{E}[Y(0)|X=x]$ and
$\mathbb{E}[Y(1)|X=x]$ are continuously differentiable for all $x\in\mathcal{X}$,
(iii) for $D(Z)=\frac{WY}{p(X)^{2}}+\frac{Y(1-W)}{\{1-p(X)\}^{2}}$,
there exist positive constants $c_{0}$ and $M_{0}$ such that $\mathbb{E}[|D(Z)|^{m}|X=x]\leq m!M_{0}^{m-2}c_{0}$
holds for all integers $m\geq2$ and every $x$, and (iv) $Y(1),Y(0)\perp W|X$
almost surely. \end{asm}

Assumption \ref{asm:4} (i)-(iii) regulates the tail behaviors of
the (conditional functions of) potential outcomes, which are necessary
for $\sqrt{N}$-consistent estimation. Assumption \ref{asm:4} (iv)
is the standard unconfoundedness assumption. Under these assumptions,
we have the following key equivalence result.

\begin{thm} \label{thm:matching_IPW} Under Assumptions \ref{asm:1}-\ref{asm:4},
the matching estimator $\hat{\tau}$ for the ATE $\tau$ using the
UC-isotonic estimator $\tilde{p}(\cdot)$ is equal to the corresponding inverse probability weighting (IPW) estimator. \end{thm}

\subsubsection*{Remark on Theorem \ref{thm:matching_IPW}}

Imbens (2004) pointed out that with $M\to\infty$ and $M/N\to0$,
the matching estimator is essentially like a regression estimator.
In comparison, we find out that with propensity scores estimated by
the UC-isotonic estimator, the (propensity score) matching estimator
is numerically equal to the weighting estimator in each finite sample.
This equivalence is tightly associated with the fact that the isotonic
estimator can be regarded as a type of partitioning estimator (e.g.,
Györfi \emph{et al.}, 2002; Cattaneo and Farrell, 2013). See Section
\ref{subsec:link-partition} below for a further comparison of isotonic
and partitioning estimators within the context of a two-stage matching
estimator of the ATE. Additionally, our method is related to the propensity
score methods of blocking, stratification, and radius matching (Rosenbaum
and Rubin, 1983, 1985; Dehejia and Wahba, 1999, 2002; among others).
See Section \ref{subsec:block-strat-radius} for a comparison with
these methods.

Moreover, as mentioned in the introduction, the equivalence result
in Theorem \ref{thm:matching_IPW} relies crucially on the implementation
of the UC-isotonic estimator \eqref{eq:uc_isoton}, which guarantees
that both the matching and IPW estimators at the second stage are
well-defined.

We notice that the threshold $\lfloor N^{2/3}\rfloor$ in the algorithm
\eqref{eq:data_transform} can be interpreted as an implicit tuning
parameter. We would like to point out that, first, it is convenient
to choose since it depends only on the sample size $N$; second, it
is aimed at correcting the boundary problem, which is also faced by
other semiparametric and even parametric matching methods. In practice,
trimming estimated propensity scores is widely adopted, and the amount
of trimming is chosen subjectively in most cases. Our proposed method
provides transparent guidance for correcting this common boundary
issue. Furthermore, to investigate the impact of different threshold
choices, we have included both theoretical analysis and simulation
evidence in Sections \ref{supp:thresh_theorey} and \ref{supp:thresh_sim}
of the supplementary material, respectively.

Our main result, consistency and asymptotic normality of the isotonic
propensity score matching estimator, is obtained as follows.

\begin{thm} \label{thm:ATE-matching} Under Assumptions \ref{asm:1}-\ref{asm:4},
it holds $\hat{\tau}\overset{p}{\to}\tau$ and
\[
\sqrt{N}(\hat{\tau}-\tau)\overset{d}{\to}N(0,\Omega),
\]
where $\Omega=\mathbb{V}(\mathbb{E}[Y(1)-Y(0)|X])+\mathbb{E}[\mathbb{V}(Y(1)|X)/p(X)]+\mathbb{E}[\mathbb{V}(Y(0)|X)/(1-p(X))]$.
\end{thm}

We note that the asymptotic variance $\Omega$ is the semiparametric
efficiency bound for $\tau$ (see e.g., Hahn, 1998, and Hirano, Imbens
and Ridder, 2003). Although we may conduct inference based on an estimator
of $\Omega$, we suggest a bootstrap inference method, which will
be discussed in Section \ref{sec:Boot}.

\section{Comparison to related propensity score methods\protect\label{sec:Comparison}}

In this section, we draw comparisons of our approach with a range
of related estimators for the ATE. The comparison with matching methods
based on propensity score estimated by partitioning estimator is presented
in Section \ref{subsec:link-partition}, the comparison with propensity
score methods of blocking, stratification, and radius matching is
presented in Section \ref{subsec:block-strat-radius}, the comparison
with matching methods based on propensity score estimated by regression
trees is presented in Section \ref{subsec:trees}, and the comparison
with the double machine learning (DML) estimator for the ATE can be
found in Section \ref{subsec:DML}.

\subsection{Propensity score estimated by partitioning estimator\protect\label{subsec:link-partition}}

One notable feature of the proposed isotonic propensity score matching
method is that it is a one-to-many matching method that provides exact
matches, as illustrated by the formula \eqref{eq:matching}. This
is attributed to the isotonic estimator being considered a special
type of partitioning estimator, in which the volume sizes of partitions
are automatically chosen by the monotonicity constraint, and a simple
average is implemented within each partition. 

The partitioning estimator is a nonparametric method for estimating
regression functions.\footnote{We refer to Györfi \emph{et al.} (2002) and Cattaneo and Farrell (2013)
for comprehensive discussions of the partitioning estimator.} It divides the domain of the running variables into disjoint partitions.
Within each partition, a local estimator is implemented by the user,
such as the sample mean, a linear estimator, or a series estimator.
Each sample point is exclusively used in the estimation within the
partition to which it belongs. This feature simplifies the complex
correlation structure of a matching estimator such that it achieves
equivalence with an IPW estimator. For the UC-isotonic estimator,
this equivalence is presented by equation \eqref{eq:IPW_tau_ATE}
in Appendix \ref{appsub:equivalent-result}. In the resulting matching
estimator of ATE, the same set of partitions serves both the first-
and the second-stage nonparametric estimation. Usually, these two
stages are not associated with each other since they have distinct
objects, the propensity score and the potential outcomes. Certainly,
a matching estimator of the ATE that utilizes propensity scores estimated
with a partitioning estimator should exhibit a similar equivalence
to the weighting estimator. However, the selection of the number of
partitions and their sizes necessitates careful consideration, as
they must meet specific undersmoothing conditions to secure the desired
asymptotic properties of the second-stage ATE estimator. The challenge
of selecting an appropriate undersmoothed bandwidth or volume size,
as mentioned in the introduction, remains a difficult open question
in the semiparametric estimation. In contrast, our proposed isotonic
matching estimator automatically chooses these tuning parameters,
leading to the efficient estimation of the ATE, as demonstrated in
Theorem \ref{thm:ATE-matching}.

\subsection{Propensity score methods of blocking, stratification, and radius
matching\protect\label{subsec:block-strat-radius}}

Our proposed isotonic matching estimator is also related to some of
the seminal ideas introduced at the outset of the propensity score
methods: blocking, stratification (Rosenbaum and Rubin, 1983; Dehejia
and Wahba, 1999, 2002), and radius matching (Rosenbaum and Rubin,
1985).

The isotonic propensity score matching shares similarities with blocking
and stratification matching on propensity scores, notably: (i) they
initially categorize data points into distinct groups (or strata,
blocks, partitions) according to estimated propensity scores, and
(ii) within each group, they calculate the conditional average treatment
effect as the simple difference in means of outcomes between the treatment
and comparison groups. The primary distinction lies in the grouping
mechanism: for isotonic propensity score matching, the groups are
determined adaptively in a data-driven manner through isotonic regression,
whereas for the stratification estimator of the ATE, the strata must
be explicitly specified by the user. Another distinction is that for
the isotonic propensity score matching method, the same set of partitions
is utilized for both the first and second stages of nonparametric
estimation. As presented by Theorem \ref{thm:matching_IPW}, this
characteristic leads to the equivalence between the matching and IPW
estimator, resulting in the efficient estimation of the ATE. In contrast,
in the case of blocking or stratification matching methods, particularly
when the propensity score is estimated using parametric models, this
equivalence cannot generally be established, and efficiency cannot
be assured without implementing some bias correction method.

The case for the radius matching estimator is similar to the stratified
matching estimator. The difference is that for stratified matching,
each unit is matched solely with units from the opposite treatment
group within the same stratum, while radius matching allows each unit
to be matched to several local balls, the centers of which belong
to the opposite treatment group. For both radius and stratified matching
estimators, the sizes of strata or the radii act as tuning parameters,
which must be chosen by the users when the propensity score is estimated
via parametric or nonparametric methods dependent on smoothing parameters
(such as kernel or series estimation). These smoothing parameters
play a key role in balancing the bias and variance, thereby significantly
affecting the second-stage estimator of ATE. In contrast, isotonic
regression distinguishes itself by automatically generating these
partitions through the application of the monotonicity constraint.

\subsection{Propensity score estimated by regression trees\protect\label{subsec:trees}}

As methods of estimating the propensity scores, the isotonic estimator
and regression trees share several similarities. First, both are nonparametric
estimators that do not impose restrictive parametric structures on
the underlying response function. Second, both approaches partition
the domain of running variables (the feature space in regression tree
terminology) into several regions and use the sample average within
each region as estimators. As a result, both estimators take the form
of piecewise-constant functions. Third, both methods form their piecewise-constant
functions in data-adaptive manners. In particular, the partitions
created by both methods depend on the dependent variable (the response),
which differentiates them from regular nonparametric methods, such
as the kernel estimator.

On the other hand, there are notable distinctions between the two
methods. First, both approaches construct their piecewise-constant
functions differently: the partitions in a regression tree are obtained
in a stepwise manner. In each step, a partition is chosen to achieve
the maximum marginal reduction of the mean square error (MSE), without
imposing any shape constraints during this process. In contrast, the
isotonic estimator employs a one-step approach that determines partitions
to minimize the MSE over the class of monotone functions. Second,
although both approaches are data-driven, the isotonic estimator is
free of smoothing parameters, whereas the regression tree depends
on the user to specify the tree's length. (When the tree length is
determined by cross-validation, the user must select the penalty parameter.)
Third, the regression tree is inherently designed for multi-dimensional
problems, whereas the canonical form of isotonic regression addresses
one-dimensional issues, given that the traditional definition of monotonicity
characterizes the relationship between two variables. Nevertheless,
the isotonic estimation can be extended to multivariate cases by being
incorporated into a partially linear model or a monotone single index
model. The latter is illustrated in Section \ref{sec:Multi} below.

To summarize, the isotonic estimator necessitates the monotonicity
assumption in the underlying response function, a requirement not
shared by regression trees. This assumption, however, enables the
isotonic estimation algorithm to automatically regulate the trade-off
between bias and variance. Conversely, when using regression trees,
practitioners are tasked with the challenge of selecting an appropriate
tree length to effectively manage the balance between bias and the
risk of overfitting. The strength of regression trees is their natural
aptitude for tackling multivariate problems. When employing regression
trees in the preliminary stage of propensity score estimation as part
of a two-stage approach to estimating the ATE, it is commonly combined
with methods for bias correction and sample splitting, as discussed
by Chernozhukov \emph{et al.} (2018). See Section \ref{subsec:DML}
below for more details about the comparison of our approach with the
double machine learning estimator. 

\subsection{Double machine learning estimator\protect\label{subsec:DML}}

The isotonic propensity score matching estimator and the DML estimator
for the ATE both share the benefit of not requiring subjective choices
of tuning parameters. For estimating the ATE, a typical example of
a DML estimator is given by applying the sample splitting to the augmented
inverse probability weighting (AIPW) estimator. In the following,
we abstract from sample splitting to simplify notation:
\begin{eqnarray}
 &  & \frac{1}{N}\sum_{i=1}^{N}\{\hat{\psi}_{1}(X_{i})-\hat{\psi}_{0}(X_{i})\}+\frac{1}{N}\sum_{i=1}^{N}\left[\frac{W_{i}(Y_{i}-\hat{\psi}_{1}(X_{i}))}{\hat{p}(X_{i})}-\frac{(1-W_{i})(Y_{i}-\hat{\psi}_{0}(X_{i}))}{1-\hat{p}(X_{i})}\right]\nonumber \\
 & = & \frac{1}{N}\sum_{i=1}^{N}\left[\frac{Y_{i}W_{i}}{\hat{p}(X_{i})}-\frac{Y_{i}(1-W_{i})}{1-\hat{p}(X_{i})}\right]-\frac{1}{N}\sum_{i=1}^{N}\left[\frac{W_{i}-\hat{p}(X_{i})}{\hat{p}(X_{i})}\hat{\psi}_{1}(X_{i})-\frac{W_{i}-\hat{p}(X_{i})}{1-\hat{p}(X_{i})}\hat{\psi}_{0}(X_{i})\right],\label{eq:AIPW}
\end{eqnarray}
where $\hat{\psi}_{1}$$(\cdot)$ and $\hat{\psi}_{0}(\cdot)$ are
estimators of $\mathbb{E}[Y(1)|X=\cdot]$ and $\mathbb{E}[Y(0)|X=\cdot]$,
respectively. The first and second lines of \eqref{eq:AIPW} present
two formulations of the DML estimator for the ATE. The first terms
in both lines correspond to the standard regression and IPW estimators,
respectively, while the subsequent terms represent their bias-correction
components.

The AIPW has been extensively studied since the seminal work of Robins,
Rotnitzky and Zhao (1995), Robins and Rotnitzky (1995); see also Newey,
Hsieh, and Robins (1998, 2004), Scharfstein, Rotnitzky and Robins
(1999), Rothe and Firpo (2019), among others. In an influential work,
Chernozhukov \emph{et al}. (2018) combined orthogonal moment functions
– of which the formula \eqref{eq:AIPW} is a specific case for the
ATE – with sample splitting, accommodating a broad array of the first-stage
machine learners that are prone to bias due to regularization or model
selection. Recent developments by Chernozhukov \emph{et al. }(2022)
and Chernozhukov, Newey and Singh (2022) have proposed methods for
constructing the correction term without requiring an explicit function
form for the bias correction.

Both estimators have their own advantages and comparative strengths.
From a practical standpoint, the isotonic propensity score matching
method stands out for its simplicity and ease of implementation: it
does not require the correction terms, thereby sparing the effort
of estimating the conditional means of potential outcomes and sidesteps
the challenges associated with their correct specification. In contrast,
the DML estimator’s efficiency relies on correctly specifying and
effectively estimating both the propensity score and the conditional
means of potential outcomes. A misstep in either leads to a consistent
yet inefficient estimator. On the other hand, the DML estimator exhibits
great flexibility: through the use of sample splitting, it supports
a variety of first-stage estimators, accommodating high dimensional
data or highly complex function classes, such as random forest, neural
networks, and other advanced machine learning technologies. 

From a technical standpoint, the isotonic propensity score matching
and the DML for the ATE represent two distinct pathways of semiparametric
estimation: undersmoothing and bias correction. Both strategies aim
for $\sqrt{N}$-consistent (or efficient in certain cases) estimators
(see Newey, 1994, for a relevant discussion). The undersmoothing strategy
depends on a first-stage estimator with reduced bias, achievable in
nonparametric estimators by selecting smoothing parameters smaller
than the MSE-optimal levels. Conversely, the bias correction method
addresses bias by incorporating an estimated correction term into
the second-stage sample moment function, rather than concentrating
on the first stage. 

The proposed isotonic propensity score matching estimator utilizes
the isotonic estimator, which achieves a similar effect of ``undersmoothing'',
and this effect is automatically rendered by enforcing monotonicity.
The isotonic estimator does not really shrink its bias to a level
lower than $N^{-1/2}$. However, when combined with the monotonicity,
it eventually achieves a deviation from the efficient influence function
that decays at a rate faster than $N^{-1/2}$ (see \eqref{eq:uni_rate_II}
in Appendix). In contrast, the DML for the ATE represents a typical
bias correction approach. The second terms in both lines of \eqref{eq:AIPW},
while achieving the ``doubly robust'' effect, also serve as bias-correction
components. At the cost of computing additional correction terms and
some efficiency loss due to sample splitting, the DML approach manages
to mitigate potential bias and prevent overfitting risks, while being
less restrictive on the first-stage estimation. It is not only less
sensitive to the choice of the smoothing parameter for the traditional
first-stage nonparametric estimator but can also accommodate many
black-box machine learning methods, whose asymptotic properties remain
to be fully understood. Consequently, the theoretical development
of the isotonic propensity score matching and the DML for the ATE
differs substantially. The DML approach significantly reduces the
effort needed to address issues arising from the complexity of function
classes, which is associated either with the correlation brought by
plug-in estimators or with the choice of smoothing parameter. In contrast,
this paper needs to address the impact of the plug-in estimator in
the theoretical development of the isotonic propensity score matching
estimator.

Finally, we would like to emphasize that our proposed method represents
a targeted advancement within the matching estimation literature,
specifically addressing several limitations present in existing matching
techniques for estimating the ATE. In contrast, the DML is a versatile
tool designed for broader semiparametric estimation tasks, which include
a wide array of econometric problems such as average derivatives,
partially linear models, and parameters of economic structural models.
Our approach, therefore, complements rather than competes with the
expansive toolkit that DML offers, by providing subtle yet significant
improvements in the specialized area of matching estimation.

\section{Multivariate covariates \protect\label{sec:Multi}}

Certainly, researchers are more interested in models with multivariate
covariates $X$. One way to balance the robustness and the curse of
dimensionality is to estimate the propensity score with the monotone
single-index model: 
\begin{equation}
W=p_{0}(X^{\prime}\alpha_{0})+\varepsilon,\qquad\mathbb{E}[\varepsilon|X]=0,\label{eq:=000020single=000020index}
\end{equation}
where $p_{0}(\cdot)$ is a monotone increasing link function of its
index $X^{\prime}\alpha_{0}$ and $X\in\mathbb{R}^{k}$. For identification,
$\text{\ensuremath{\alpha}}_{0}$ is a $k$-dimensional vector normalized
with $||\text{\ensuremath{\alpha}}_{0}||=1$.\footnote{In the estimation, the constraint $||\text{\ensuremath{\alpha}}_{0}||=1$
can be dealt with reparametrization or the augmented Lagrange method
by Balabdaoui and Groeneboom (2021). In this section, we study our
model without discussing those technical details. See BGH for more
details.}

For a binary dependent variable, this model can be derived from \eqref{eq:binary},
and $p_{0}(\cdot)$ is by nature monotone increasing. It was studied
by Cosslett (1983, 1987, 2007), Han (1987), Matzkin (1992), Sherman
(1993), Klein and Spady (1993), among others. In the case where $p_{0}(\cdot)$
is estimated with isotonic regression, Balabdaoui, Durot and Jankowski
(2019) studied \eqref{eq:=000020single=000020index} with the monotone
least square method, and Groeneboom and Hendrickx (2018), Balabdaoui,
Groeneboom and Hendrickx (2019), and Balabdaoui and Groeneboom (2021)
(BGH) estimated $\alpha_{0}$ and $p_{0}(\cdot)$ by solving a score-type
sample moment condition of
\begin{equation}
\mathbb{E}[X\{W-p_{0}(X^{\prime}\alpha_{0})\}]=0.\label{eq:=000020moment-single=000020index}
\end{equation}

To estimate $p_{0}$ and $\text{\ensuremath{\alpha}}_{0}$, we can
apply the method of BGH. For a fixed $\alpha$, define
\begin{equation}
\hat{p}_{\alpha}=\arg\min_{p\in\mathcal{M}}\frac{1}{N}\sum_{i=1}^{N}\{W_{i}-p(X_{i}^{\prime}\alpha)\}^{2},\label{eq:link_mono_index}
\end{equation}
where $\mathcal{M}$ is the set of monotone increasing functions defined
on $\mathbb{R}$. Note that $\hat{p}_{\alpha}(u)$ can be solved with
isotonic regression of $W_{i}$ on the data points $\{X_{i}^{\prime}\alpha\}_{i=1}^{N}$.
Then, $\alpha_{0}$ can be estimated by minimizing the squared sum
of a score function. For example, the simple score estimator in Balabdaoui
and Groeneboom (2021) is given by solving
\begin{equation}
\hat{\alpha}=\text{arg}\underset{\alpha}{\text{min}}\left\Vert \frac{1}{N}\sum_{i=1}^{N}X_{i}^{\prime}\{W_{i}-\hat{p}_{\alpha}(X_{i}^{\prime}\alpha)\}\right\Vert ^{2}.\label{eq:para_mono_index}
\end{equation}

BGH showed that under certain assumptions, $\hat{\alpha}$ is a $\sqrt{N}$-consistent
estimator for $\alpha_{0}$,\footnote{BGH proposed solving a ``zero-crossing'' root of $\frac{1}{N}\sum_{i=1}^{N}X\{W_{i}-\hat{p}_{\alpha}(X_{i}^{\prime}\alpha)\}=0$.
Then they realized that there is an issue with the existence of the
zero-crossing root for a finite sample (due to the discreteness of
$\hat{p}_{\alpha}$). To fix this problem, Balabdaoui and Groeneboom
(2021) replaced this objective function with \eqref{eq:para_mono_index},
where a minimizer always exists. If there are multiple minimizers,
any of them is a $\sqrt{N}$-consistent estimator for $\alpha_{0}$.
(See a discussion on p.1426 of Groeneboom and Hendrickx, 2018). BGH
also proposed an efficient estimator of $\alpha_{0}$ by solving a
kernel-adjusted score function. Since our aim is the second-stage
ATE $\tau$ instead of the first-stage propensity score $p$, we do
not apply BGH's efficient estimator. It will introduce additional
tuning parameters without improving the second-stage ATE.} and $\mathbb{E}[\hat{p}_{\hat{\alpha}}(X^{\prime}\hat{\alpha})-p_{0}(X^{\prime}\text{\ensuremath{\alpha}}_{0})]=O_{P}((\log N)N^{-2/3})$.
We apply their method to estimate the propensity score with multi-dimensional
control variables $X$.

In this section, $\tilde{\tau}$ denotes the ATE estimator based on
the multi-dimensional covariates $X$. Similarly to Section \ref{subsec:Uniformly-consistent-isotonic},
to solve the boundary problem of the isotonic estimator to ensure
that each observation has a non-empty matched set, we develop a uniformly
consistent monotone single-index (hereafter, UC-iso-index) estimator,
which is denoted by $\tilde{p}_{\tilde{\alpha}}$. The matching procedure
can be implemented as follows.
\begin{enumerate}
\item Compute $\hat{\alpha}$ by \eqref{eq:link_mono_index} and \eqref{eq:para_mono_index}.
\item Define $\tilde{\alpha}=\hat{\alpha}$, and transform the sample $\{Y_{i},W_{i},X_{i}\}_{i=1}^{N}$
indexed by $X_{1}^{\prime}\tilde{\alpha}<\cdots<X_{N}^{\prime}\tilde{\alpha}$
into $\{Y_{i},\tilde{W}_{i},X_{i}\}_{i=1}^{N}$ with \eqref{eq:data_transform}.
\item Compute the UC-iso-index estimator $\tilde{p}_{\tilde{\alpha}}$ by
\[
\tilde{p}_{\tilde{\alpha}}=\arg\min_{p\in\mathcal{M}}\frac{1}{N}\sum_{i=1}^{N}\{\tilde{W}_{i}-p(X_{i}^{\prime}\tilde{\alpha})\}^{2}.
\]
\item For each $i=1,\ldots,N$, compute the matching counterparts
\begin{equation}
\mathcal{J}(i)=\left\{ j=1,\dots,N:W_{j}=1-W_{i}\text{ and }\tilde{p}_{\tilde{\alpha}}(X_{j}^{\prime}\tilde{\alpha})=\tilde{p}_{\tilde{\alpha}}(X_{i}^{\prime}\tilde{\alpha})\right\} .\label{eq:matching-multi}
\end{equation}
\item Calculate the matching estimator for the ATE $\tau$ by
\begin{equation}
\tilde{\tau}=\frac{1}{N}\sum_{i=1}^{N}(2W_{i}-1)\left(Y_{i}-\frac{1}{M_{i}}\sum_{j\in\mathcal{J}(i)}Y_{j}\right),\label{eq:ATE=000020sample-multi}
\end{equation}
where $M_{i}=|\mathcal{J}(i)|$ is the number of matches for $i$.
\end{enumerate}
We modify Assumptions \ref{asm:1}-\ref{asm:4} in Section \ref{sec:main}
as follows.

\begin{asm1'} {[}Sampling{]} $\{Y_{i},W_{i},X_{i}\}_{i=1}^{N}$ is
an iid sample of $(Y,W,X)\in\mathbb{R}\times\{0,1\}\times\mathcal{X}$,
where the space $\mathcal{X}$ is a convex subset of $\mathbb{R}^{k}$
with a nonempty interior. There exists $R>0$ such that $\mathcal{X}\subset\mathcal{B}(0,R)=\{x:||x||\leq R\}$.
\end{asm1'}

Given $\alpha$, we define the true link function of \eqref{eq:link_mono_index}:
\[
p_{\alpha}(u)=\mathbb{E}[W|X^{\prime}\alpha=u].
\]
Obviously, $p_{\alpha_{0}}=p_{0}$. Let $a_{0}$ and $b_{0}$ be the
minimum and the maximum of the interval $I_{\alpha_{0}}=\{x^{\prime}\alpha_{0}:x\in\mathcal{X}\}$,
respectively.

\begin{asm2'}{[}Monotonicity and continuity{]} (i) There exists $\delta_{0}>0$
such that for each $\alpha\in\mathcal{B}(\alpha_{0},\delta_{0})$,
the function $u\mapsto\mathbb{E}[W|X^{\prime}\alpha=u]$ is monotone
increasing in $u$ and differentiable in $\alpha$; (ii) $p_{0}(\cdot)$
is continuously differentiable with its first derivative $p^{(1)}(u)>0$
on $u\in(a_{0}-\delta_{0}R,b_{0}+\delta_{0}R)$, and (iii) $X$ has
a continuous density $f(x)$ satisfying that for some positive constants
$\underline{f}$ and $\overline{f}$ , it holds $\underline{f}<f(x)<\overline{f}$
all $x\in\mathcal{X}$. \end{asm2'}

\begin{asm3'} {[}Strict overlaps{]} There exist positive constants
$\underline{p}$ and $\bar{p}$ such that $0<\underline{p}\leq p_{0}(x^{\prime}\alpha_{0})\leq\bar{p}<1$
for all $x\in\mathcal{X}$.\emph{ }\end{asm3'}

\begin{asm4'} {[}Data generating process{]} (i) $\mathbb{E}[Y(0)^{2}]<\infty$
and $\mathbb{E}[Y(1)^{2}]<\infty$, (ii) $u\mapsto\mathbb{E}[Y(1)|X=x]$
are continuously differentiable for all $x\in\mathcal{X}$ and $\alpha\in\mathcal{B}(\alpha_{0},\delta_{0})$,
(iii) for $D(Z)=\frac{WY}{p_{0}(X^{\prime}\text{\ensuremath{\alpha}}_{0})^{2}}+\frac{Y(1-W)}{\{1-p_{0}(X^{\prime}\text{\ensuremath{\alpha}}_{0})\}^{2}}$,
there exist positive constants $c_{0}$ and $M_{0}$ such that $\mathbb{E}[|D(Z)|^{m}|X=x]\leq m!M_{0}^{m-2}c_{0}$
holds for all integers $m\geq2$ and every $x$, and (iv) $Y(1),Y(0)\perp W|X$
almost surely. \end{asm4'}

Let $Z$ denote the triple $(Y,W,X)$, and $\mathcal{Z}$ denote the
space of the random vector $Z$. For each $\alpha\in\mathcal{B}(\alpha_{0},\delta_{0})$,
$u\in I_{\alpha}=\{x^{\prime}\alpha:x\in\mathcal{X}\}$, and a function
$f(\cdot)$ defined on $\mathcal{Z}$, we define $\mathbb{E}_{\alpha}[f(Z)|u]=\mathbb{E}[f(Z)|X^{\prime}\alpha=u]$.
Similarly, we define the conditional covariance $\text{Cov}_{\alpha_{0}}(f(Z),X|u)$.
The following two assumptions are adapted from BGH, which ensure that
the score estimators \eqref{eq:link_mono_index} and \eqref{eq:para_mono_index}
have desirable properties.

\begin{asm} \label{asm:5} For all $\alpha\ne\alpha_{0}$ such that
$\alpha\in\mathcal{B}(\alpha_{0},\delta_{0})$, the random variable
$\text{Cov}[(\alpha-\alpha_{0})^{\prime}X,p_{0}(X^{\prime}\alpha_{0})|X^{\prime}\alpha]$
is not equal to 0 almost surely. \end{asm}

\begin{asm} \label{asm:6} {[}Potential outcomes{]} Let $p_{0}^{(1)}(u)$
denote the first derivative of $p_{0}(u)$. The matrix $\mathbb{E}[p_{0}^{(1)}(X^{\prime}\alpha_{0})\text{Cov}(X|X^{\prime}\alpha_{0})]$
has rank $k-1$. \end{asm}

Based on Assumptions 1', 2', \ref{asm:5}, and \ref{asm:6}, we have
a result similar to Proposition \ref{prop:matching-score-modified},
but the numbering is according to $X_{1}^{\prime}\tilde{\alpha}<\cdots<X_{N}^{\prime}\tilde{\alpha}$.
The uniform convergence rate of the UC-iso-index estimator is obtained
as follows.

\begin{thm} \label{thm:uc_isoton-m} Under Assumptions 1', 2', \ref{asm:5},
and \ref{asm:6}, it holds
\[
\sup_{x\in\mathcal{X}}|\tilde{p}_{\tilde{\alpha}}(x^{\prime}\tilde{\alpha})-p_{0}(x^{\prime}\alpha_{0})|=O_{p}\left(\frac{\log N}{N}\right)^{1/3}.
\]
\end{thm}

The existence of matching counterparts is guaranteed by an argument
similar to Corollary \ref{cor:non-empty-matched-set}. Finally, let
$\mathbf{B}^{-}$ denote the Moore-Penrose inverse of a square matrix
$\mathbf{B}$. The asymptotic properties of the isotonic propensity
score matching estimator are obtained as follows.

\begin{thm} \label{thm:ATE-matching-m} Under Assumptions \ref{asm:1}'-\ref{asm:4}',
\ref{asm:5}, and \ref{asm:6}, it holds $\tilde{\tau}\overset{p}{\to}\tau$
and
\[
\sqrt{N}(\tilde{\tau}-\tau)\overset{d}{\to}N(0,\Sigma),
\]
where $\Sigma=\mathbb{E}[\{m(Z)+M(Z)+A(Z)\}\{m(Z)+M(Z)+A(Z)\}]$,
and
\begin{eqnarray}
m(Z) & = & \frac{YW}{p_{0}(X^{\prime}\alpha_{0})}-\frac{Y(1-W)}{1-p_{0}(X^{\prime}\alpha_{0})}-\tau,\qquad D(Z)=\frac{YW}{p_{0}(X^{\prime}\alpha_{0})^{2}}+\frac{Y(1-W)}{(1-p_{0}(X^{\prime}\alpha_{0}))^{2}},\nonumber \\
M(Z) & = & -\mathbb{E}_{\alpha_{0}}[D(Z)|X^{\prime}\alpha_{0}]\{W-p_{0}(X^{\prime}\alpha_{0})\},\nonumber \\
A(Z) & = & -\mathbb{E}[\mathrm{Cov}_{\alpha_{0}}(D(Z),X|X^{\prime}\alpha_{0})p_{0}^{(1)}(X^{\prime}\alpha_{0})],\nonumber \\
 &  & \times\mathbb{E}[p_{0}^{(1)}(X^{\prime}\alpha_{0})\mathrm{Cov}_{\alpha_{0}}(X|X^{\prime}\alpha_{0})]^{-}\{X-\mathbb{E}_{\alpha_{0}}[X|X^{\prime}\alpha_{0}]\}\{W-p_{0}(X^{\prime}\alpha_{0})\}.\label{eq:variance=000020component}
\end{eqnarray}
 \end{thm}

Note that the semiparametric efficiency bound for estimating $\tau$
with known $\alpha_{0}$ is given by $\mathbb{E}[\{m(Z)+M(Z)\}\{m(Z)+M(Z)\}^{\prime}]$
(see, e.g., Newey, 1994). The additional term $A(Z)$ can be interpreted
as the influence of estimating the index coefficients $\alpha_{0}$.
This influence is also faced by parametric matching estimators. In
general, our proposed method uses the matched sets, in which the number
of matches increases to infinite, so it better balances the variance
and bias in the second stage and should asymptotically outperform
any matching method with fixed numbers of matches. In Section \ref{subsec:Monte-Carlo-m}
below, we present simulation results to illustrate that the proposed
ATE estimator $\tilde{\tau}$ outperforms the probit matching estimator
in every sample size, even in the case that the true propensity score
is a probit (the correct specification). 

Theoretically, the additional term $A(Z)$ can be avoided by using
a semiparametric weighting estimator. However, the costs are strong
assumptions on the smoothness of the propensity scores (typically,
$7\cdot\dim(X)$-th continuous differentiability; see Hirano, Imbens
and Ridder, 2003) and a proper choice of smoothing parameters. Our
proposed method only requires the propensity score to be once continuously
differentiable, and it does not involve smoothing parameters, such
as bandwidths or series lengths.

\section{Bootstrap inference\protect\label{sec:Boot}}

The asymptotic variances in Theorems \ref{thm:ATE-matching} and \ref{thm:ATE-matching-m}
contain conditional mean and variance functions, such as $\mathbb{V}(Y(1)|X)$
and $\mathbb{E}[X|X^{\prime}\alpha_{0}]$, which need to be estimated.
If we use nonparametric methods to estimate them, we still have to
choose some smoothing parameters even though the point estimators
are free from smoothing. To avoid the estimation of such nonparametric
components, we employ a bootstrap method to approximate the asymptotic
distribution of the proposed isotonic propensity score matching estimator. 

After Abadie and Imbens (2008) showed that the nonparametric bootstrap
of the fixed-number matching estimator is invalid in the presence
of continuous covariates, much work tried to solve this problem by
proposing modified wild bootstraps, including Otsu and Rai (2017)
for covariates matching estimators, and Bodory \emph{et al.} (2016)
and Adusumilli (2020) for propensity score matching estimators. In
contrast, the nonparametric bootstrap of our one-to-many matching
method is valid, which is an interesting implication of Theorem \ref{thm:matching_IPW}.
In this section, we discuss an asymptotically valid bootstrap procedure
for the estimator $\hat{\tau}$ in Theorem \ref{thm:ATE-matching}.
This result can be similarly adapted to $\tilde{\tau}$ in Theorem
\ref{thm:ATE-matching-m}.

The nonparametric bootstrap is implemented as follows.
\begin{enumerate}
\item $\{Y_{i}^{*},W_{i}^{*},X_{i}^{*}\}_{i=1}^{N}$ is a bootstrap sample
from $\{Y_{i},W_{i},X_{i}\}_{i=1}^{N}$, and the numbering is according
to $X_{1}^{*}\leq\cdots\leq X_{N}^{*}.$ 
\item $\tilde{p}^{*}(\cdot)$ is the UC-isotonic estimator based on $\{Y_{i}^{*},W_{i}^{*},X_{i}^{*}\}_{i=1}^{N}$.
\item The bootstrap counterpart $\hat{\tau}^{*}$ of $\hat{\tau}$ is given
by 
\begin{eqnarray*}
\hat{\tau}^{*} & = & \frac{1}{N}\sum_{i=1}^{N}(2W_{i}^{*}-1)\left(Y_{i}^{*}-\frac{1}{M_{i}^{*}}\sum_{j\in\mathcal{J^{*}}(i)}Y_{j}^{*}\right),\\
\mathcal{J^{*}}(i) & = & \left\{ j=1,\dots,N:W_{j}^{*}=1-W_{i}^{*}\text{ and }\tilde{p}^{*}(X_{j}^{*})=\tilde{p}^{*}(X_{i}^{*})\right\} ,
\end{eqnarray*}
where $M_{i}^{*}=|\mathcal{J^{*}}(i)|$ is the number of matches for
the $i$-th observation in the bootstrap sample.
\item After repeating Step (1)-(3) for $B$ times and obtaining estimator
$\hat{\tau}_{1}^{*},\hat{\tau}_{2}^{*},\dots,\hat{\tau}_{B}^{*},$
we can conduct inference for $\tau$.
\end{enumerate}
The asymptotic validity of this bootstrap approximation is obtained
as follows.

\begin{thm}\label{thm:=000020boot} Let $\mathbb{P}^{*}$ be the
bootstrap distribution conditional on the data, and $c_{1-\alpha}^{*}$
be the $(1-\alpha)$-th sample quantile of $(\sqrt{N}\left(\hat{\tau}_{1}^{*}-\hat{\tau}\right),\sqrt{N}\left(\hat{\tau}_{2}^{*}-\hat{\tau}\right),\dots,\sqrt{N}\left(\hat{\tau}_{B}^{*}-\hat{\tau}\right))$.
Under Assumptions \ref{asm:1}-\ref{asm:4}, it holds

\begin{itemize} \item[(i)] $\sup_{t\in\mathbb{R}}|\mathbb{P}^{*}\{\sqrt{N}(\hat{\tau}^{*}-\hat{\tau})\le t\}-\mathbb{P}\{\sqrt{N}(\hat{\tau}-\tau)\le t\}|\overset{p}{\to}0;$

\item[(ii)] $\mathbb{P}\{\sqrt{N}(\hat{\tau}-\tau)\le c_{1-\alpha}^{*}\}\overset{p}{\to}1-\alpha$.\end{itemize}
\end{thm}

\section{Monte-Carlo simulations \protect\label{sec:Monte-Carlo}}

In this section, we use three simulation studies to assess the finite
sample properties of our isotonic propensity score matching estimator. 

\subsection{Univariate case\protect\label{subsec:sim_uni}}

Let $X=0.15+0.7Z$, where $Z$ and $\nu$ are independently uniformly
distributed on $[0,1]$, and
\begin{eqnarray}
W & = & \begin{cases}
0 & \text{if }X<\nu\\
1 & \text{if }X\geq\nu
\end{cases},\nonumber \\
Y & = & 0.5W+2X+\varepsilon,\nonumber \\
\varepsilon & \sim & N(0,1).\label{sim:uni_setup}
\end{eqnarray}
The true ATE is the coefficient of $W$, which is 0.5. The simulation
results are presented in Table \ref{tab:ATE-model}, where $\hat{\mu}_{\tau}$
is the Monte-Carlo mean, and the mean square errors (MSE) are rescaled
by $N$. The number of Monte-Carlo simulations is 5000 for each sample
size. 

\begin{table}[H]
\caption{\protect\label{tab:ATE-model}Matching estimators of ATE: the univariate
case}

\centering{}{\small{}%
\begin{tabular}{cr@{\extracolsep{0pt}.}lr@{\extracolsep{0pt}.}lr@{\extracolsep{0pt}.}lr@{\extracolsep{0pt}.}lr@{\extracolsep{0pt}.}lr@{\extracolsep{0pt}.}l}
\toprule 
\multicolumn{5}{c}{with UC-isotonic} & \multicolumn{2}{c}{} & \multicolumn{6}{c}{with logit and $M=1$}\tabularnewline
\midrule
$N$ & \multicolumn{2}{c}{$\hat{\mu}_{\tau}$} & \multicolumn{2}{c}{$MSE$} & \multicolumn{2}{c}{} & \multicolumn{2}{c}{$N$} & \multicolumn{2}{c}{$\hat{\mu}_{\tau}$} & \multicolumn{2}{c}{$MSE$}\tabularnewline
\cmidrule{1-5}\cmidrule{8-13}
100 & 0&4977 & 5&2723 & \multicolumn{2}{c}{} & \multicolumn{2}{c}{100} & 0&4997 & 7&1068\tabularnewline
1000 & 0&4934 & 5&2589 & \multicolumn{2}{c}{} & \multicolumn{2}{c}{1000} & 0&5009 & 7&0630\tabularnewline
2000 & 0&4946 & 5&2158 & \multicolumn{2}{c}{} & \multicolumn{2}{c}{2000} & 0&4999 & 7&0816\tabularnewline
5000 & 0&4963 & 4&9418 & \multicolumn{2}{c}{} & \multicolumn{2}{c}{5000} & 0&4995 & 6&8376\tabularnewline
10000 & 0&4974 & 4&9785 & \multicolumn{2}{c}{} & \multicolumn{2}{c}{10000} & 0&5000 & 6&8238\tabularnewline
\midrule
$\infty$ & 0&5 & 4&96 & \multicolumn{2}{c}{} & \multicolumn{2}{c}{$\infty$} & 0&5 & 4&96\tabularnewline
\bottomrule
\end{tabular}}{\small\par}
\end{table}

The left panel shows the simulation results of the proposed matching
method based on propensity scores estimated by the UC-isotonic estimator,
and the right panel shows those of the one-to-one matching estimator
based on propensity scores estimated with the logit model $\mathbb{P}(W=1|X=x)=\frac{\text{exp}(a+bx)}{\text{exp}(a+bx)+1}.$
The last row shows the true value of ATE and the semiparametric efficiency
bound of this problem calculated according to Hahn (1998): 
\begin{align*}
\Omega_{\text{SEB}} & =\text{Var}(\mathbb{E}[Y(1)-Y(0)|X])+\mathbb{E}[\text{Var}(Y(1)|X)/p_{0}(X)]+\mathbb{E}[\text{Var}(Y(0)|X)/(1-p_{0}(X))]\\
 & =\text{Var}(0.5)+\mathbb{E}[1/p_{0}(X)]+\mathbb{E}[1/(1-p_{0}(X))]\\
 & =0+\int_{0.15}^{0.85}\frac{1}{x}\frac{1}{0.7}dx+\int_{0.15}^{0.85}\frac{1}{1-x}\frac{1}{0.7}dx\approx4.96.
\end{align*}
In comparison, the logit matching estimator has a slightly smaller
bias, and it seems that both estimators are asymptotically unbiased.
The MSEs of the isotonic propensity score matching estimator are considerably
smaller than those of the logit matching estimator in every sample
size. With the sample size growing, the MSEs of isotonic propensity
score matching estimator approaches to the semiparametric efficiency
bound. 

\subsection{Multivariate case\protect\label{subsec:Monte-Carlo-m}}

Consider the following setting:
\begin{eqnarray*}
Y & = & X^{\prime}\gamma_{0}+W\tau_{0}+\varepsilon,\\
W & = & \begin{cases}
0 & \text{if }X^{\prime}\alpha_{0}<\nu\\
1 & \text{if }X^{\prime}\alpha_{0}\geq\nu
\end{cases},\\
\varepsilon & \sim & N(0,1),\qquad\nu\sim N(0,1),\qquad\varepsilon\perp\nu,
\end{eqnarray*}
where $X\sim U[-1,1]^{3}$, and the true parameters are set as $\alpha_{0}=(1,1,1)^{\prime}/\sqrt{3}$,
and $\gamma_{0}=(0.1,0.2,0.3)^{\prime}$, and the ATE is $\tau_{0}=0.5$.
Under this setting, we have $\mathbb{P}(W=1|X=x)=p_{0}(x)=\Phi(x^{\prime}\alpha_{0})$,
where $\Phi$ is the CDF of the standard normal distribution, i.e.,
the propensity score is correctly specified in probit estimation.

The simulation results are presented in Table \ref{tab:ATE-model-m},
where $\hat{\mu}_{\tau}$ is the Monte-Carlo mean, and the MSEs are
rescaled by $N$. The number of Monte-Carlo simulations is 5000 for
each sample size. The left panel shows the simulation results of the
proposed matching method based on propensity scores estimated by the
UC-iso-index estimator, and the right panel shows those of the one-to-one
matching estimator based on propensity scores estimated with the correctly
specified probit model. 

\begin{table}[H]
\caption{\protect\label{tab:ATE-model-m}Matching estimators of ATE: the multivariate
case}

\centering{}{\small{}%
\begin{tabular}{cr@{\extracolsep{0pt}.}lr@{\extracolsep{0pt}.}lr@{\extracolsep{0pt}.}lr@{\extracolsep{0pt}.}lr@{\extracolsep{0pt}.}lr@{\extracolsep{0pt}.}l}
\toprule 
\multicolumn{5}{c}{with UC-iso-index} & \multicolumn{2}{c}{} & \multicolumn{6}{c}{with probit and $M=1$}\tabularnewline
\midrule
$N$ & \multicolumn{2}{c}{$\hat{\mu}_{\tau}$} & \multicolumn{2}{c}{$MSE$} & \multicolumn{2}{c}{} & \multicolumn{2}{c}{$N$} & \multicolumn{2}{c}{$\hat{\mu}_{\tau}$} & \multicolumn{2}{c}{$MSE$}\tabularnewline
\cmidrule{1-5}\cmidrule{8-13}
100 & 0&5080 & 5&0442 & \multicolumn{2}{c}{} & \multicolumn{2}{c}{100} & 0&5114 & 7&3459\tabularnewline
1000 & 0&5016 & 5&0014 & \multicolumn{2}{c}{} & \multicolumn{2}{c}{1000} & 0&5030 & 6&9813\tabularnewline
2000 & 0&4991 & 5&0727 & \multicolumn{2}{c}{} & \multicolumn{2}{c}{2000} & 0&4997 & 7&2275\tabularnewline
5000 & 0&5003 & 5&2115 & \multicolumn{2}{c}{} & \multicolumn{2}{c}{5000} & 0&5010 & 7&2640\tabularnewline
10000 & 0&5001 & 5&0161 & \multicolumn{2}{c}{} & \multicolumn{2}{c}{10000} & 0&5002 & 7&0509\tabularnewline
\bottomrule
\end{tabular}}{\small\par}
\end{table}

The pattern is similar to the univariate case. The biases of both
estimators are small and converge to zero. The isotonic matching estimator
outperforms the probit matching estimator in every sample size in
terms of MSE. 

\subsection{Bootstrap}

Table \ref{tab:coverage} shows the bootstrap coverage rates. We draw
2000 Monte-Carlo simulations, and for each simulation, we draw 500
bootstrap samples. The coverage rates are calculated with these 2000
sets of confidence intervals for both 90\% and 95\% confidence levels.
From Table \ref{tab:coverage}, we see clear trends that the bootstrap
coverage rates are converging to their theoretical limits. 

\begin{table}[H]
\caption{\protect\label{tab:coverage}Bootstrap coverage rates}

\centering{}{\small{}%
\begin{tabular}{ccrr}
\toprule 
\multirow{1}{*}{$n$} &  & 90\% CI & 95\% CI\tabularnewline
\cmidrule{1-1}\cmidrule{3-4}
100 &  & 0.860 & 0.918\tabularnewline
1000 &  & 0.889 & 0.938\tabularnewline
2000 &  & 0.881 & 0.940\tabularnewline
5000 &  & 0.901 & 0.945\tabularnewline
10000 &  & 0.891 & 0.948\tabularnewline
\cmidrule{1-1}\cmidrule{3-4}
$\infty$ &  & 0.90 & 0.95\tabularnewline
\bottomrule
\end{tabular}}{\small\par}
\end{table}

Overall, the simulation outcomes of the univariate case, the multivariate
case, and the bootstrap encourage the proposed isotonic propensity
score matching method. Additionally, for further simulation comparisons
of our approach with propensity score methods of one-to-many matching
and radius matching, as well as the impact of thresholds for averaging
treatment variables at boundaries, see Section \ref{supp:add_Monte_Carlo}
in the supplementary material.

\section{Conclusion}

We develop a one-to-many matching estimator of ATE based on propensity
scores estimated by modified isotonic regression. We reveal that the
nature of the isotonic estimator can help us to fix many problems
of existing matching methods, including efficiency, choice of the
number of matches, choice of tuning parameter, robustness to the propensity
score misspecification, and bootstrap validity. As by-products, a
uniformly consistent isotonic estimator and a uniformly consistent
monotone single-index estimator, for both univariate and multivariate
cases, are designed for our proposed isotonic matching estimator,
and we study their asymptotic properties. The method can be further
extended to other causal estimators based on propensity scores, such
as blocking on propensity scores and regression on propensity scores.

\appendix

\section{Proofs}

\subsection{Proof of Proposition \ref{prop:disjoint-grouping}\protect\label{appsub:Proof-P-disjoint}}

The proof is based on the following lemma.

\begin{lem} {[}Groeneboom and Jongbloed (2014, Lemma 2.1){]}\emph{
}The vector $\hat{p}=(\hat{p}_{1},\ldots,\hat{p}_{N})$ minimizes
$Q(p)=\frac{1}{2}\sum_{i=1}^{N}(W_{i}-p_{i})^{2}$ over the closed
convex cone $\mathcal{C}=\{p\in\mathbb{R}^{N}:p_{1}\leq p_{2}\leq\cdots\leq p_{N}\}$
if and only if 
\begin{equation}
\sum_{j=1}^{i}\hat{p}_{j}\text{ }\begin{cases}
\leq\sum_{j=1}^{i}W_{j}\\
=\sum_{j=1}^{i}W_{j} & \text{if }\hat{p}_{i+1}>\hat{p}_{i}\text{ or }i=N
\end{cases}.\label{pf:isot_reg}
\end{equation}
\end{lem}

We now prove Proposition \ref{prop:disjoint-grouping}. For any $k=1,\dots,K$,
we have $\hat{p}_{n_{k}}>\hat{p}_{n_{k}-1}$ and $\hat{p}_{n_{k+1}}>\hat{p}_{n_{k+1}-1}$
by \eqref{eq:unique}. By \eqref{pf:isot_reg}, we have
\begin{equation}
\sum_{j=1}^{n_{k}-1}\hat{p}_{j}=\sum_{j=1}^{n_{k}-1}W_{j},\qquad\sum_{j=1}^{n_{k+1}-1}\hat{p}_{j}=\sum_{j=1}^{n_{k+1}-1}W_{j},\qquad\hat{p}_{n_{k}}=\hat{p}_{n_{k}+1}=\cdots=\hat{p}_{n_{k}+N_{k}-1}.\label{pf:group_k}
\end{equation}
Since $n_{k}-1=n_{k-1}+N_{k-1}-1$ and $n_{k+1}-1=n_{k}+N_{k}-1$,
\eqref{pf:group_k} implies
\begin{equation}
\sum_{j=n_{k}}^{n_{k}+N_{k}-1}\hat{p}_{j}=\sum_{j=n_{k}}^{n_{k}+N_{k}-1}W_{j}.\label{pf:within_g_equal}
\end{equation}
Combining \eqref{pf:group_k} and \eqref{pf:within_g_equal}, it holds
that for any $i=n_{k},\dots,n_{k}+N_{k}-1$,
\[
\hat{p}_{i}=\frac{1}{N_{k}}\sum_{j=n_{k}}^{n_{k}+N_{k}-1}\hat{p}_{j}=\frac{1}{N_{k}}\sum_{j=n_{k}}^{n_{k}+N_{k}-1}W_{j}.
\]

\subsection{Proof of Proposition \ref{prop:matching-score}\protect\label{appsub:Proof-P-matching-score}}

Part (i) is a direct implication of Proposition \ref{prop:disjoint-grouping}
and the definitions of $N_{k,1}$, $N_{k}$, and $n_{k}$. By $W_{i}\in\{0,1\}$,
$0<\hat{p}(X_{i})<1$, and \eqref{eq:iso_estimate_k}, we must have
$W_{i}=1$ and $W_{j}=0$ for some $i,j\in\{n_{k},\ldots,(n_{k}+N_{k}-1)\}$.
Thus, Part (ii) follows.

\subsection{Proof of Proposition \ref{prop:matching-score-modified}\protect\label{appsub:Proof-L-matching-score-modi}}

\subsubsection*{Proof of (i)\protect\label{appsub:Proof-P-matching-score-1}}

Since the proof is similar, we focus on the proof of the first statement,
$N_{1}\geq\lfloor N^{2/3}\rfloor$. The isotonic estimator can be
written as (see, Barlow and Brunk, 1972):
\begin{equation}
\hat{p}(X_{i})=\max_{s\le i}\min_{t\ge i}\sum_{j=s}^{t}\frac{W_{j}}{t-s+1}.\label{pf:B=000026B}
\end{equation}
Let
\begin{equation}
\bar{W}_{l}=\frac{1}{\lfloor N^{2/3}\rfloor}\sum_{i=1}^{\lfloor N^{2/3}\rfloor}W_{i},\qquad\bar{W}_{u}=\frac{1}{\lfloor N^{2/3}\rfloor}\sum_{i=N-\lfloor N^{2/3}\rfloor+1}^{N}W_{i}.\label{eq:left_end}
\end{equation}
For any $i$ with $1\leq i\leq\lfloor N^{2/3}\rfloor$, \eqref{eq:data_transform},
\eqref{eq:uc_isoton}, and \eqref{pf:B=000026B} imply
\begin{eqnarray}
\tilde{p}(X_{i}) & = & \max_{s\le i}\min_{t\ge i}\sum_{j=s}^{t}\frac{\tilde{W}_{j}}{t-s+1}\nonumber \\
 & = & \max_{s\le i}\min_{t\ge i}\frac{\sum_{j=s}^{t\land\lfloor N^{2/3}\rfloor}\bar{W}_{l}+\mathbb{I}\{t>\lfloor N^{2/3}\rfloor\}\sum_{j=\lfloor N^{2/3}\rfloor+1}^{t}W_{j}}{t-s+1}\nonumber \\
 & = & \max_{s\le i}\min_{t\ge i}\frac{\sum_{j=s}^{t}\bar{W}_{l}+\mathbb{I}\{t>\lfloor N^{2/3}\rfloor\}\left(\sum_{j=\lfloor N^{2/3}\rfloor+1}^{t}W_{j}-\sum_{j=\lfloor N^{2/3}\rfloor+1}^{t}\bar{W}_{l}\right)}{t-s+1}\nonumber \\
 & = & \bar{W}_{l}+\max_{s\le i}\min_{t\ge i}\left[\mathbb{I}\{t>\lfloor N^{2/3}\rfloor\}\frac{t-\lfloor N^{2/3}\rfloor}{t-s+1}\left(\frac{1}{t-\lfloor N^{2/3}\rfloor}\sum_{j=\lfloor N^{2/3}\rfloor+1}^{t}W_{j}-\bar{W}_{l}\right)\right].\label{pf:PAVA_bound}
\end{eqnarray}
Since $\mathbb{I}\{t>\lfloor N^{2/3}\rfloor\}\frac{t-\lfloor N^{2/3}\rfloor}{t-s+1}\ge0$,
the minimizer with respect to $t$ is determined by the sign of $\left(\frac{1}{t-\lfloor N^{2/3}\rfloor}\sum_{j=\lfloor N^{2/3}\rfloor+1}^{t}W_{j}-\bar{W}_{l}\right)$,
and we discuss two cases: 

(I) $\min_{t>\lfloor N^{2/3}\rfloor}\frac{1}{t-\lfloor N^{2/3}\rfloor}\sum_{j=\lfloor N^{2/3}\rfloor+1}^{t}W_{j}>\bar{W}_{l}$,

(II) $\min_{t>\lfloor N^{2/3}\rfloor}\frac{1}{t-\lfloor N^{2/3}\rfloor}\sum_{j=\lfloor N^{2/3}\rfloor+1}^{t}W_{j}\leq\bar{W}_{l}$.

For Case (I), adding any terms after $\lfloor N^{2/3}\rfloor$ cannot
make the average smaller Thus, we have $\tilde{p}(X_{i})=\bar{W}_{l}$
for all $1\leq i\leq\lfloor N^{2/3}\rfloor$, and it holds $N_{1}=\lfloor N^{2/3}\rfloor$.

For Case (II), it makes sense to add more terms after $\lfloor N^{2/3}\rfloor$
since for any fixed $s$, adding more items after $\lfloor N^{2/3}\rfloor$
will lower the overall level of the sample mean \eqref{pf:PAVA_bound}.
Define 
\begin{equation}
t_{s}=\arg\min_{t\geq\lfloor N^{2/3}\rfloor}\mathbb{I}\{t>\lfloor N^{2/3}\rfloor\}\frac{t-\lfloor N^{2/3}\rfloor}{t-s+1}\left(\frac{1}{t-\lfloor N^{2/3}\rfloor}\sum_{j=\lfloor N^{2/3}\rfloor+1}^{t}W_{j}-\bar{W}_{l}\right).\label{pf:mini_ts}
\end{equation}
After minimizers are chosen for each $s$, the maxmin operator requires
to choose the maximum across different $s$. For any $i$ smaller
than $\lfloor N^{2/3}\rfloor$ and any $j\leq i$, we have $\tilde{W}_{j}=\bar{W}_{l}\geq\min_{t>\lfloor N^{2/3}\rfloor}\frac{1}{t-\lfloor N^{2/3}\rfloor}\sum_{m=\lfloor N^{2/3}\rfloor+1}^{t}W_{m}$.
Therefore, adding more terms before $i$ will increase the overall
level of the sample mean \eqref{pf:PAVA_bound}, so we must have $s=1$.
(This is also justified by \eqref{pf:mini_ts}: for $s<t$, the smaller
$s$, the greater $-\frac{t-\lfloor N^{2/3}\rfloor}{t-s+1}$. Note
that $\min_{t>\lfloor N^{2/3}\rfloor}\frac{1}{t-\lfloor N^{2/3}\rfloor}\sum_{j=\lfloor N^{2/3}\rfloor+1}^{t}W_{j}-\bar{W}_{l}\leq0$
by the setup of Case (II).)

Consequently, \eqref{pf:PAVA_bound} can be written as
\begin{equation}
\tilde{p}(X_{i})=\bar{W}_{l}+\frac{t_{1}-\lfloor N^{2/3}\rfloor}{t_{1}}\left(\frac{1}{t_{1}-\lfloor N^{2/3}\rfloor}\sum_{j=\lfloor N^{2/3}\rfloor+1}^{t_{1}}W_{j}-\bar{W}_{l}\right)=\sum_{j=1}^{t_{1}}\frac{\tilde{W}_{j}}{t_{1}},\label{pf:N1_value}
\end{equation}
with $t_{1}>\lfloor N^{2/3}\rfloor$. \eqref{pf:N1_value} gives a
common value of $\tilde{p}(X_{i})$ for all $i=1,\ldots,\lfloor N^{2/3}\rfloor$.
By \eqref{eq:unique}, we have $N_{1}=t_{1}>\lfloor N^{2/3}\rfloor$,
which implies the conclusion.

\subsubsection*{Proof of (ii)}

Part (i) shows that all the changed treatment variables are clustered
in the first and the last group. Therefore, for $k=2,3,\dots,K-1$,
Part (ii) holds by the same arguments for Propositions \ref{prop:disjoint-grouping}
and \ref{prop:matching-score} (i), so it remains to show the cases
for for $k=1$ and $K$. Since the proof is similar, we only present
the proof for $k=1$.

By using $N_{1}\geq\lfloor N^{2/3}\rfloor$ from Part (i), it holds
that for each $i=1,\ldots,N_{1}$, 
\begin{eqnarray*}
\tilde{p}(X_{i}) & = & \sum_{j=1}^{N_{1}}\frac{\tilde{W}_{j}}{N_{1}}=\sum_{j=1}^{\lfloor N^{2/3}\rfloor}\frac{\tilde{W}_{j}}{N_{1}}+\mathbb{I}\{N_{1}>\lfloor N^{2/3}\rfloor\}\sum_{j=\lfloor N^{2/3}\rfloor+1}^{N_{1}}\frac{W_{j}}{N_{1}}\\
 & = & \frac{1}{N_{1}}\left(\sum_{j=1}^{\lfloor N^{2/3}\rfloor}\left(\frac{1}{\lfloor N^{2/3}\rfloor}\sum_{i=1}^{\lfloor N^{2/3}\rfloor}W_{i}\right)+\mathbb{I}\{N_{1}>\lfloor N^{2/3}\rfloor\}\sum_{j=\lfloor N^{2/3}\rfloor+1}^{N_{1}}W_{j}\right)\\
 & = & \frac{1}{N_{1}}\left(\sum_{i=1}^{\lfloor N^{2/3}\rfloor}W_{i}+\mathbb{I}\{N_{1}>\lfloor N^{2/3}\rfloor\}\sum_{j=\lfloor N^{2/3}\rfloor+1}^{N_{1}}W_{j}\right)\\
 & = & \sum_{j=1}^{N_{1}}\frac{W_{j}}{N_{1}}=\frac{N_{1,1}}{N_{1}}.
\end{eqnarray*}

\subsubsection*{Proof of (iii)}

Since the proofs are similar, we focus on the first statement, $N_{1}=O_{p}(N^{2/3})$.
The idea of this proof is based on the intuition that under Assumption
\ref{asm:2} (ii), the treatment propensity should be higher after
$\lfloor N^{2/3}\rfloor$ than before this point. As a result, it
becomes increasingly unlikely for the points to the right of $\lfloor N^{2/3}\rfloor$
to be allocated to the first partition.

By definition, it is equivalent to show that for any $\nu>0$, there
exists $c>0$ such that $\mathbb{P}(c_{1}>c)<\nu$, where
\begin{equation}
c_{1}=\frac{N_{1}}{N^{2/3}}.\label{eq:c1}
\end{equation}
 Without loss of generality, we choose $N$ such that $N^{2/3}=\lfloor N^{2/3}\rfloor$;
and we can set $c$ to be a positive integer and $c>2$; Note that
$N_{1}=c_{1}N^{2/3}$, and we have {\small
\begin{eqnarray}
\mathbb{P}(c_{1}>c) & = & \mathbb{P}\left(\bar{W}_{l}\geq\frac{1}{c_{1}N^{2/3}-N^{2/3}}\sum_{j=N^{2/3}+1}^{c_{1}N^{2/3}}W_{j},c_{1}>c\right)\nonumber \\
 & = & \mathbb{P}\left(\frac{1}{N^{2/3}}\sum_{i=1}^{N^{2/3}}W_{i}>\frac{1}{c_{1}N^{2/3}-N^{2/3}}\sum_{j=N^{2/3}+1}^{c_{1}N^{2/3}}W_{j},c_{1}>c\right)\nonumber \\
 & = & \mathbb{P}\bigg(\frac{1}{N^{2/3}}\sum_{i=1}^{N^{2/3}}\{W_{i}-p(X_{i})\}+\frac{1}{N^{2/3}}\sum_{i=1}^{N^{2/3}}p(X_{i})\nonumber \\
 &  & \qquad>\frac{1}{c_{1}N^{2/3}-N^{2/3}}\sum_{j=N^{2/3}+1}^{c_{1}N^{2/3}}\{W_{j}-p(X_{j})\}+\frac{1}{c_{1}N^{2/3}-N^{2/3}}\sum_{j=N^{2/3}+1}^{c_{1}N^{2/3}}p(X_{j}),\text{ }c_{1}>c\bigg)\nonumber \\
 & = & \mathbb{P}\bigg(\frac{1}{N^{1/3}}\sum_{i=1}^{N^{2/3}}\{W_{i}-p(X_{i})\}-\frac{N^{1/3}}{c_{1}N^{2/3}-N^{2/3}}\sum_{j=N^{2/3}+1}^{c_{1}N^{2/3}}\{W_{j}-p(X_{j})\}\nonumber \\
 &  & \qquad>\frac{N^{1/3}}{c_{1}N^{2/3}-N^{2/3}}\sum_{j=N^{2/3}+1}^{c_{1}N^{2/3}}p(X_{j})-\frac{1}{N^{1/3}}\sum_{i=1}^{N^{2/3}}p(X_{i}),\text{ }c_{1}>c\bigg)\nonumber \\
 & = & \mathbb{P}\left(\sum_{i=1}^{c_{1}N^{2/3}}B_{i}>a,\text{ }c_{1}>c\right),\label{eq:probabiliy_chain}
\end{eqnarray}
}where the first equality follows from $c_{1}>c>2$ and the implication
of Case (II) of the proof of Proposition \ref{prop:matching-score-modified}
(i), the second equality follows from the definition of $\bar{W}_{l}$
in \eqref{eq:left_end} and $N^{2/3}=\lfloor N^{2/3}\rfloor$, the
third equality follows by centering $W$ around $p(X)$, the fourth
equality follows from a rearrangement and multiplying both sides by
$N^{1/3}$, and the last equality is given by the definitions: 
\begin{eqnarray}
B_{i} & = & \begin{cases}
\frac{W_{i}-p(X_{i})}{N^{1/3}} & \text{for }1\leq i\leq N^{2/3}\\
-\frac{N^{1/3}\{W_{i}-p(X_{i})\}}{c_{1}N^{2/3}-N^{2/3}} & \text{for }\lfloor N^{2/3}\rfloor+1\leq i\leq c_{1}N^{2/3},
\end{cases}\nonumber \\
a & = & \frac{N^{1/3}}{c_{1}N^{2/3}-N^{2/3}}\sum_{j=N^{2/3}+1}^{c_{1}N^{2/3}}p(X_{j})-\frac{1}{N^{1/3}}\sum_{i=1}^{N^{2/3}}p(X_{i}).\label{eq:berstein_ab}
\end{eqnarray}
Note that $\sum_{i=1}^{c_{1}N^{2/3}}B_{i}$ is an average centered
around zero; the term $a$, due to Assumption \ref{asm:2} (ii), should
be strictly positive. We now apply the following Bernstein inequality
(see, e.g., van de Geer, 2000) to \eqref{eq:probabiliy_chain}. 

\textbf{Bernstein inequality. }\emph{Let $B_{1,},\dots,B_{n}$ be
independent random variables satisfying
\begin{eqnarray}
\mathbb{E}[B_{i}] & = & 0,\qquad\mathbb{E}[|B_{i}|^{m}]\leq\frac{m!}{2}A^{m-2}\mathbb{V}(B_{i})\quad\text{for each }m=2,3,\dots,\nonumber \\
b^{2} & = & \sum_{i=1}^{n}\mathbb{V}(B_{i}),\label{eq:bernstein}
\end{eqnarray}
for some constant $A$. Then,}
\[
\mathbb{P}\left(\sum_{i=1}^{n}B_{i}\geq a\right)\leq\exp\left(-\frac{a^{2}}{2aA+2b^{2}}\right).
\]
Let $q_{\alpha}$ denote the $\alpha$-th quantile of $X$. For any
positive integer $c>2$, {\small{\small
\begin{eqnarray}
 &  & \frac{N^{1/3}}{cN^{2/3}-N^{2/3}}\sum_{j=N^{2/3}+1}^{cN^{2/3}}p(X_{j})-\frac{1}{N^{1/3}}\sum_{i=1}^{N^{2/3}}p(X_{i})\nonumber \\
 & = & N^{1/3}\left(\frac{\int_{q_{N^{-1/3}}}^{q_{c\cdot N^{-1/3}}}p(x)f(x)dx}{\int_{q_{N^{-1/3}}}^{q_{c\cdot N^{-1/3}}}f(x)dx}-\frac{\int_{x_{L}}^{q_{N^{-1/3}}}p(x)f(x)dx}{\int_{x_{L}}^{q_{N^{-1/3}}}f(x)dx}+O_{p}((N^{2/3})^{-1/2})\right)\nonumber \\
 & = & N^{1/3}\bigg(\frac{\int_{q_{N^{-1/3}}}^{q_{c\cdot N^{-1/3}}}\{p(x_{L})+p^{(1)}(x_{L})\cdot(x-x_{L})+o(x-x_{L})\}f(x)dx}{\int_{q_{N^{-1/3}}}^{q_{c\cdot N^{-1/3}}}f(x)dx}\nonumber \\
 &  & -\frac{\int_{x_{L}}^{q_{c\cdot N^{-1/3}}}\{p(x_{L})+p^{(1)}(x_{L})\cdot(x-x_{L})+o(x-x_{L})\}f(x)dx}{\int_{x_{L}}^{q_{N^{-1/3}}}f(x)dx}+O_{p}(N^{-1/3})\bigg)\nonumber \\
 & \geq & N^{1/3}\bigg[p(x_{L})+p^{(1)}(x_{L})\underline{f}(q_{c\cdot N^{-1/3}}-x_{L})\nonumber \\
 &  & -\{p(x_{L})+p^{(1)}(x_{L})p^{(1)}(x_{L})\overline{f}(q_{N^{-1/3}}-x_{L})+o(N^{-1/3})\}\bigg]+O_{p}(1)\nonumber \\
 & = & p^{(1)}(x_{L})N^{1/3}\{\underline{f}q_{c\cdot N^{-1/3}}(q_{c\cdot N^{-1/3}}-x_{L})-\overline{f}(q_{N^{-1/3}}-x_{L})\}+O_{p}(1)\nonumber \\
 & =: & a_{c}+O_{p}(1),\label{eq:a_decompose}
\end{eqnarray}
}}where the first equality follows from the fact that the sample
mean of a sample of size $N^{2/3}$ can estimate the population mean
at the $O_{p}((N^{2/3})^{-1/2})$ rate, the second equality follows
from an extension of $p(x)$ around $x_{L}$, the first inequality
follows from Assumption \ref{asm:2} (iii), and the last equality
is given by the definition 
\begin{equation}
a_{c}=p^{(1)}(x_{L})N^{1/3}\{\underline{f}(q_{c\cdot N^{-1/3}}-x_{L})-\overline{f}(q_{N^{-1/3}}-x_{L})\}.\label{eq:ac}
\end{equation}

Now we show 
\begin{equation}
a_{c}\to\infty\quad\text{as }c\rightarrow\infty.\label{eq:growing_a}
\end{equation}
 To this end, it is enough to show $\lim_{c\rightarrow\infty}N^{1/3}\cdot\underline{f}(q_{c\cdot N^{-1/3}}-x_{L})=\infty$.
By Assumption \ref{asm:2} (iii), for or any $c\in\mathbb{N}$, we
have
\begin{equation}
N^{-1/3}/\overline{f}\leq q_{(c+1)\cdot N^{-1/3}}-q_{(c)\cdot N^{-1/3}}\leq N^{-1/3}/\underline{f}.\label{eq:length_N1/3}
\end{equation}
Combining \eqref{eq:ac} and \eqref{eq:length_N1/3} yields $a_{c}\geq p^{(1)}(x_{L})\cdot c(\underline{f}/\overline{f})+O(1)$,
which implies \eqref{eq:growing_a}. 

Furthermore, by \eqref{eq:ac}, we have
\begin{equation}
c_{1}>c\Rightarrow a_{c_{1}}>a_{c}.\label{eq:ac1}
\end{equation}
 On the other hand, for $a$ defined in \eqref{eq:berstein_ab}, applying
\eqref{eq:a_decompose} to $a$ yields
\begin{equation}
a=a_{c_{1}}+O_{p}(1).\label{eq:=000020order_a}
\end{equation}

Now, we study $b$ in \eqref{eq:bernstein}. Note that for a binary
$W$ and $p(X)=\mathbb{E}(W|X)$, we have $\mathbb{V}(W-p(X))=\{1-p(X)\}p(X)$.
Thus, for any $c>2$,
\begin{eqnarray*}
\sum_{i=1}^{cN^{2/3}}\mathbb{V}(B_{i}) & = & \frac{N^{2/3}}{(cN^{2/3}-N^{2/3})^{2}}\sum_{j=N^{2/3}+1}^{cN^{2/3}}\{1-p(X_{j})\}p(X_{j})+\frac{1}{N^{2/3}}\sum_{i=1}^{N^{2/3}}\{1-p(X_{i})\}p(X_{i})\\
 & = & \frac{1}{c-1}\frac{1}{(c-1)N^{2/3}}\sum_{j=N^{2/3}+1}^{cN^{2/3}}\{1-p(X_{j})\}p(X_{j})+\frac{1}{N^{2/3}}\sum_{i=1}^{N^{2/3}}\{1-p(X_{i})\}p(X_{i})\\
 & = & \frac{c}{c-1}\{1-p(x_{L})\}p(x_{L})+o_{p}(1)\\
 & =: & b_{c}^{2}+o_{p}(1),
\end{eqnarray*}
where the second equality follows from the consistency of the sample
mean to the population mean, and the last equality follows by the
definition $b_{c}^{2}=\frac{c}{c-1}\{1-p(x_{L})\}p(x_{L}).$ Thus,
we have 
\begin{equation}
\sum_{i=1}^{N_{1}}\mathbb{V}(B_{i})=\sum_{i=1}^{c_{1}N^{2/3}}\mathbb{V}(B_{i})=b_{c_{1}}^{2}+o_{p}(1).\label{eq:order=000020b}
\end{equation}
 Due to \eqref{eq:growing_a}, \eqref{eq:ac1}, and \eqref{eq:=000020order_a},
for any $\nu>0$, we can choose a large enough $N$ and $c$ such
that
\begin{equation}
\mathbb{P}\left(a<\frac{1}{2}a_{c},\text{ }c_{1}>c\right)\leq\mathbb{P}\left(a<\frac{1}{2}a_{c},\text{ }a_{c_{1}}>a_{c}\right)<\frac{\nu}{4}.\label{eq:a_bound}
\end{equation}

Further, note that $b_{c}^{2}=\frac{c}{c-1}\{1-p(x_{L})\}p(x_{L})$
is decreasing in $c$ when $c>1$. By \eqref{eq:order=000020b}, we
have
\begin{equation}
\mathbb{P}\left(\text{ }\sum_{i=1}^{c_{1}N^{2/3}}\mathbb{V}(B_{i})\geq2b_{c}^{2},\text{ }c_{1}>c\right)\leq\mathbb{P}\left(\text{ }\sum_{i=1}^{c_{1}N^{2/3}}\mathbb{V}(B_{i})\geq2b_{c_{1}}^{2},\text{ }c_{1}>c\right)<\frac{\nu}{4}.\label{eq:b_bound}
\end{equation}
Now we use the Bernstein inequality. Since $B_{i}$ defined in \eqref{eq:berstein_ab}
is a centered and normalized binary variable, we can simply choose
$A=1$ in \eqref{eq:bernstein}, then
\begin{eqnarray}
\mathbb{P}(c_{1}>c) & \le & \mathbb{P}(\sum_{i=1}^{c_{1}N^{2/3}}B_{i}>a,\text{ }c_{1}>c)\nonumber \\
 & \le & \mathbb{P}\left(\sum_{i=1}^{N_{1}}B_{i}\geq a,\text{ }a\geq\frac{1}{2}a_{c},\text{ }c_{1}>c\right)+\frac{\nu}{4}\nonumber \\
 & \le & \mathbb{P}\left(\sum_{i=1}^{N_{1}}B_{i}\geq a,\text{ }a\geq\frac{1}{2}a_{c},\text{ }c_{1}>c,\sum_{i=1}^{c_{1}N^{2/3}}\mathbb{V}(B_{i})<2b_{c}^{2}\right)+\frac{\nu}{2}\nonumber \\
 & \le & \mathbb{P}\left(\sum_{i=1}^{N_{1}}B_{i}\geq\frac{1}{2}a_{c},\sum_{i=1}^{c_{1}N^{2/3}}\mathbb{V}(B_{i})<2b_{c}^{2}\right)+\frac{\nu}{2}\nonumber \\
 & \le & \exp\left(-\frac{\frac{1}{4}a_{c}^{2}}{a_{c}+4b_{c}^{2}}\right)+\frac{\nu}{2}\le\nu.\label{eq:c_1=000020and=000020c}
\end{eqnarray}
After choosing a large enough $N$ and $c$, the second inequality
follows from \eqref{eq:a_bound}; the third inequality follows from
\eqref{eq:b_bound}; the fourth inequality follows from $\sum_{i=1}^{N_{1}}B_{i}\geq a,\text{ }a\geq\frac{1}{2}a_{c}\Rightarrow\sum_{i=1}^{N_{1}}B_{i}\geq\frac{1}{2}a_{c}$;
the fifth inequality follows from Bernstein inequality. Consequently,
the conclusion $N_{1}=O_{p}(N^{2/3})$ follows.

\subsection{Proof of Theorem \ref{thm:uc_isoton}\protect\label{appsub:Proof-T-uc-iso}}

Let $q_{\alpha}$ denote the $\alpha$-th quantile of $X$. We define
the following sequences of positive numbers
\begin{eqnarray*}
a_{N} & = & X_{N_{1}},\qquad b_{N}=q_{(c_{1}+1)N^{-1/3}},\\
c_{N} & = & q_{1-(c_{K}+1)N^{-1/3}},\qquad d_{N}=X_{n_{K}},
\end{eqnarray*}
where $n_{K}$ is defined in Section \ref{subsec:setup}, which is
the first element of partition $K$ (the last partition). $c_{1}$
is defined in \eqref{eq:c1}. $c_{K}$ is defined similarly to $c_{1}$
by $c_{K}N^{2/3}=N_{K}$, where $N_{K}$ is the number of elements
partition $K.$ Without a loss of generality, we assume that $N$
is large enough to ensure that both quantiles $b_{N}$ and $c_{N}$
are well defined. For given $N$, the UC-isotonic estimator estimates
the following function
\[
p_{N}(x)=\begin{cases}
\frac{\mathbb{E}[p(X)\mathbb{I}(X<a_{N})]}{\mathbb{P}(X<a_{N})} & \text{if }x\in[x_{L},a_{N})\\
p(x) & \text{if }x\in[a_{N},d_{N}]\\
\frac{\mathbb{E}[p(X)\mathbb{I}(X>d_{N})]}{\mathbb{P}(X>d_{N})} & \text{if }x\in(d_{N},x_{U}].
\end{cases}
\]
It is shown in the following figure. 

\begin{figure}[H]
\caption{\protect\label{fig:uc_iso} UC-isotonic estimator}

\centering

\includegraphics[clip,scale=0.55]{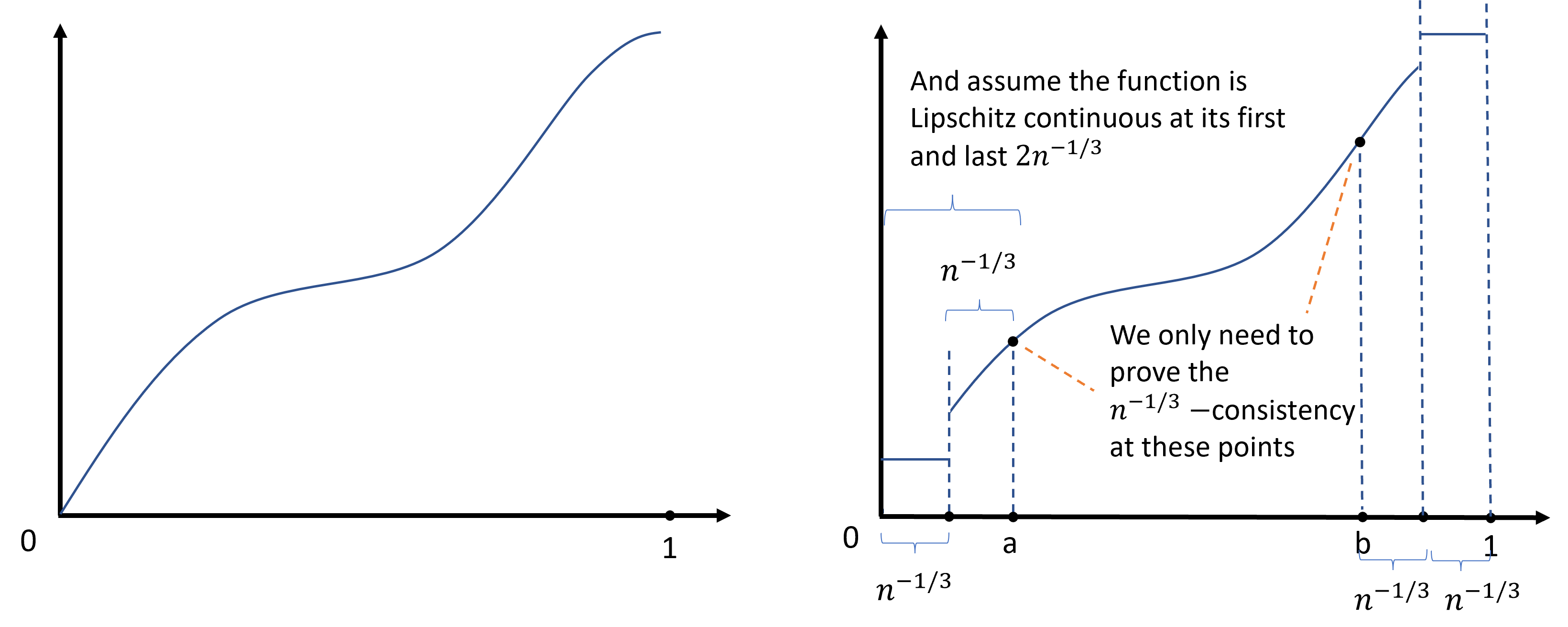}

The left panel is $p(x)$, and the right panel is $p_{N}(x)$.
\end{figure}

The conclusion of Theorem \ref{thm:uc_isoton} follows by showing
these steps.

\textbf{Step 1}: $\sup_{x\in[x_{L},a_{N}]}|\tilde{p}(x)-p(x)|=O_{p}(N^{-1/3})$
and $\sup_{x\in[d_{N},x_{U}]}|\tilde{p}(x)-p(x)|=O_{p}(N^{-1/3}).$

\textbf{Step 2}: $\sup_{x\in[b_{N},c_{N}]}|\tilde{p}(x)-p(x)|=O_{p}\left(\frac{\log N}{N}\right)^{1/3}.$

\textbf{Step 3}: $\sup_{x\in(a_{N},b_{N})}|\tilde{p}(x)-p(x)|=O_{p}\left(\frac{\log N}{N}\right)^{1/3}$
and $\sup_{x\in(c_{N},d_{N})}|\tilde{p}(x)-p(x)|=O_{p}\left(\frac{\log N}{N}\right)^{1/3}.$

\paragraph*{Step 1}

Note that $\tilde{p}(a_{N})=\tilde{p}(x)$ for each $x\in[x_{L},a_{N}]$.
Therefore, for $\sup_{x\in[x_{L},a_{N}]}|\tilde{p}(x)-p(x)|=O_{p}(N^{-1/3})$,
it is enough to show that 
\[
\sup_{x\in[x_{L},a_{N}]}|\tilde{p}(a_{N})-p(x)|=O_{p}(N^{-1/3}).
\]
First, by Assumption \ref{asm:2} (iii) and Proposition \ref{prop:matching-score-modified}
(iii), we have $a_{N}-x_{L}=O_{p}(N^{-1/3}),$ which implies
\begin{equation}
x-x_{L}=O_{p}(N^{-1/3})\quad\text{for }x\in[x_{L},a_{N}].\label{eq:x_to_xL}
\end{equation}
Thus, we have
\begin{eqnarray*}
\tilde{p}(a_{N}) & = & \frac{1}{N_{1}}\sum_{i=1}^{N_{1}}W_{i}=\frac{1}{c_{1}N^{2/3}}\sum_{i=1}^{c_{1}N^{2/3}}W_{i}\\
 & = & \frac{\int_{x_{L}}^{q_{c_{1}\cdot N^{-1/3}}}p(x)f(x)dx}{\int_{x_{L}}^{q_{c_{1}\cdot N^{-1/3}}}f(x)dx}+O_{p}((c_{1}N^{2/3})^{-1/2})+O_{p}(N^{-1/2})\\
 & = & \frac{\int_{x_{L}}^{q_{c_{1}\cdot N^{-1/3}}}\{p(x_{L})+p^{(1)}(x_{L})\cdot(x_{c_{1}}-x_{L})\}f(x)dx}{\int_{x_{L}}^{q_{c_{1}\cdot N^{-1/3}}}f(x)dx}+O_{p}(N^{-1/3})\\
 & = & p(x_{L})+O_{p}(c_{1}\cdot N^{-1/3})+O_{p}(N^{-1/3})=p(x)+O_{p}(N^{-1/3}),
\end{eqnarray*}
where the first equality follows from Proposition \ref{prop:matching-score-modified}
(ii); the $O_{p}(N^{-1/2})$ term in the third equality follows from
that $X_{c_{1}N^{2/3}}$ can estimate $q_{c_{1}\cdot N^{-1/3}}$ at
the rate of $O_{p}(N^{-1/2})$; $x_{c_{1}}$ in the fourth equality
is a number within the interval $(x_{L},q_{c_{1}\cdot N^{-1/3}})$;
the fifth equality follows from $x_{c_{1}}-x_{L}=O_{p}(c_{1}\cdot N^{-1/3})$;
the sixth equality follows from \eqref{eq:x_to_xL} and Assumption
\ref{asm:2}; and the last equality follows from $c_{1}=O_{p}(1)$,
which is implied by \eqref{eq:c_1=000020and=000020c}. 

Similarly, we can show $\sup_{x\in[d_{N},x_{U}]}|\tilde{p}(x)-p(x)|=O_{p}(N^{-1/3}).$

\paragraph*{Step 2}

The result $\sup_{x\in[b_{N},c_{N}]}|\tilde{p}(x)-p(x)|=O_{p}\left(\frac{\log N}{N}\right)^{1/3}$
follows by Durot, Kulikov, and Lopuhaä (2012, Theorem 2.1). We can
adapt the domain from $[0,1]$ in their theorem to $[x_{L},x_{U}]$
in our problem, and their conditions (A1)-(A3) hold under Assumptions
\ref{asm:1} and \ref{asm:2}.

\paragraph*{Step 3}

The statement follows directly by combining the results from Steps
1 and 2 with Assumption \ref{asm:2}, $b_{N}-a_{N}=O_{p}(N^{-1/3})$,
and $d_{N}-c_{N}=O_{p}(N^{-1/3})$.

Combining these steps, the conclusion of this theorem follows.

\subsection{Proof of Theorem \ref{thm:matching_IPW}\protect\label{appsub:equivalent-result}}

If $X_{i}$ is in the $k$-th partition given by the UC-isotonic estimator
(i.e., $i\in\{n_{k},\ldots,(n_{k}+N_{k}-1)\}$), then we have $M_{i}=N_{k,1-W_{i}}$.
Thus, the matching estimator $\hat{\tau}$ is written as{\small

\begin{eqnarray}
\hat{\tau} & = & \frac{1}{N}\sum_{i=1}^{N}(2W_{i}-1)\left(Y_{i}-\frac{1}{M_{i}}\sum_{j\in\mathcal{J}(i)}Y_{j}\right)=\frac{1}{N}\sum_{k=1}^{K}\left\{ \sum_{i=n_{k}}^{n_{k}+N_{k}-1}(2W_{i}-1)\left(Y_{i}-\frac{1}{N_{k,1-W_{i}}}\sum_{j\in\mathcal{J}(i)}Y_{j}\right)\right\} \nonumber \\
 & = & \frac{1}{N}\sum_{k=1}^{K}\left\{ \sum_{i\in\{n_{k},\ldots,(n_{k}+N_{k}-1)\},W_{i}=1}\left(Y_{i}-\frac{1}{N_{k,0}}\sum_{j\in\{n_{k},\ldots,(n_{k}+N_{k}-1)\},W_{j}=0}Y_{j}\right)\right.\nonumber \\
 &  & \qquad\left.+\sum_{i\in\{n_{k},\ldots,(n_{k}+N_{k}-1)\},W_{i}=0}\left(\frac{1}{N_{k,1}}\sum_{j\in\{n_{k},\ldots,(n_{k}+N_{k}-1)\},W_{j}=1}Y_{j}-Y_{i}\right)\right\} \nonumber \\
 & = & \frac{1}{N}\sum_{k=1}^{K}\left\{ \left(1+\frac{N_{k,0}}{N_{k,1}}\right)\sum_{i\in\{n_{k},\ldots,(n_{k}+N_{k}-1)\},W_{i}=1}Y_{i}-\left(\frac{N_{k,1}}{N_{k,0}}+1\right)\sum_{i\in\{n_{k},\ldots,(n_{k}+N_{k}-1)\},W_{i}=0}Y_{i}\right\} \nonumber \\
 & = & \frac{1}{N}\sum_{k=1}^{K}\left\{ \frac{N_{k,1}+N_{k,0}}{N_{k,1}}\sum_{i\in\{n_{k},\ldots,(n_{k}+N_{k}-1)\},W_{i}=1}Y_{i}-\frac{N_{k,1}+N_{k,0}}{N_{k,0}}\sum_{i\in\{n_{k},\ldots,(n_{k}+N_{k}-1)\},W_{i}=0}Y_{i}\right\} \nonumber \\
 & = & \frac{1}{N}\sum_{k=1}^{K}\left\{ \sum_{i\in\{n_{k},\ldots,(n_{k}+N_{k}-1)\},W_{i}=1}\frac{Y_{i}}{N_{k.1}/N_{k}}-\sum_{i\in\{n_{k},\ldots,(n_{k}+N_{k}-1)\},W_{i}=0}\frac{Y_{i}}{N_{k.0}/N_{k}}\right\} \nonumber \\
 & = & \frac{1}{N}\sum_{k=1}^{K}\left\{ \sum_{i\in\{n_{k},\ldots,(n_{k}+N_{k}-1)\}}\left(\frac{W_{i}Y_{i}}{\tilde{p}(X_{i})}-\frac{(1-W_{i})Y_{i}}{1-\tilde{p}(X_{i})}\right)\right\} \nonumber \\
 & = & \frac{1}{N}\sum_{i=1}^{N}\left(\frac{W_{i}Y_{i}}{\tilde{p}(X_{i})}-\frac{(1-W_{i})Y_{i}}{1-\tilde{p}(X_{i})}\right),\label{eq:IPW_tau_ATE}
\end{eqnarray}
}where the first equality is the formula of matching estimator for
ATE (see, e.g., Abadie and Imbens, 2016), with a changing matched
set of size $M_{i}$, the fourth equality follows from the fact that
with $w\in\{0,1\}$ and $i\ne j$, we have $\sum_{i\in n_{k}:(n_{k}+N_{k}-1),W_{i}=w}Y_{j}=N_{k,w}\cdot Y_{j}$,
the second last equality follows from Lemma \ref{prop:matching-score-modified},
and the last equality follows from $\sum_{k=1}^{K}N_{k}=N$. 

\subsection{Proof of Theorem \ref{thm:ATE-matching}\protect\label{appsub:Proof-of-T_univariate}}

Given Theorem \ref{thm:matching_IPW}, it is sufficient to show that
the last line of \eqref{eq:IPW_tau_ATE} has the desired properties.
By the Taylor extension, 
\begin{eqnarray}
\frac{W_{i}Y_{i}}{\tilde{p}(X_{i})}-\frac{(1-W_{i})Y_{i}}{1-\tilde{p}(X_{i})} & = & \left(\frac{W_{i}Y_{i}}{p(X_{i})}-\frac{(1-W_{i})Y_{i}}{1-p(X_{i})}\right)-\left(\frac{W_{i}Y_{i}}{p(X_{i})^{2}}+\frac{Y_{i}(1-W_{i})}{\{1-p(X_{i})\}^{2}}\right)\{\tilde{p}(X_{i})-p(X_{i})\}\nonumber \\
 &  & +2\left(\frac{W_{i}Y_{i}}{\check{p}_{i}^{3}}-\frac{Y_{i}(1-W_{i})}{\{1-\check{p}_{i}\}^{3}}\right)\{\tilde{p}(X_{i})-p(X_{i})\}^{2},\label{eq:ATE_taylor}
\end{eqnarray}
where the random variable $\check{p}_{i}$ takes values in the interval
between $\tilde{p}(X_{i})$ and $p(X_{i})$. For $Z=(Y,X,W)$, we
define 
\begin{equation}
D(Z)=\frac{WY}{p(X)^{2}}+\frac{Y(1-W)}{\{1-p(X)\}^{2}}.\label{eq:D(z)}
\end{equation}
For any $x\in\mathcal{X}$, we denote $\mathbb{E}[D(Z)|X=x]$ by $\mathbb{E}[D(Z)|x]$.
Then, the second term in the first line of \eqref{eq:ATE_taylor}
can be written as:

\begin{eqnarray*}
 &  & -D(Z_{i})\{\tilde{p}(X_{i})-p(X_{i})\}\\
 & = & -\mathbb{E}[D(Z_{i})|X_{i}]\{\tilde{p}(X_{i})-p(X_{i})\}-\{D(Z_{i})-\mathbb{E}[D(Z_{i})|X_{i}]\}\{\tilde{p}(X_{i})-p(X_{i})\}\\
 & = & -\mathbb{E}[D(Z_{i})|X_{i}][\{W_{i}-p(X_{i})\}-\{W_{i}-\tilde{p}(X_{i})\}]\\
 &  & -\{D(Z_{i})-\mathbb{E}[D(Z_{i})|X_{i}]\}\{\tilde{p}(X_{i})-p(X_{i})\}\\
 & = & -\mathbb{E}[D(Z_{i})|X_{i}]\{W_{i}-p(X_{i})\}+\mathbb{E}[D(Z_{i})|X_{i}]\{W_{i}-\tilde{p}(X_{i})\}\\
 &  & -\{D(Z_{i})-\mathbb{E}[D(Z_{i})|X_{i}]\}\{\tilde{p}(X_{i})-p(X_{i})\}.
\end{eqnarray*}
Plugging it back to \eqref{eq:ATE_taylor}, we have
\begin{eqnarray}
\hat{\tau}-\tau & = & \frac{1}{N}\sum_{i=1}^{N}\left[\left(\frac{W_{i}Y_{i}}{p(X_{i})}-\frac{(1-W_{i})Y_{i}}{1-p(X_{i})}-\tau\right)-\mathbb{E}[D(Z_{i})|X_{i}]\{W_{i}-p(X_{i})\}\right]\nonumber \\
 &  & +\frac{1}{N}\sum_{i=1}^{N}\mathbb{E}[D(Z_{i})|X_{i}]\{W_{i}-\tilde{p}(X_{i})\}\nonumber \\
 &  & -\frac{1}{N}\sum_{i=1}^{N}\{D(Z_{i})-\mathbb{E}[D(Z_{i})|X_{i}]\}\{\tilde{p}(X_{i})-p(X_{i})\}\nonumber \\
 &  & +\frac{2}{N}\sum_{i=1}^{N}\left(\frac{W_{i}Y_{i}}{\check{p}_{i}^{3}}-\frac{Y_{i}(1-W_{i})}{\{1-\check{p}_{i}\}^{3}}\right)\{\tilde{p}(X_{i})-p(X_{i})\}^{2}\nonumber \\
 & =: & I+II-III+IV.\label{eq:ATE_s_decom}
\end{eqnarray}
We will show the asymptotic properties of these four terms in the
subsequent subsections.

\subsubsection{The limit of $I$\protect\label{subsec:The-limit-of-I}}

Note that $\mathbb{E}[D(Z_{i})|X_{i}]=\mathbb{E}\left[\left.\frac{W_{i}Y_{i}}{^{p(X_{i})^{2}}}+\frac{(1-W_{i})Y_{i}}{\{1-p(X_{i})\}^{2}}\right|X_{i}\right]=\frac{\mathbb{E}[Y(1)]}{^{p(X_{i})}}+\frac{\mathbb{E}[Y(0)]}{1-p(X_{i})}.$
Therefore, by Theorem 1 of Hirano, Imbens and Ridder (2003) (see also
their equations (12) and (38)), it holds 
\begin{equation}
\sqrt{N}\cdot I\overset{d}{\to}N(0,\Omega),\label{eq:uni_rate_I}
\end{equation}
where $\Omega=\mathbb{V}(\mathbb{E}[Y(1)-Y(0)|X])+\mathbb{E}[\mathbb{V}(Y(1)|X)/p(X)]+\mathbb{E}[\mathbb{V}(Y(0)|X)/(1-p(X))]$. 

\subsubsection{\protect\label{subsec:Uni-rate-of=000020II}The rate of $II$}

Since $\tilde{p}(\cdot)$ is the isotonic estimator of regressing
$\{\tilde{W}_{i}\}_{i=1}^{N}$ on $\{X_{i}\}_{i=1}^{N}$, by the construction
of the isotonic estimator (see, e.g., Lemmas 2.1 and 2.3 in Groeneboom
and Jongbloed, 2014; see also Barlow and Brunk, 1972), we have $\sum_{i=n_{k}}^{n_{k+1}-1}\{\tilde{W}_{i}-\tilde{p}(X_{i})\}=0$
for each $k=1,\ldots,K$. (For the last summand, we can simply set
$n_{K+1}=N+1$.) By Proposition \ref{prop:matching-score-modified}
(i) and the construction of $\tilde{W}$ (given by \eqref{eq:data_transform}),
it holds that $\sum_{i=n_{k}}^{n_{k+1}-1}\{W_{i}-\tilde{p}(X_{i})\}=0$
for each $k=1,\ldots,K$. As a result, 
\begin{equation}
\sum_{k=1}^{K}m_{k}\sum_{i=n_{k}}^{n_{k+1}-1}\{W_{i}-\tilde{p}(X_{i})\}=0\label{eq:lemma1}
\end{equation}
holds for any weights $\{m_{k}\}_{k=1}^{K}$. To proceed, we define
the function $\delta(x)$ as
\begin{equation}
\delta(x)=\mathbb{E}[D(Z)|X=x],\label{eq:E_D(x)}
\end{equation}
and its corresponding step function $\bar{\delta}_{N}(x)$ as
\[
\bar{\delta}_{N}(x)=\begin{cases}
\delta(X_{n_{k}}) & \text{if }p(x)>\tilde{p}(X_{n_{k}})\text{ for all }x\in(X_{n_{k}},X_{n_{k+1}})\\
\delta(s) & \text{if }p(s)=\tilde{p}(s)\text{ for some }s\in(X_{n_{k}},X_{n_{k+1}})\\
\delta(X_{n_{k+1}}) & \text{if }p(x)<\tilde{p}(X_{n_{k}})\text{ for all }x\in(X_{n_{k}},X_{n_{k+1}})
\end{cases},
\]
for each $x\in[X_{n_{k}},X_{n_{k+1}})$ with $k=1,\ldots,K$ (if $k=K$,
set $X_{n_{k+1}}=\text{\ensuremath{\underset{k}{\text{max }}}}X_{n_{k}}$).
Further, we define $z=(y,w,x)$ as a given vector belonging to the
domain of the random vector $Z=(Y,W,X)$. By \eqref{eq:lemma1},
it holds
\[
\int\bar{\delta}_{N}(x)\{w-\tilde{p}(x)\}d\mathbb{P}_{N}(z)=0,
\]
 where $\mathbb{P}_{N}$ is the empirical measure. Thus, we have 
\begin{eqnarray}
II & = & \frac{1}{N}\sum_{i=1}^{N}\mathbb{E}[D(Z_{i})|X_{i}]\{W_{i}-\tilde{p}(X_{i})\}=\int\delta(x)\{w-\tilde{p}(x)\}d\mathbb{P}_{N}(z)\nonumber \\
 & = & \int\{\delta(x)-\bar{\delta}_{N}(x)\}\{w-\tilde{p}(x)\}d\mathbb{P}_{N}(z).\label{eq:e-bar1}
\end{eqnarray}
By definition, $\delta(x)$ is a bounded function with a finite total
variation, so is $\bar{\delta}_{N}(x)$. For $\mathbb{P}_{0}$ denoting
the joint probability measure of $(Y,W,X)$, the last row of \eqref{eq:e-bar1}
can be decomposed as:
\begin{eqnarray*}
 &  & \int\{\delta(x)-\bar{\delta}_{N}(x)\}\{w-\tilde{p}(x)\}d\mathbb{P}_{n}(z).\\
 & = & \int\{\delta(x)-\bar{\delta}_{N}(x)\}\{w-\tilde{p}(x)\}d(\mathbb{P}_{n}(z)-\mathbb{P}_{0}(z))\\
 &  & +\int\{\delta(x)-\bar{\delta}_{N}(x)\}\{w-p(x)\}d\mathbb{P}_{0}(z)+\int\{\delta(x)-\bar{\delta}_{N}(x)\}\{p(x)-\tilde{p}(x)\}d\mathbb{P}_{0}(z)\\
 & =: & II_{1}+II_{2}+II_{3}.
\end{eqnarray*}
It shall be noted that by Assumption \ref{asm:4}, $\delta(x)-\bar{\delta}_{N}(x)$
is a bounded function with a finite total variation, and $\delta(x)$
is continuously differentiable in $x$. Under Assumption \ref{asm:2}
(ii) and similar arguments following (10.64) of Groeneboom and Jongbloed
(2014), it holds that for some $C_{0}>0$ and all $x\in\mathcal{X}$,
\begin{equation}
|\delta(x)-\bar{\delta}_{N}(x)|\leq C_{0}|p(x)-\tilde{p}(x)|.\label{eq:ebar-C}
\end{equation}

\subsubsection*{The rate of $II_{1}$}

For $R:=\max\{|x_{L}|,|x_{u}|\}$ and a positive constant $K$, let
us define
\begin{eqnarray}
\mathcal{M}_{RK} & = & \{\text{monotone increasing functions on \ensuremath{[-R,R]} and bounded by \ensuremath{K}}\},\nonumber \\
\mathcal{G}_{RK} & = & \{g:g(x)=p(x),x\in\mathcal{X},p\in\mathcal{M}_{RK}\},\nonumber \\
\mathcal{D}_{RKv} & = & \{d:d(x)=g_{1}(x)-g_{2}(x),\text{ }(g_{1},g_{2})\in\mathcal{\mathcal{G}}_{RK}^{2},\text{ }\left\Vert d(\cdot)\right\Vert _{\mathbb{P}_{0}}\leq v\},\nonumber \\
\mathcal{H}_{RKv} & = & \{h:h(y,x)=wd_{1}(x)-d_{2}(x),\text{ }(d_{1},d_{2})\in\mathcal{D}_{RKv}^{2},\text{ }z\in\mathcal{Z}\},\label{pf:H}
\end{eqnarray}
where $\left\Vert f\right\Vert _{\mathbb{P}}=\sqrt{\int|f(x)|^{2}d\mathbb{P}(x)}$
denotes the $L_{2}(\mathbb{P})$ norm of function $f$, given the
probability measure $\mathbb{P}$. Then, the integrand of $II_{1}$
can be written as
\begin{equation}
\{\delta(x)-\bar{\delta}_{N}(x)\}\{w-\tilde{p}(x)\}=\{\delta(x)-\bar{\delta}_{N}(x)\}w-\{\delta(x)-\bar{\delta}_{N}(x))\}\tilde{p}(x).\label{eq:integrand}
\end{equation}
Furthermore, we define
\[
\mathcal{F}_{a}=\big\{ f:f(z)=\{\delta(x)-\bar{\delta}_{N}(x)\}w-\{\delta(x)-\bar{\delta}_{n}(x)\}\tilde{p}(x),\text{ }z\in\mathcal{Z}\big\}.
\]

We note the following points:

(i) By Assumption \ref{asm:2} and the construction of $\bar{\delta}_{N}(x)$,
$\{\delta(x)-\bar{\delta}_{N}(x)\}$ is a bounded function of $x$
with a finite total variation. 

(ii) $\tilde{W}\in[0,1]$ implies that $\sup_{x\in\mathcal{X}}|\tilde{p}(x)|\leq1$.
Therefore, for any constant $K_{1}>1$, it holds that $\tilde{p}(x)\in\mathcal{G}_{RK_{1}}$. 

(iii) By Theorem \ref{thm:uc_isoton} and \eqref{eq:ebar-C}, there
exists some constant $C_{1}>0$, such that $\left\Vert \delta(x)-\bar{\delta}_{N}(x)\right\Vert _{\mathbb{P}}\leq C_{1}\left(\frac{\log N}{N}\right)^{1/3}$
holds with the probability arbitrarily close to one (hereafter denoted
as w.p.a.1) if we choose a large enough $C_{1}$. Therefore, by (i)
and Lemma 21 of the supplemental material of BGH (BGH-supp hereafter),\footnote{The lemma states that a function $f$, which is bounded and has a
finite total variation, can be decomposed as $f=f_{1}-f_{2}$, where
both $f_{1}$ and $f_{2}$ are bounded and monotone increasing.} for a constant $C_{2}$ that is larger than twice the bound of $\delta(x)$
(which is guaranteed by Assumption \ref{asm:4}) and $v_{1}=C_{1}\left(\frac{\log N}{N}\right)^{1/3}$,
it holds that $\{\delta(x)-\bar{\delta}_{N}(x)\}\in\mathbb{\mathcal{D}}_{RC_{2}v_{1}}$,
w.p.a.1. 

(iv) By (i), (ii), a similar argument to that (iii), Theorem \ref{thm:uc_isoton},
the Jensen's inequality, and the fact that the product of two monotone
increasing functions remains monotone increasing, we have $\{\delta(x)-\bar{\delta}_{n}(x)\}\tilde{p}(x)\in\mathcal{D}_{RKv}$
holds for constants $K=C_{2}K_{1}$ and $v=v_{1}K_{1}$, w.p.a.1. 

(v) By definition of function classes presented by \eqref{pf:H},
we have $\mathcal{F}_{a}\subseteq\mathcal{H}_{RKv}$, w.p.a.1. Let
$N_{[]}(\epsilon,\mathcal{F},\left\Vert \cdot\right\Vert )$ be the
$\epsilon$-bracketing number of the function class $\mathcal{F}$
under the norm $\left\Vert \cdot\right\Vert $, and
\[
H_{B}(\epsilon,\mathcal{F},\left\Vert \cdot\right\Vert )=\text{log}N_{[]}(\epsilon,\mathcal{F},\left\Vert \cdot\right\Vert )
\]
 be the entropy of $N_{[]}(\varepsilon,\mathcal{F},\left\Vert \cdot\right\Vert )$.
Furthermore, let us define
\[
J_{B}(\delta,\mathcal{F},\left\Vert \cdot\right\Vert ):=\int_{0}^{\delta}\sqrt{1+H_{B}(\epsilon,\mathcal{F},\left\Vert \cdot\right\Vert )}d\epsilon.
\]
By Theorem 2.7.5 in van der Vaart and Wellner (1996) and Lemma 11
in BGH-supp, given a positive constant $C$ and a bracket size $\epsilon$,
there exists a constant $A>0$, such that the entropy of the function
class $\mathcal{D}_{RCv}$ satisfies
\[
H_{B}(\epsilon,\mathcal{D}_{RCv},\left\Vert \cdot\right\Vert _{\mathbb{P}_{0}})\leq\frac{AC}{\epsilon}.
\]

Let $(d_{1}^{L},d_{1}^{U})$ and $(d_{2}^{L},d_{2}^{U})$ be two $\epsilon$-brackets
for the function class $\mathcal{D}_{RKv}$. Now we calculate the
entropy of $\mathcal{H}_{RKv}$, by using a set of brackets derived
from the $\epsilon$-brackets of $\mathcal{D}_{RKv}$. Note that $w\in\{0,1\}$
is non-negative. Then, we can define a bracket $(h^{L},h^{U})$ within
$\mathcal{H}_{RKv}$ as
\[
h^{L}=wd_{1}^{L}(x)-d_{2}^{U}(x),\qquad h^{U}=wd_{1}^{U}(x)-d_{2}^{L}(x),
\]
and its size is
\begin{eqnarray*}
 &  & \sqrt{\int\{h^{U}(z)-h^{L}(z)\}^{2}d\mathbb{P}_{0}(z)}\\
 & \leq & \sqrt{2\left[\int w^{2}\{d^{U}(x)-d^{L}(x)\}^{2}d\mathbb{P}_{0}(z)+\int\{d^{U}(x)-d^{L}(x)\}^{2}d\mathbb{P}_{0}(z)\right]}\\
 & \leq & \sqrt{4\int\{d^{U}(x)-d^{L}(x)\}^{2}d\mathbb{P}_{0}(x)}\leq2\epsilon,
\end{eqnarray*}
where the last inequality follows from the definition of $\epsilon$-bracket
with respect to (w.r.t.) the $L_{2}(\mathbb{P}_{0})$ norm. As a result,
for a constant $\tilde{A}>0$, it holds
\begin{equation}
H_{B}(\epsilon,\mathcal{H}_{RCv},\left\Vert \cdot\right\Vert _{\mathbb{P}_{0}})\leq\frac{\tilde{A}C}{\epsilon}.\label{eq:H_B_number}
\end{equation}
Combining the point (v) above and \eqref{eq:H_B_number}, there exists
$C_{3}>0$, such that 
\begin{equation}
H_{B}(\epsilon,\mathcal{F}_{a},\left\Vert \cdot\right\Vert _{\mathbb{P}_{0}})\leq\frac{C_{3}}{\epsilon}\label{pf:HB}
\end{equation}
holds w.p.a.1. For $f_{a}\in\mathcal{F}_{a}$, points (iii) and (iv)
imply that
\begin{equation}
\left\Vert f_{a}\right\Vert _{\mathbb{P}_{0}}\leq C_{1}\left(\frac{\log N}{N}\right)^{1/3}\label{pf:deltan}
\end{equation}
holds w.p.a.1. 

We have defined $\mathbb{P}_{0}$ to be the joint probability measure
of $Z=(Y,W,X)$. With some abuse of notation, in the following, $\mathbb{P}$
will be used to denote $\mathbb{P}_{0}$ whenever the context permits
without risk of confusion. Further, we use $\mathscr{E}$ to denote
the event that both \eqref{pf:HB} and \eqref{pf:deltan} happen.
Note that we can select sufficiently large constants to ensure that
$\mathbb{P}(\mathscr{E})$ approaches as close to one as desired. 

In the following, we define $\left\Vert \mathbb{G}_{N}\right\Vert _{\mathcal{F}}=\sup_{f\in\mathcal{F}}|\sqrt{N}(\mathbb{P}_{N}-\mathbb{P}_{0})f|$,
and we use $J_{B}(\delta)$ to denote $J_{B}(\delta,\mathcal{F}_{a},\left\Vert \cdot\right\Vert _{B,\mathbb{P}_{0}})$.
Let $\eta_{N}:=C_{1}\left(\frac{\log N}{N}\right)^{1/3}$. For any
positive constants $B$ and $\nu$, there exist positive constants
$B_{1}$, $B_{2}$, and $C_{2}$, such that for all $N$ large enough,
\begin{eqnarray}
 &  & \mathbb{P}(|II_{1}|>BN^{-1/2})\le\mathbb{P}(|II_{1}|>BN^{-1/2},\mathscr{E})+\mathbb{P}(\mathscr{E}^{c})\nonumber \\
 & \le & \mathbb{P}\left(\left\Vert \mathbb{G}_{N}\right\Vert _{\mathcal{F}_{a}}>B,\mathscr{E}\right)+\frac{\nu}{2}\le\frac{\mathbb{E}\left(\left\Vert \mathbb{G}_{N}\right\Vert _{\mathcal{F}_{a}}|\mathscr{E}\right)}{B}+\frac{\nu}{2}\nonumber \\
 & \lesssim & \frac{1}{B}J_{B}(\eta_{N})\left(1+\frac{J_{B}(\eta_{N})}{\sqrt{N}\eta_{N}^{2}}\right)+\frac{\nu}{2}\lesssim\frac{1}{B}(\eta_{N}+2B_{1}^{1/2}\eta_{N}^{1/2})\left(1+\frac{\eta_{N}+2B_{1}^{1/2}\eta_{N}^{1/2}}{\sqrt{N}\eta_{N}^{2}}\right)+\frac{\nu}{2}\nonumber \\
 & \lesssim & C_{2}\left(\frac{\log N}{N}\right)^{1/6}\times\left(1+\frac{B_{2}}{\left(\log N\right)^{\frac{1}{2}}}\right)+\frac{\nu}{2}\leq\nu,\label{pf:entro}
\end{eqnarray}
where the third inequality follows from the Markov inequality; the
first wave inequality ($\lesssim$) follows from Lemma 3.4.2 of van
der Vaart and Wellner (1996); the second wave inequality follows from
(\ref{pf:HB}) and equation (.2) in BGH-supp;\footnote{Equation (.2) in BGH-supp states that $J_{B}(\delta)\leq\delta+2C^{1/2}\delta^{1/2}$
holds for some positive constant $C$.} the third wave inequality follows from $\ensuremath{\eta_{N}\lesssim\eta_{N}^{1/2}}$
and the definition of $\eta_{N}$. Since $\nu$ can be chosen arbitrarily
small, we obtain
\begin{equation}
II_{1}=o_{p}(N^{-1/2}).\label{pf:IIa}
\end{equation}

\subsubsection*{The rate of $II_{2}$}

By the Law of iterated expectation and $p(x)=\mathbb{E}[W|X=x]$,
\begin{eqnarray*}
II_{2} & = & \int\{\delta(x)-\bar{\delta}_{N}(x)\}\{w-p(x)\}d\mathbb{P}_{0}(z)\\
 & = & \int\{\delta(x)-\bar{\delta}_{N}(x)\}\mathbb{E}[\{W-p(X)\}|X=x]d\mathbb{P}_{0}(x)=0.
\end{eqnarray*}

\subsubsection*{The rate of $II_{3}$}

The inequality \eqref{eq:ebar-C} implies:
\begin{eqnarray*}
II_{3} & = & \int\{\delta(x)-\bar{\delta}_{N}(x)\}\{p(x)-\tilde{p}(x)\}d\mathbb{P}_{0}(z)\\
 & \lesssim & \int\{p(x)-\tilde{p}(x)\}^{2}d\mathbb{P}_{0}(z)=O_{p}\left(\left(\frac{\log N}{N}\right)^{2/3}\right)=o_{p}(N^{-1/2}),
\end{eqnarray*}
where the wave inequality follows from \eqref{eq:ebar-C}, and the
second equality follows from Theorem \ref{thm:uc_isoton}. Combining
the rates for $II_{1}$, $II_{2}$, and $II_{3}$, we have 
\begin{equation}
II=\frac{1}{N}\sum_{i=1}^{N}\mathbb{E}[D(Z_{i})|X_{i}]\{W_{i}-\tilde{p}(X_{i})\}=o_{p}(N^{-1/2}).\label{eq:uni_rate_II}
\end{equation}

\subsubsection{The rate of $III$\protect\label{subsec:Uni-rate-of-III}}

By defining $V(\cdot)=D(\cdot)-\mathbb{E}[D(Z)|X=\cdot]$, we have
\begin{eqnarray*}
III & = & \int V(z)\{\tilde{p}(x)-p(x)\}d\mathbb{P}_{N}(z)\\
 & = & \int V(z)\{\tilde{p}(x)-p(x)\}d(\mathbb{P}_{n}(z)-\mathbb{P}_{0}(z))+\int V(z)\left[\tilde{p}(x)-p(x)\right]d\mathbb{P}_{0}(z)\\
 & =: & III_{1}+III_{2}.
\end{eqnarray*}
By definition, $\mathbb{E}[V(Z)|X=x]=0$ holds for all $x\in\mathcal{X}$.
Then, $III_{2}=0$ follows from the Law of iterated expectation. 

The derivation of the rate of $III_{1}$ is similar to that of $II_{1}$.
The main difference is $D(z)$ not being assumed to be uniformly bounded
over $\mathcal{Z}$. Consequently, we shall use Lemma 3.4.3 from van
der Vaart and Wellner (1996), instead of Lemma 3.4.2. The former is
formulated w.r.t. the Bernstein norm and is suitable for unbounded
function classes. To simplify our discussion and avoid cumbersome
notation, we will reuse some notations previously introduced in Appendix
\ref{subsec:Uni-rate-of=000020II} (e.g., those notations denoting
various constants), provided it does not lead to confusion in the
context. 

Recall the function classes $\mathcal{M}_{RK}$, $\mathcal{G}_{RK}$,
and $\mathcal{D}_{RKv}$ defined in \eqref{pf:H}. Here, we define
additionally
\begin{eqnarray*}
\mathcal{H}_{RKv}^{(2)} & = & \{h:h(z)=V(z)d(x),\text{ }d(\cdot)\in\mathcal{D}_{RKv},\text{ }z\in\mathcal{Z}\},\\
\mathcal{F}_{b} & = & \big\{ f:f(z)=V(z)\{\tilde{p}(x)-p(x)\},\text{ }z\in\mathcal{Z}\big\}
\end{eqnarray*}
Let $(d^{L},d^{U})$ be any $\epsilon$-bracket within the function
class $\mathcal{D}_{RCv}$. Define
\begin{eqnarray*}
h^{L} & = & \begin{cases}
D(z)d^{L}(x) & \text{if }D(z)\geq0\\
D(z)d^{U}(x) & \text{if }D(z)<0
\end{cases},\\
h^{U} & = & \begin{cases}
D(z)d^{U}(x) & \text{if }D(z)\geq0\\
D(z)d^{L}(x) & \text{if }D(z)<0
\end{cases}.
\end{eqnarray*}
Note that $(h^{L},h^{U})$ is a bracket within $\mathcal{H}_{RKv}^{(2)}$,
and its size is
\begin{eqnarray*}
 &  & \sqrt{\int\{h^{U}(z)-h^{L}(z)\}^{2}d\mathbb{P}_{0}(z)}=\sqrt{\int D(z)^{2}\{d^{U}(x)-d^{L}(x)\}^{2}d\mathbb{P}_{0}(z)}\\
 & = & \sqrt{\int\mathbb{E}\left[D(Z)^{2}|X=x\right]\{d^{U}(x)-d^{L}(x)\}^{2}d\mathbb{P}_{0}(x)}\leq A_{1}\epsilon,
\end{eqnarray*}
for some $A_{1}>0$. The last inequality follows from Assumption \ref{asm:4}
and the definition of $\epsilon$-bracket w.r.t. the $L_{2}(\mathbb{P}_{0})$
norm. As a result, for some $\tilde{A}>0$, we have 
\begin{equation}
H_{B}(\epsilon,\mathcal{H}_{RCv}^{(2)},\left\Vert \cdot\right\Vert _{\mathbb{P}_{0}})\leq\frac{\tilde{A}C}{\epsilon}.\label{eq:H2class}
\end{equation}

We now switch to the Bernstein norm because we prefer not to impose
a bound on $D(z)$. Let $\left\Vert \cdot\right\Vert _{B,\mathbb{P}}$
be the Bernstein norm under a measure $\mathbb{P}$. By the definition,
\begin{align*}
\left\Vert h\right\Vert _{B,\mathbb{P}}^{2} & =2\mathbb{P}(\text{exp}(|h|)-|h|-1)=2\int\sum_{k=2}^{\infty}\frac{1}{k!}|h|^{k}d\mathbb{P}(z),
\end{align*}
where the second equality follows by the extension of the natural
exponential function. Next, we attempt to bound the Bernstein norm
of $H^{-1}h(\cdot)$, where $H$ is a positive number that we will
select in subsequent steps to establish a finite upper bound. For
a constant $C$ and any $h\in\mathcal{H}_{RCv}^{(2)}$, it holds
\begin{eqnarray*}
\left\Vert H^{-1}h^{U}-H^{-1}h^{L}\right\Vert _{B,\mathbb{P}_{0}}^{2} & = & 2\int\sum_{k=2}^{\infty}\frac{1}{H^{k}}\frac{1}{k!}|D(z)\{d^{U}(x)-d^{L}(x)\}|^{k}d\mathbb{P}_{0}(z)\\
 & \leq & 2\int\sum_{k=2}^{\infty}\frac{1}{H^{k}}\frac{1}{k!}|D(z)|^{k}|d^{U}(x)-d^{L}(x)|^{k}d\mathbb{P}_{0}(z)\\
 & \leq & 2\sum_{k=2}^{\infty}\frac{1}{H^{k}}\frac{(4C)^{k-2}}{k!}k!M_{0}^{k-2}c_{0}\int|d^{U}(x)-d^{L}(x)|^{2}d\mathbb{P}_{0}(x)\\
 & = & \frac{2}{H^{2}}\sum_{k=2}^{\infty}\frac{(4M_{0}C)^{k-2}}{H^{k-2}}c_{0}\int|d^{U}(x)-d^{L}(x)|^{2}d\mathbb{P}_{0}(x)\\
 & = & \frac{2}{H^{2}}\sum_{k=2}^{\infty}\left(\frac{4M_{0}C}{H}\right)^{k-2}c_{0}\epsilon^{2}=2c_{0}\left(\frac{\epsilon}{H}\right)^{2},
\end{eqnarray*}
where the second inequality follows from Assumption \ref{asm:4} and
$d(\cdot)\leq2C$ (implied by $d(\cdot)\in\mathcal{D}_{RCv}$); $c_{0}$
and $M_{0}$ are the same constants defined in Assumption \ref{asm:4}
(iii); the third equality follows from the definition of the $\epsilon$-bracket;
and the last equality follows from choosing $H=8M_{0}C$. As a result,
we have for some positive constant $C_{1}$,
\begin{equation}
H_{B}(\epsilon,\mathcal{H}_{RCv}^{(2)},\left\Vert \cdot\right\Vert _{B,\mathbb{P}_{0}})\leq\frac{C_{1}}{\epsilon}.\label{eq:Hb2-1}
\end{equation}
Further, by similar arguments, for any $h\in\mathcal{H}_{RCv}^{(2)}$,
it holds
\begin{align*}
\left\Vert H^{-1}h\right\Vert _{B,\mathbb{P}_{0}}^{2} & \leq2\int\sum_{k=2}^{\infty}\frac{1}{H^{k}}\frac{1}{k!}|D(z)|^{k}|d(x)|^{k}d\mathbb{P}_{0}(z)\\
 & =\frac{2}{H^{2}}\sum_{k=2}^{\infty}\frac{\left(2M_{0}C\right)^{k-2}}{H^{k-2}}c_{0}\int|d(x)|^{2}d\mathbb{P}_{0}(x)=\frac{2}{H^{2}}\sum_{k=2}^{\infty}\left(\frac{2M_{0}C}{H}\right)^{k-2}c_{1}v^{2}=\left(\frac{2}{H}\right)^{2}c_{1}v^{2}.
\end{align*}
where the third equality follows from $d(\cdot)\in\mathcal{D}_{RCv}.$
Thus,
\begin{equation}
\left\Vert H^{-1}h\right\Vert _{B,\mathbb{P}_{0}}\lesssim\frac{v}{H}.\label{eq:v/H}
\end{equation}

Based on these results, we now study the function class $\mathcal{F}_{b}$.
Set $C\geq1$ and $v:=C_{2}\left(\frac{\log N}{N}\right)^{1/3}$ for
some $C_{2}>0$. Then, by Theorem \ref{thm:uc_isoton} and the definitions
of $\mathcal{H}_{RCv}^{(2)}$ and $\mathcal{F}_{b}$, 
\begin{equation}
\mathcal{F}_{b}\subseteq\mathcal{H}_{RCv}^{(2)}\label{eq:F_b=000020and=000020H_2}
\end{equation}
holds w.p.a.1. Furthermore, we define
\begin{equation}
\mathcal{\tilde{F}}_{b}=H^{-1}\mathcal{F}_{b}.\label{eq:tilde_Fb}
\end{equation}
By combining \eqref{eq:Hb2-1}, \eqref{eq:F_b=000020and=000020H_2},
and \eqref{eq:tilde_Fb}, there exists $C_{2}>0$, such that
\begin{equation}
H_{B}(\epsilon,\tilde{\mathcal{F}}_{b},\left\Vert \cdot\right\Vert _{B,\mathbb{P}_{0}})\leq\frac{C_{2}}{\epsilon}\label{eq:Hb2}
\end{equation}
holds w.p.a.1. Further, for $f_{b}\in\mathcal{F}_{a}$ we define $\tilde{f}_{b}=H^{-1}f_{b}$.
By \eqref{eq:v/H} and $v=C_{2}\left(\frac{\log N}{N}\right)^{1/3}$,
there exists some $C_{3}>0$, such that
\begin{equation}
\text{ }\left\Vert \tilde{f}_{b}\right\Vert _{B,\mathbb{P}_{0}}\leq C_{3}\left(\frac{\log N}{N}\right)^{1/3}\label{eq:f_tilde2}
\end{equation}
 holds w.p.a.1. 

Again, we use $\mathscr{E}$ to denote the event that both \eqref{eq:Hb2}
and \eqref{eq:f_tilde2} happen. With sufficiently large constants
selected, $\mathbb{P}(\mathscr{E})$ can approach as close to one
as desired. For $\eta_{N}:=C_{3}\left(\frac{\log N}{N}\right)^{1/3}$
and any positive constants $B$ and $\nu$, there exist constants
$B_{1}$, $B_{2}$, and $C_{4}$, such that for all $N$ large enough,
it holds that
\begin{eqnarray}
 &  & \mathbb{P}(|III_{1}|>BN^{-1/2})\le\mathbb{P}(|II_{1}|>BN^{-1/2},\mathscr{E})+\mathbb{P}(\mathscr{E}^{c})\nonumber \\
 & \leq & \mathbb{P}\left(\left\Vert \mathbb{G}_{N}\right\Vert _{\mathcal{F}_{b}}>B,\mathscr{E}\right)+\frac{\nu}{2}\le\frac{H}{B}\mathbb{E}\left(\left\Vert \mathbb{G}_{N}\right\Vert _{\mathcal{\tilde{F}}_{b}}|\mathscr{E}\right)+\frac{\nu}{2}\nonumber \\
 & \lesssim & \frac{H}{B}J_{B}(\eta_{N})\left(1+\frac{J_{B}(\eta_{N})}{\sqrt{N}\eta_{N}^{2}}\right)+\frac{\nu}{2}\lesssim\frac{H}{B}(\eta_{N}+2B_{1}^{1/2}\eta_{N}^{1/2})\left(1+\frac{\eta_{N}+2B_{1}^{1/2}\eta_{N}^{1/2}}{\sqrt{N}\eta_{N}^{2}}\right)+\frac{\nu}{2}\nonumber \\
 & \lesssim & C_{4}\left(\frac{\log N}{N}\right)^{1/6}\times\left(1+\frac{B_{2}}{\left(\log N\right)^{\frac{1}{2}}}\right)+\frac{\nu}{2}\leq\nu,\label{pf:entro-2}
\end{eqnarray}
where the third inequality follows from the Markov inequality and
the definition of $\mathcal{\tilde{F}}_{b}$ in \eqref{eq:tilde_Fb};
the first wave inequality ($\lesssim$) comes from Lemma 3.4.3 of
van der Vaart and Wellner (1996); the second wave inequality comes
from \eqref{eq:Hb2} and equation (.2) in BGH-supp; the third wave
inequality follows from $\ensuremath{\eta_{N}\lesssim\eta_{N}^{1/2}}$
and the definition of $\eta_{N}$. Since $\nu$ can be chosen arbitrarily
small, we obtain $III_{1}=o_{p}(N^{-1/2}).$ Combined with $III_{2}=0$
yields
\begin{equation}
III=o_{p}(N^{-1/2}).\label{eq:uni_rate_III}
\end{equation}

\subsubsection{The rate of $IV$\protect\label{subsec:uni-rate-of-IV}}

Note that 
\begin{eqnarray*}
IV & = & \frac{2}{N}\sum_{i=1}^{N}\left(\frac{W_{i}}{\check{p}_{i}^{3}}-\frac{1-W_{i}}{\{1-\check{p}_{i}\}^{3}}\right)Y_{i}\{\tilde{p}(X_{i})-p(X_{i})\}^{2}\\
 & = & \frac{2}{N}\sum_{i=1}^{N}\left(\frac{W_{i}}{\check{p}_{i}^{3}}-\frac{1-W_{i}}{\{1-\check{p}_{i}\}^{3}}\right)\{Y_{i}\mathbb{I}(Y_{i}>0)\}\{\tilde{p}(X_{i})-p(X_{i})\}^{2}\\
 &  & +\frac{2}{N}\sum_{i=1}^{N}\left(\frac{W_{i}}{\check{p}_{i}^{3}}-\frac{1-W_{i}}{\{1-\check{p}_{i}\}^{3}}\right)\{Y_{i}\mathbb{I}(Y_{i}\leq0)\}\{\tilde{p}(X_{i})-p(X_{i})\}^{2}\\
 & =: & IV_{+}+IV_{-}.
\end{eqnarray*}
For $IV_{+}$, we have 
\begin{eqnarray*}
\left|IV_{+}\right| & \leq & 2\sup_{i\in\{1:N\}}\left|\frac{W_{i}}{\check{p}_{i}^{3}}-\frac{1-W_{i}}{\{1-\check{p}_{i}\}^{3}}\right|\frac{1}{N}\sum_{i=1}^{N}\{Y_{i}\mathbb{I}(Y_{i}>0)\}\{\tilde{p}(X_{i})-p(X_{i})\}^{2}\\
 & \leq & 2\sup_{i\in\{1:N\}}\left|\frac{1}{\check{p}_{i}^{3}}+\frac{1}{\{1-\check{p}_{i}\}^{3}}\right|\sup_{i\in\{1:N\}}\{\tilde{p}(X_{i})-p(X_{i})\}^{2}\frac{1}{N}\sum_{i=1}^{N}\{Y_{i}\mathbb{I}(Y_{i}>0)\}\\
 & \leq & 2\cdot\left|\frac{1}{\underline{p}^{3}}+\frac{1}{\{1-\bar{p}\}^{3}}+o_{p}(1)\right|\cdot O_{p}\left(\frac{\log N}{N}\right)^{2/3}\cdot O_{p}(1)\\
 & = & o_{p}(N^{-1/2}),
\end{eqnarray*}
where the second inequality follows from $W_{i}\in\{0,1\}$, and $\check{p}_{i}$
being non-negative (because both $\tilde{p}(X_{i})$ and $p(X_{i})$
are non-negative); the third inequality follows from Assumptions \ref{asm:3},
\ref{asm:4}, and Theorem \ref{thm:uc_isoton}. 

By a similar argument, we have
\begin{eqnarray*}
\left|IV_{-}\right| & \leq & 2\max_{1\le i\le N}\left|\frac{1}{\check{p}_{i}^{3}}+\frac{1}{\{1-\check{p}_{i}\}^{3}}\right|\max_{1\le i\le N}\{\tilde{p}(X_{i})-p(X_{i})\}^{2}\frac{1}{N}\sum_{i=1}^{N}\{-Y_{i}\mathbb{I}(Y_{i}\leq0)\}\\
 & = & o_{p}(N^{-1/2}),
\end{eqnarray*}
and thus,
\begin{equation}
IV=o_{p}(N^{-1/2}).\label{eq:uni_rate_IV}
\end{equation}

\begin{rem} {[}The role of the UC-isotonic estimator{]} We observe
the critical role played by the UC-isotonic estimator in establishing
the rate of $IV$: it allows us to uniformly bound $\frac{W_{i}}{\check{p}_{i}^{3}}-\frac{1-W_{i}}{\{1-\check{p}_{i}\}^{3}}$
from above (with probability approaching one). However, if we were
to use a standard isotonic estimator instead, there is no guarantee
that those $\frac{W_{i}}{\check{p}_{i}^{3}}-\frac{1-W_{i}}{\{1-\check{p}_{i}\}^{3}}$
at boundaries are bounded. This lack of boundedness occurs because
the bias inherent in the standard isotonic estimator at boundaries
is not mitigated by increasing the sample size. Although this bias
affects only the summands at the two shrinking boundaries, the bias
is towards zero and will be disproportionately amplified by the reciprocal
structure, considerably impacting the overall moment estimator. The
consequence is partially exemplified by column (d) of Table \ref{tab:iso_thresh_MSEs-nearly01}
in the supplementary material. This column presents the case with
a conservative per-averaging (only averaging the first and the last
$\lfloor N^{1/3}\rfloor$), which closely approximates the scenario
without pre-processing the data. \end{rem}

\subsubsection{Summary of Appendix \ref{appsub:Proof-of-T_univariate}}

Combining \eqref{eq:uni_rate_I}, \eqref{eq:uni_rate_II}, \eqref{eq:uni_rate_III}
and \eqref{eq:uni_rate_IV} yields
\[
\sqrt{N}(\hat{\tau}-\tau)\overset{d}{\to}N(0,\Omega),
\]
where $\Omega=\mathbb{V}(\mathbb{E}[Y(1)-Y(0)|X])+\mathbb{E}[\mathbb{V}(Y(1)|X)/p(X)]+\mathbb{E}[\mathbb{V}(Y(0)|X)/(1-p(X))]$. 

\subsection{Proof of Theorem \ref{thm:uc_isoton-m}}

The proof is similar to that of Theorem \ref{thm:uc_isoton}. Since
we have $\tilde{\alpha}-\alpha_{0}=O_{p}(N^{-1/2})=o_{p}(N^{-1/3})$,
the estimation of $\alpha_{0}$ does not affect the uniform convergence
rate. All the steps in Appendix \ref{appsub:Proof-T-uc-iso} can be
similarly applied with plugged-in $\tilde{\alpha}$.

\subsection{Proof of Theorem \ref{thm:ATE-matching-m}\protect\label{subsec:Proof-of-Th_m}}

Note that Proposition \ref{prop:matching-score-modified} and Corollary
\ref{cor:non-empty-matched-set} also hold for the UC-iso-index estimator
$\tilde{p}_{\tilde{\alpha}}.$ By a similar argument for the proof
of Theorem \ref{thm:matching_IPW} (in Appendix \ref{appsub:Proof-of-T_univariate}),
we have
\begin{equation}
\tilde{\tau}=\frac{1}{N}\sum_{i=1}^{N}(2W_{i}-1)\left(Y_{i}-\frac{1}{M_{i}}\sum_{j\in\mathcal{J}(i)}Y_{j}\right)=\frac{1}{N}\sum_{i=1}^{N}\left(\frac{W_{i}Y_{i}}{\tilde{p}_{\tilde{\alpha}}(X_{i}^{\prime}\tilde{\alpha})}-\frac{(1-W_{i})Y_{i}}{1-\tilde{p}_{\tilde{\alpha}}(X_{i}^{\prime}\tilde{\alpha})}\right).\label{eq:IPW_m}
\end{equation}
It remains to derive the asymptotic properties of \eqref{eq:IPW_m}.
Recall that $\mathbb{E}_{\alpha}[D(Z)|u]:=\mathbb{E}[D(Z)|X^{\prime}\alpha=u]$.
Similarly to \eqref{eq:ATE_s_decom},
\begin{eqnarray*}
\tilde{\tau}-\tau & = & \frac{1}{N}\sum_{i=1}^{N}\left[\left(\frac{W_{i}Y_{i}}{p_{0}(X_{i}^{\prime}\alpha_{0})}-\frac{(1-W_{i})Y_{i}}{1-p_{0}(X_{i}^{\prime}\alpha_{0})}-\tau\right)-\mathbb{E}_{\tilde{\alpha}}[D(Z)|X_{i}^{\prime}\tilde{\alpha}]\{W_{i}-p_{0}(X_{i}^{\prime}\alpha_{0})\}\right]\\
 &  & +\frac{1}{N}\sum_{i=1}^{N}\mathbb{E}_{\tilde{\alpha}}[D(Z)|X_{i}^{\prime}\tilde{\alpha}]\{W_{i}-\tilde{p}_{\tilde{\alpha}}(X_{i}^{\prime}\tilde{\alpha})\}\\
 &  & -\frac{1}{N}\sum_{i=1}^{N}\{D(Z_{i})-\mathbb{E}_{\tilde{\alpha}}[D(Z)|X_{i}^{\prime}\tilde{\alpha}]\}\{\tilde{p}_{\tilde{\alpha}}(X_{i}^{\prime}\tilde{\alpha})-p_{0}(X_{i}^{\prime}\alpha_{0})\}\\
 &  & +\frac{2}{N}\sum_{i=1}^{N}\left(\frac{W_{i}Y_{i}}{\check{p}_{i}^{3}}-\frac{Y_{i}(1-W_{i})}{\{1-\check{p}_{i}\}^{3}}\right)\{\tilde{p}_{\tilde{\alpha}}(X_{i}^{\prime}\tilde{\alpha})-p_{0}(X_{i}^{\prime}\alpha_{0})\}^{2}\\
 & =: & I_{m}+II_{m}-III_{m}+IV_{m}.
\end{eqnarray*}
Among these terms, the convergence rates of $I_{m}$, $II_{m}$,
and $IV_{m}$ can be derived similarly as that in Section \ref{appsub:Proof-of-T_univariate},
while $III_{m}$ behaves differently than $III$. We will discuss
these rates in the following subsections.

\subsubsection{The limit of $I_{m}$.}

It holds that 
\begin{eqnarray*}
I_{m} & = & \frac{1}{N}\sum_{i=1}^{N}\left[\left(\frac{W_{i}Y_{i}}{p_{0}(X_{i}^{\prime}\alpha_{0})}-\frac{(1-W_{i})Y_{i}}{1-p_{0}(X_{i}^{\prime}\alpha_{0})}-\tau\right)-\mathbb{E}_{\alpha_{0}}[D(Z_{i})|X_{i}^{\prime}\alpha_{0}]\{W_{i}-p_{0}(X_{i}^{\prime}\alpha_{0})\}\right]\\
 &  & +\frac{1}{N}\sum_{i=1}^{N}\{\mathbb{E}_{\alpha_{0}}[D(Z_{i})|X_{i}^{\prime}\alpha_{0}]-\mathbb{E}_{\tilde{\alpha}}[D(Z_{i})|X_{i}^{\prime}\tilde{\alpha}]\}\{W_{i}-p_{0}(X_{i}^{\prime}\alpha_{0})\}\\
 & =: & I_{m,1}+I_{m,2}.
\end{eqnarray*}
Similarly to \eqref{eq:E_D(x)}, we define $\delta_{\alpha}(u)=\mathbb{E}[D(Z)|X^{\prime}\alpha=u]$.
Then, 
\begin{eqnarray*}
I_{m,2} & = & \int\{\delta_{\alpha_{0}}(x^{\prime}\alpha_{0})-\delta_{\tilde{\alpha}}(x^{\prime}\tilde{\alpha})\}\{w-p_{0}(x^{\prime}\alpha_{0})\}d(\mathbb{P}_{n}(z)-\mathbb{P}_{0}(z))\\
 &  & +\int\{\delta_{\alpha_{0}}(x^{\prime}\alpha_{0})-\delta_{\tilde{\alpha}}(x^{\prime}\tilde{\alpha})\}\{w-p_{0}(x^{\prime}\alpha_{0})\}d\mathbb{P}_{0}(z)\\
 & =: & I_{m,2,1}+I_{m,2,2}.
\end{eqnarray*}

Let us first study the rate of $I_{m,2,1}$. Given that the function
class of $\big\{\delta_{\alpha}(u)=\mathbb{E}_{\alpha}[D(Z)|X^{\prime}\alpha=u]:\alpha\in\mathcal{B}(\alpha_{0},\delta_{0}),u\in I_{\alpha}=\{x^{\prime}\alpha:x\in\mathcal{X}\}\big\}$
is parameterized by a $k$-dimensional parameter $\alpha$, the $\epsilon$-bracket
number of this class is of the order of $\frac{1}{\epsilon}$ (see,
e.g., Example 19.7 of van der Vaart and Wellner, 2000). Its corresponding
entropy is smaller than that presented in \eqref{pf:HB}. 

Furthermore, BGH shows that $\tilde{\alpha}$ is $\sqrt{N}$-consistent
of $\alpha_{0}$ (recall that we have defined $\tilde{\alpha}=\hat{\alpha}$),
and we know that $\delta_{\alpha}(u)$ is by construction differentiable
w.r.t $\alpha$ for all $\alpha\in\mathcal{B}(\alpha_{0},\delta_{0})$
and $u\in I_{\alpha}=\{x^{\prime}\alpha:x\in\mathcal{X}\}$. As a
result, we have $\left\Vert \delta_{\alpha_{0}}(\cdot^{\prime}\alpha_{0})-\delta_{\tilde{\alpha}}(\cdot^{\prime}\tilde{\alpha})\right\Vert _{\mathbb{P}_{0}}=O_{p}(N^{-1/2})$.
Therefore, we can apply similar arguments for the term $II_{1}$ in
Appendix \ref{subsec:Uni-rate-of=000020II} to show that $I_{m,2,1}=o_{p}(N^{-1/2})$. 

Finally, $I_{m,2,2}=0$ by the Law of iterated expectation. Thus,
we have shown that $I_{m,2}=o_{p}(N^{-1/2}).$ Consequently, {\small
\begin{align}
I_{m} & =\frac{1}{N}\sum_{i=1}^{N}\left[\left(\frac{W_{i}Y_{i}}{p_{0}(X_{i}^{\prime}\alpha_{0})}-\frac{(1-W_{i})Y_{i}}{1-p_{0}(X_{i}^{\prime}\alpha_{0})}-\tau\right)-\mathbb{E}_{\alpha_{0}}[D(Z_{i})|X_{i}^{\prime}\alpha_{0}]\{W_{i}-p_{0}(X_{i}^{\prime}\alpha_{0})\}\right]+o_{p}(N^{-1/2})\nonumber \\
 & =\frac{1}{N}\sum_{i=1}^{N}\{m(Z_{i})+M(Z_{i})\}+o_{p}(N^{-1/2}),\label{eq:mul_rate_Im}
\end{align}
} where functions $m(\cdot)$ and $M(\cdot)$ are defined in Theorem
\ref{thm:ATE-matching-m}.

\subsubsection{The rates of $II_{m}$ and $IV_{m}$}

By the consistency of $\tilde{\alpha}$ and Assumption \ref{asm:2}',
it holds that for all large $N$, the function $p_{\tilde{\alpha}}(u)=\mathbb{E}[W|X^{\prime}\tilde{\alpha}=u]$
is monotone increasing, w.p.a.1. As a result, we can apply Theorem
\ref{thm:uc_isoton-m} and the same arguments presented Appendix \ref{subsec:Uni-rate-of=000020II}
to show 
\begin{equation}
II_{m}=o_{p}(N^{-1/2}).\label{eq:mul_rate_IIm}
\end{equation}
See pp.17-20 of BGH-supp for a similar case concerning the monotone
single index model.

By Theorem \ref{thm:uc_isoton-m} and the same arguments as presented
in Appendix \ref{subsec:uni-rate-of-IV}, it holds that 
\begin{equation}
IV_{m}=o_{p}(N^{-1/2}).\label{eq:mul_rate_IVm}
\end{equation}

\subsubsection{The rates of $III_{m}$ }

The term $III_{m}$ can be decomposed as
\begin{eqnarray}
III_{m} & = & \frac{1}{N}\sum_{i=1}^{N}\{D(Z_{i})-\mathbb{E}_{\tilde{\alpha}}[D(Z_{i})|X_{i}^{\prime}\tilde{\alpha}]\}\{\tilde{p}_{\tilde{\alpha}}(X_{i}^{\prime}\tilde{\alpha})-p_{0}(X_{i}^{\prime}\alpha_{0})\}\nonumber \\
 & = & \frac{1}{N}\sum_{i=1}^{N}\{D(Z_{i})-\mathbb{E}_{\tilde{\alpha}}[D(Z_{i})|X_{i}^{\prime}\tilde{\alpha}]\}\{\tilde{p}_{\tilde{\alpha}}(X_{i}^{\prime}\tilde{\alpha})-p_{\tilde{\alpha}}(X_{i}^{\prime}\tilde{\alpha})\}\nonumber \\
 &  & +\frac{1}{N}\sum_{i=1}^{N}\{D(Z_{i})-\mathbb{E}_{\tilde{\alpha}}[D(Z_{i})|X_{i}^{\prime}\tilde{\alpha}]\}\{p_{\tilde{\alpha}}(X_{i}^{\prime}\tilde{\alpha})-p_{0}(X_{i}^{\prime}\alpha_{0})\}\nonumber \\
 & =: & III_{m,1}+III_{m,2}.\label{eq:III_m}
\end{eqnarray}
By the consistency of $\tilde{\alpha}$, Assumption \ref{asm:2}',
it holds that for all large $N$, the function $p_{\tilde{\alpha}}(u)=\mathbb{E}[W|X^{\prime}\tilde{\alpha}=u]$
is monotone increasing, w.p.a.1. Therefore, we can apply Theorem \ref{thm:uc_isoton-m}
and the same arguments presented Appendix \ref{subsec:Uni-rate-of-III}
to show 
\begin{equation}
III_{m,1}=o_{p}(N^{-1/2}).\label{eq:III_m1}
\end{equation}
For $III_{m,2}$, by Lemma 17 of BGH-supp, it holds that 
\begin{align*}
\frac{\partial}{\partial\alpha^{(j)}}p_{\alpha}(x^{\prime}\alpha)\bigg|_{\alpha=\alpha_{0}} & =\{x^{(j)}-\mathbb{E}_{\alpha_{0}}[X^{(j)}|X^{\prime}\alpha_{0}=x^{\prime}\alpha_{0}]\}p_{0}^{(1)}(x^{\prime}\alpha_{0}),
\end{align*}
where $\alpha^{(j)}$ and $x^{(j)}$ are $j$-th elements of vectors
$\alpha$ and $x$. Extending $III_{m,2}$ around $\alpha_{0}$ yields{\small
\begin{eqnarray}
III_{m,2} & = & \frac{1}{N}\sum_{i=1}^{N}\{D(Z_{i})-\mathbb{E}_{\tilde{\alpha}}[D(Z_{i})|X_{i}^{\prime}\tilde{\alpha}]\}\{p_{\tilde{\alpha}}(X_{i}^{\prime}\tilde{\alpha})-p_{0}(X_{i}^{\prime}\alpha_{0})\}\nonumber \\
 & = & \frac{1}{N}\sum_{i=1}^{N}[D(Z_{i})-\mathbb{E}_{\tilde{\alpha}}[D(Z_{i})|X_{i}^{\prime}\tilde{\alpha}]][X_{i}-\mathbb{E}_{\alpha_{0}}[X_{i}|X_{i}^{\prime}\alpha_{0}]]^{\prime}p_{0}^{(1)}(X_{i}^{\prime}\alpha_{0})(\tilde{\alpha}-\alpha_{0})+o_{p}(\tilde{\alpha}-\alpha_{0})\nonumber \\
 & = & (\tilde{\alpha}-\alpha_{0})\frac{1}{N}\sum_{i=1}^{N}\{D(Z_{i})-\mathbb{E}_{\alpha_{0}}[D(Z_{i})|X_{i}^{\prime}\alpha_{0}]\}[X_{i}-\mathbb{E}_{\alpha_{0}}[X_{i}|X_{i}^{\prime}\alpha_{0}]]^{\prime}p_{0}^{(1)}(X_{i}^{\prime}\alpha_{0})+o_{p}(\tilde{\alpha}-\alpha_{0}),\nonumber \\
\label{eq:III_m2}
\end{eqnarray}
}where the last equality follows from $\mathbb{E}_{\alpha_{0}}[D(Z_{i})|X_{i}^{\prime}\alpha_{0}]-\mathbb{E}_{\tilde{\alpha}}[D(Z_{i})|X_{i}^{\prime}\tilde{\alpha}]=o_{p}(1)$.
By the Law of large numbers and the Law of iterated expectation, 
\begin{eqnarray}
 &  & \frac{1}{N}\sum_{i=1}^{N}[D(Z_{i})-\mathbb{E}_{\alpha_{0}}[D(Z_{i})|X_{i}^{\prime}\alpha_{0}]][X_{i}-\mathbb{E}_{\alpha_{0}}[X_{i}|X_{i}^{\prime}\alpha_{0}]]^{\prime}p_{0}^{(1)}(X_{i}^{\prime}\alpha_{0})\nonumber \\
 &  & \to_{p}\mathbb{E}[\mathrm{Cov}_{\alpha_{0}}(D(Z),X|X^{\prime}\alpha_{0})p_{0}^{(1)}(X^{\prime}\alpha_{0})].\label{eq:III_m2_w}
\end{eqnarray}
where $\mathrm{Cov}_{\alpha_{0}}(D(Z),X|\cdot):=\mathbb{E}\left\{ [D(Z)-\mathbb{E}_{\alpha_{0}}[D(Z)|X^{\prime}\alpha_{0}]][X_{i}-\mathbb{E}_{\alpha_{0}}[X|X^{\prime}\alpha_{0}]]^{\prime}|X^{\prime}\alpha_{0}=\cdot\right\} $. 

Furthermore, by Theorem 5 of BGH and $\tilde{\alpha}\equiv\hat{\alpha}$,
we have 
\begin{equation}
\tilde{\alpha}-\alpha_{0}=\mathbb{E}[p_{0}^{(1)}(X^{\prime}\alpha_{0})\mathrm{Cov}_{\alpha_{0}}(X|X^{\prime}\alpha_{0})]^{-}\frac{1}{N}\sum_{i=1}^{N}\{X_{i}-\mathbb{E}_{\alpha_{0}}[X_{i}|X_{i}^{\prime}\alpha_{0}]\}\{W_{i}-p_{0}(X_{i}^{\prime}\alpha_{0})\}+o_{p}(\tilde{\alpha}-\alpha_{0}),\label{eq:BGH_alpha}
\end{equation}
where $\mathbf{B}^{-}$ represents the Moore-Penrose inverse of a
matrix $\mathbf{B}$. 

For each $i=1,\ldots,N$, define
\begin{eqnarray*}
A(Z_{i}) & = & -\mathbb{E}[\mathrm{Cov}_{\alpha_{0}}(D(Z),X|X^{\prime}\alpha_{0})p_{0}^{(1)}(X^{\prime}\alpha_{0})]\mathbb{E}[p_{0}^{(1)}(X^{\prime}\alpha_{0})\text{\ensuremath{\mathrm{Cov}_{\alpha_{0}}}}(X|X^{\prime}\alpha_{0})]^{-}\\
 &  & \times\{X_{i}-\mathbb{E}_{\alpha_{0}}[X_{i}|X_{i}^{\prime}\alpha_{0}]\}\{W_{i}-p_{0}(X_{i}^{\prime}\alpha_{0})\}.
\end{eqnarray*}
Then, combining \eqref{eq:III_m}, \eqref{eq:III_m1}, \eqref{eq:III_m2},
\eqref{eq:BGH_alpha}, \eqref{eq:III_m2_w}, and $\tilde{\alpha}-\alpha_{0}=O_{p}(N^{-1/2})$
yields 
\begin{align}
III_{m} & =\frac{1}{N}\sum_{i=1}^{N}-A(Z_{i})+o_{p}(N^{-1/2}).\label{eq:mul_rate_IIIm}
\end{align}

\subsubsection{Summary of Appendix \ref{subsec:Proof-of-Th_m}}

Combining \eqref{eq:mul_rate_Im}, \eqref{eq:mul_rate_IIm}, \eqref{eq:mul_rate_IVm},
and \eqref{eq:mul_rate_IIIm}, we obtain 
\begin{eqnarray*}
\sqrt{N}(\tilde{\tau}-\tau) & = & \sqrt{N}(I_{m}+II_{m}-III_{m}+IV_{m})\\
 & = & \frac{1}{\sqrt{N}}\sum_{i=1}^{N}\{m(Z_{i})+M(Z_{i})+A(Z_{i})\}+o_{p}(1).
\end{eqnarray*}

\subsection{Proof of Theorem \ref{thm:=000020boot}}

The proof is adapted from Groeneboom and Hendrickx (2017). By Theorem
\ref{thm:matching_IPW}, it is sufficient to show the validity of
the bootstrap approximation for $\hat{\tau}=\frac{1}{N}\sum_{i=1}^{N}\left(\frac{W_{i}Y_{i}}{\tilde{p}(X_{i})}-\frac{(1-W_{i})Y_{i}}{1-\tilde{p}(X_{i})}\right)$. 

Let $\{Z_{i}\}_{i=1}^{N}=\{Y_{i},W_{i},X_{i}\}_{i=1}^{N}$ be the
original sample and $\{Z_{i}^{*}\}_{i=1}^{N}$ be its bootstrap resample.
Define $\tilde{p}^{*}(\cdot)$ and $\hat{\tau}^{*}$ as the UC-isotonic
estimator of the propensity score and the corresponding ATE estimator
with the resample $\{Z_{i}^{*}\}_{i=1}^{N}$. By the same arguments
for \eqref{eq:ATE_s_decom}, we have
\begin{eqnarray}
\hat{\tau}^{*}-\tau & = & \frac{1}{N}\sum_{i=1}^{N}\left[\left(\frac{W_{i}^{*}Y_{i}^{*}}{p(X_{i}^{*})}-\frac{(1-W_{i}^{*})Y_{i}^{*}}{1-p(X_{i}^{*})}-\tau\right)-\mathbb{E}[D(Z_{i})|X_{i}^{*}]\{W_{i}^{*}-p(X_{i}^{*})\}\right]\nonumber \\
 &  & +\frac{1}{N}\sum_{i=1}^{N}\mathbb{E}[D(Z_{i})|X_{i}^{*}]\left[W_{i}^{*}-\tilde{p}(X_{i}^{*})\right]-\frac{1}{N}\sum_{i=1}^{N}\{D(Z_{i})-\mathbb{E}[D(Z_{i})|X_{i}^{*}]\}\{\tilde{p}(X_{i}^{*})-p(X_{i}^{*})\}\nonumber \\
 &  & +\frac{2}{N}\sum_{i=1}^{N}\left(\frac{W_{i}^{*}Y_{i}^{*}}{\check{p}_{i}^{3}}-\frac{(1-W_{i}^{*})Y_{i}^{*}}{\{1-\check{p}_{i}\}^{3}}\right)\{\tilde{p}(X_{i}^{*})-p(X_{i}^{*})\}^{2}:=I^{*}+II^{*}-III^{*}+IV^{*}.\label{eq:ATE_s_decom_boot}
\end{eqnarray}
For the term $IV^{*}$, with some abuse of notation, we use the same
$\check{p}_{i}^{3}$ to denote a random value between $\tilde{p}(X_{i}^{*})$
and $p(X_{i}^{*})$. By similar arguments in Appendices \ref{subsec:Uni-rate-of=000020II},
\ref{subsec:Uni-rate-of-III}, and \ref{subsec:uni-rate-of-IV}, we
have 
\[
II^{*}=o_{p_{M}}(N^{-1/2}),\quad III^{*}=o_{p_{M}}(N^{-1/2}),\quad IV^{*}=o_{p_{M}}(N^{-1/2}),
\]
where $P_{M}$ is the probability measure in the bootstrap world defined
in p.3450 of Groeneboom and Hendrickx (2017). As a result, we have
\[
\hat{\tau}^{*}-\tau=\frac{1}{N}\sum_{i=1}^{N}\left[\left(\frac{W_{i}^{*}Y_{i}^{*}}{p(X_{i}^{*})}-\frac{(1-W_{i}^{*})Y_{i}^{*}}{1-p(X_{i}^{*})}-\tau\right)-\mathbb{E}[D(Z)|X_{i}^{*}]\{W_{i}^{*}-p(X_{i}^{*})\}\right]+o_{p_{M}}(N^{-1/2}).
\]
Define
\begin{eqnarray*}
m(Z,\tau,p(\cdot)) & = & \frac{YW}{p(X)}-\frac{Y(1-W)}{1-p(X)}-\tau,\qquad M(Z)=-\mathbb{E}[D(Z)|X]\{W-p(X)\}.
\end{eqnarray*}
Then, we can write
\begin{eqnarray}
\hat{\tau}^{*}-\tau & = & \left[\text{\ensuremath{\frac{1}{N}\sum_{i=1}^{N}}}m(Z_{i}^{*},\tau,p(X_{i}^{*}))-\text{\ensuremath{\frac{1}{N}\sum_{i=1}^{N}}}m(Z_{i},\tau,p(X_{i}))\right]+\left[\frac{1}{N}\sum_{i=1}^{N}M(Z_{i}^{*})-\text{\ensuremath{\frac{1}{N}\sum_{i=1}^{N}}}M(Z_{i})\right]\nonumber \\
 &  & +\frac{1}{N}\sum_{i=1}^{N}\{m(Z_{i},\tau,p(X_{i})+M(Z_{i})\}+o_{p_{M}}(N^{-1/2}).\label{eq:for=000020CLT=000020boot}
\end{eqnarray}
From Appendix \ref{subsec:The-limit-of-I}, we have
\begin{equation}
\hat{\tau}-\tau=\frac{1}{N}\sum_{i=1}^{N}\{m(Z_{i},\tau,p(X_{i}))+M(Z_{i})\}+o_{p}(N^{-1/2}).\label{eq:for=000020CLT}
\end{equation}
 Subtracting \eqref{eq:for=000020CLT} from \eqref{eq:for=000020CLT=000020boot},
\begin{eqnarray*}
\hat{\tau}^{*}-\hat{\tau} & = & \left\{ \frac{1}{N}\sum_{i=1}^{N}m(Z_{i}^{*},\tau,p(X_{i}^{*}))-\frac{1}{N}\sum_{i=1}^{N}m(Z_{i},\tau,p(X_{i}))\right\} +\left\{ \frac{1}{N}\sum_{i=1}^{N}M(Z_{i}^{*})-\frac{1}{N}\sum_{i=1}^{N}M(Z_{i})\right\} \\
 &  & +o_{p_{M}}(N^{-1/2})+o_{p}(N^{-1/2}).
\end{eqnarray*}
Note that $\mathbb{E}_{P_{M}}[m(Z_{i}^{*},\tau,p(X_{i}^{*}))]=\frac{1}{N}\sum_{i=1}^{N}m(Z_{i},\tau,p(X_{i}))$
and $\mathbb{E}_{P_{M}}[M(Z_{i}^{*})]=\frac{1}{N}\sum_{i=1}^{N}M(Z_{i})$,
where $\mathbb{E}_{P_{M}}[\cdot]$ is the expectation under $P_{M}$.
Consequently, a central limit theorem yields $\sqrt{N}(\hat{\tau}^{*}-\hat{\tau})\overset{d}{\to}N(0,\Omega)$,
where $\Omega$ is defined in Theorem \ref{thm:ATE-matching}. This
proves the conclusion (i) of Theorem \ref{thm:=000020boot}, i.e.,
\begin{equation}
\sup_{t\in\mathbb{R}}|\mathbb{P}^{*}\{\sqrt{N}(\hat{\tau}^{*}-\hat{\tau})\le t\}-\mathbb{P}\{\sqrt{N}(\hat{\tau}-\tau)\le t\}|\overset{p}{\to}0.\label{eq:thm_6}
\end{equation}

For (ii), note that by the definition of $c_{1-\alpha}^{*}$,
it holds that $\mathbb{P}^{*}\{\sqrt{N}(\hat{\tau}^{*}-\hat{\tau})\le c_{1-\alpha}^{*}\}=1-\alpha+o_{p}(1)$.
Combining this result with \eqref{eq:thm_6} yields
\begin{eqnarray*}
 &  & |\mathbb{P}\{\sqrt{N}(\hat{\tau}-\tau)\le c_{1-\alpha}^{*}\}-(1-\alpha)|\\
 & = & |\mathbb{P}\{\sqrt{N}(\hat{\tau}-\tau)\le c_{1-\alpha}^{*}\}-\mathbb{P}^{*}\{\sqrt{N}(\hat{\tau}^{*}-\hat{\tau})\le c_{1-\alpha}^{*}\}|+o_{p}(1)\\
 & \le & \sup_{t\in\mathbb{R}}|\mathbb{P}\{\sqrt{N}(\hat{\tau}-\tau)\le t\}-\mathbb{P}^{*}\{\sqrt{N}(\hat{\tau}^{*}-\hat{\tau})\le t\}|+o_{p}(1)\\
 & \overset{p}{\to} & 0,
\end{eqnarray*}
and thus the conclusion (ii) of Theorem \ref{thm:=000020boot} follows.
\renewcommand{\thesection}{S\arabic{section}}\renewcommand{\thetable}{S\arabic{table}}

\section{Thresholds for averaging treatment variables\protect\label{supp:thresh_theorey}}

To construct our UC-isotonic estimator, we average the first and the
last $\lfloor N^{2/3}\rfloor$-th elements of the treatment variable.
In this section, we study how the choice of this threshold will affect
the estimation performance. The following discussion is organized
into three parts: (i) the role of the $N^{-1/3}$-th quantile (which
is associated with our choice, the $\lfloor N^{2/3}\rfloor$-th element
of an arranged sample) for the isotonic estimator; (ii) different
choices characterized by changing powers of $N$; and (iii) different
choices characterized by changing constants.
\begin{enumerate}
	\item The $\lfloor N^{2/3}\rfloor$-th element corresponds approximately
	to the $N^{-1/3}$-th quantile of the covariate $X$. This $N^{-1/3}$-th
	quantile is a critical threshold in the asymptotic theory of the isotonic
	estimator at the boundary. Starting from the $N^{-1/3}$-th quantile
	(up to a constant factor; see the third point below), the isotonic
	estimator converges at the $N^{-1/3}$ rate. This boundary property
	was given by Kulikov and Lopuhaä (2006, Theorem 3.1) for the Grenander
	estimator. The result was further refined by Durot, Kulikov and Lopuhaä
	(2013; henceforth DKL), who showed that for a monotone increasing
	function $f(\cdot)$ supported on $[0,1]$ and its isotonic estimator
	$\hat{f}_{N}(\cdot)$, it holds
	\[
	\sup_{t\in[\alpha_{N},1-\beta_{N}]}|\hat{f}_{N}(t)-f(t)|=O_{p}\left(\frac{\log N}{N}\right)^{1/3},
	\]
	where $\alpha_{N}\geq\frac{K_{1}}{(\log N)^{2/3}}N^{-1/3}$ and $\beta_{N}\geq\frac{K_{2}}{(\log N)^{2/3}}N^{-1/3}$
	for some $K_{1},K_{2}>0$.
	\item Now we consider different choices for the power of $N$. To achieve
	$\sqrt{N}$-consistency in a two-stage semiparametric estimation process,
	it is generally required that the first-stage estimator uniformly
	converges at a rate faster than $N^{-1/4}$ (hereafter referred to
	as the UC-$N^{-1/4}$ condition; see, e.g., Assumption 5.1 (ii) of
	Newey, 1994). Note that the UC-$N^{-1/4}$ condition should be considered
	as a general guidance of constructing a $\sqrt{N}$-consistent plug-in
	estimator, rather than a necessary condition. There are cases that
	the UC-$N^{-1/4}$ condition is not satisfied, yet the resulting semiparametric
	estimator still remains $\sqrt{N}$-consistent.\\
	In what follows, we use the parameter $\alpha\in(0,1)$ defined by
	Kulikov and Lopuhaä (2006), to examine five cases. Based on the modification
	strategy that averages the first and the last $\lfloor N^{1-\alpha}\rfloor$
	sample points, which approximately correspond to the first and last
	$\frac{N^{1-\alpha}}{N}=N^{-\alpha}$-th quantile, we subsequently
	refer to the segment from $N^{-\alpha}$-th quantile to the $1-N^{-\alpha}$-th
	quantile of the support of $X$ as ``the middle part''. For both
	sides at boundaries, a simple average (which corresponds to the isotonic
	estimator at boundaries) exhibits a bias of order $N^{-\alpha}$ and
	a variance of order $\frac{1}{N^{1-\alpha}}=N^{\alpha-1}$. 
	\begin{enumerate}
		\item {[}Case of $\alpha=\frac{1}{3}${]} The proposal in our paper, corresponding
		to $\alpha=\frac{1}{3}$, ensures that the UC-isotonic estimator is
		uniformly consistent at the left and right boundaries at a rate of
		$O_{p}(N^{-1/3})$: the biases of the simple averages are of order
		$O(N^{-1/3})$, and the variances are of order $O_{p}(N^{-2/3})$.
		The uniform convergence rate of the middle part is $\left(\frac{\log N}{N}\right)^{1/3}$
		by Theorem 2.1 of DKL. Therefore, the UC-$N^{-1/4}$ condition is
		satisfied. 
		\item {[}Case of $\frac{1}{3}<\alpha<\frac{1}{2}${]} For the simple averages
		at both boundaries, the order of bias, $N^{-\alpha}$, is smaller
		than $N^{-1/4}$, and the order of variance, $N^{\alpha-1}$, is smaller
		than $N^{-1/2}$; for the middle part, Theorem 3.1 of Kulikov and
		Lopuhaä (2006) shows that the convergence rates at both boundaries
		are $O_{p}(N^{-(1-a)/2})$, which is faster than $O_{p}(N^{-1/4})$
		for $\alpha<\frac{1}{2}$. Furthermore, this theorem also shows that,
		towards the interior of the middle part, the convergence rate gradually
		reverts to the conventional rate of $O_{p}(N^{-1/3})$ associated
		with the isotonic estimator. By using similar arguments to those for
		Theorem 2.1 of DKL, it can be shown that within the middle part, the
		isotonic estimator is uniformly consistent at a rate faster than $N^{-1/4}$.
		Thus, the UC-$N^{-1/4}$ condition is satisfied. 
		\item {[}Case of $\frac{1}{4}<\alpha<\frac{1}{3}${]} The simple averages
		at both boundaries meet the UC-$N^{-1/4}$ condition, as detailed
		in (b); the convergence rate of the middle part is uniformly $\left(\frac{\log N}{N}\right)^{1/3}$
		by Theorem 2.1 of DKL, thereby satisfying the UC-$N^{-1/4}$ condition. 
		\item {[}Case of $\alpha\geq\frac{1}{2}${]} The variances of simple averages
		at both boundaries are of order $N^{\alpha-1}$, which converges at
		a rate less than or equal to $N^{-1/2}$. Consequently, the UC-$N^{-1/4}$
		condition is not satisfied.
		\item {[}Case of $\alpha\leq\frac{1}{4}${]} The biases of the simple averages
		are of order $N^{-\alpha}$, converging at a rate less than or equal
		to $N^{-1/4}$. Also, the UC-$N^{-1/4}$ condition is not satisfied.
	\end{enumerate}
	\item The asymptotic properties in the above-mentioned cases will not be
	changed if we multiply $\lfloor N^{1-\alpha}\rfloor$ by a positive
	integer. 
\end{enumerate}
To summarize, for an arbitrary positive integer $c$ and $\frac{1}{4}<\alpha<\frac{1}{2}$,
modified isotonic estimators of the propensity score function, which
average the first and last $c\lfloor N^{1-\alpha}\rfloor$ treatment
variables, yield matching estimators that share the same asymptotic
properties. In our paper, we select $\alpha=\frac{1}{3}$ to directly
utilize the result from DKL in our proof, and we choose the tuning
constant $c=1$ to facilitate convenient implementation.

\section{Additional Monte-Carlo simulations \protect\label{supp:add_Monte_Carlo}}

In this section, we conduct additional simulation comparisons of the
isotonic propensity score matching to other popular propensity score
matching methods for the ATE. Section \ref{supp:one-to-many} presents
comparisons with the one-to-many propensity score matching, Section
\ref{supp:radius} presents comparisons with the radius propensity
score matching, and Section \ref{supp:thresh_sim} examines how the
performance is sensitive to the choice of different thresholds for
averaging treatment variables. 

For the majority of this section, we follow the setup \eqref{sim:uni_setup}
described in Section \ref{subsec:sim_uni} of the main paper, which
is replicated here:

Let $X=0.15+0.7Z$, where $Z$ and $\nu$ are independently uniformly
distributed on $[0,1]$, and
\begin{eqnarray*}
	W & = & \begin{cases}
		0 & \text{if }X<\nu\\
		1 & \text{if }X\geq\nu
	\end{cases},\\
	Y & = & 0.5W+2X+\varepsilon,\\
	\varepsilon & \sim & N(0,1),\qquad\qquad\qquad\eqref{sim:uni_setup}
\end{eqnarray*}
and the true ATE is the coefficient of $W$, which is 0.5. In the
tables of the following subsections, $\hat{\mu}_{\tau}$ represents
the Monte-Carlo mean, and the mean square errors (MSE) are rescaled
by $N$. The number of Monte-Carlo simulations is 5000 for each sample
size. Furthermore, the one-to-many matching and the radius matching
estimators are based on propensity scores estimated with the logit
model, $\mathbb{P}(W=1|X=x)=\frac{\text{exp}(a+bx)}{\text{exp}(a+bx)+1}.$ 

\subsection{One-to-many propensity score matching\protect\label{supp:one-to-many}}

The simulation comparisons with one-to-many propensity score matching
are presented in Tables \ref{tab:radius_means} and \ref{tab:radius_MSEs}.
The left panels of both tables display the Monte-Carlo means and MSEs
for the isotonic propensity score matching estimator, respectively.
The right panels present Monte-Carlo means and MSEs of the one-to-$M$
propensity matching estimator, with radius $M=1,2,10,50,100$ and
500.

\begin{table}[H]
	\caption{\protect\label{tab:one-to-many=000020means}Monte-Carlo means (the
		true $\tau=0.5$)}
	
	\centering{}{\small{}%
		\begin{tabular}{ccr@{\extracolsep{0pt}.}lr@{\extracolsep{0pt}.}lr@{\extracolsep{0pt}.}lcccc}
			\toprule 
			\multicolumn{2}{c}{with UC-isotonic} & \multicolumn{2}{c}{} & \multicolumn{8}{c}{one-to-$M$ matching with logit }\tabularnewline
			\midrule
			$N$ & $\hat{\mu}_{\tau}$ & \multicolumn{2}{c}{} & \multicolumn{2}{c}{$M=1$} & \multicolumn{2}{c}{$M=2$} & $M=10$ & $M=50$ & $M=100$ & $M=500$\tabularnewline
			\cmidrule{1-2}\cmidrule{5-12}
			100 & 0.4977 & \multicolumn{2}{c}{} & 0&4997 & 0&5083 & 0.5593 & N/A & N/A & N/A\tabularnewline
			1000 & 0.4934 & \multicolumn{2}{c}{} & 0&5009 & 0&5011 & 0.5019 & 0.5195 & 0.5535 & N/A\tabularnewline
			2000 & 0.4946 & \multicolumn{2}{c}{} & 0&4999 & 0&4996 & 0.4997 & 0.5053 & 0.5178 & 0.6739\tabularnewline
			5000 & 0.4963 & \multicolumn{2}{c}{} & 0&4995 & 0&5001 & 0.5002 & 0.5014 & 0.5042 & 0.5524\tabularnewline
			10000 & 0.4974 & \multicolumn{2}{c}{} & 0&5000 & 0&4999 & 0.5000 & 0.5003 & 0.5012 & 0.5184\tabularnewline
			\bottomrule
	\end{tabular}}{\small\par}
\end{table}

As expected, Table \ref{tab:one-to-many=000020means} indicates that
for one-to-$M$ matching, estimates using larger $M$ tend to return
greater bias. Although the bias decreases for each choice of $M$
as the sample size grows, we have noticed that for large $M$ values,
the one-to-$M$ matching struggles to consistently identify matched
groups of the pre-specified size. This issue leads to N/A values in
the reported Monte-Carlo means. In contrast, the isotonic propensity
score matching estimator can always identify matched groups and yields
small biases without the need to adjust smoothing parameters.
\begin{table}[H]
	\caption{\protect\label{tab:one-to-many=000020MSEs}Monte-Carlo MSEs}
	
	\centering{}{\small{}%
		\begin{tabular}{ccr@{\extracolsep{0pt}.}lr@{\extracolsep{0pt}.}lr@{\extracolsep{0pt}.}lcccc}
			\toprule 
			\multicolumn{2}{c}{with UC-isotonic} & \multicolumn{2}{c}{} & \multicolumn{7}{c}{one-to-$M$ matching with logit } & \tabularnewline
			\midrule
			$N$ & $N\cdot\text{MSE}$ & \multicolumn{2}{c}{} & \multicolumn{2}{c}{$M=1$} & \multicolumn{2}{c}{$M=2$} & $M=10$ & $M=50$ & $M=100$ & $M=500$\tabularnewline
			\cmidrule{1-2}\cmidrule{5-12}
			100 & 5.2723 & \multicolumn{2}{c}{} & 7&1068 & 6&0312 & 5.3143 & N/A & N/A & N/A\tabularnewline
			1000 & 5.2589 & \multicolumn{2}{c}{} & 7&0630 & 5&9692 & 5.1740 & 5.2932 & 7.6373 & N/A\tabularnewline
			2000 & 5.2158 & \multicolumn{2}{c}{} & 7&0816 & 5&8025 & 4.9733 & 4.9090 & 5.4510 & 64.9621\tabularnewline
			5000 & 4.9418 & \multicolumn{2}{c}{} & 6&8376 & 5&9635 & 5.1878 & 5.0063 & 5.0779 & 18.5199\tabularnewline
			10000 & 4.9785 & \multicolumn{2}{c}{} & 6&8238 & 6&0263 & 5.2125 & 5.0533 & 5.0506 & 8.3006\tabularnewline
			\bottomrule
	\end{tabular}}{\small\par}
\end{table}
As the sample size increases, the isotonic matching estimator yields
small MSEs that converge to the semiparametric efficiency bound (SEB)
of this problem (approximately 4.96, as detailed in Section \ref{subsec:sim_uni}),
exhibiting superior performance than those one-to-many matching estimators
that employ small $M$ values. With $M$ growing, the MSEs of one-to-many
matching estimators approach the SEB for large sample sizes, yet face
difficulties with small sample sizes due to the simultaneous increase
in bias. Also, if $M$ is excessively large relative to a comparatively
small $N$, the one-to-many matching algorithm struggles to consistently
produce estimators—a concern not shared by users of the isotonic propensity
score matching estimator.

\subsection{Radius matching\protect\label{supp:radius}}

The simulation comparisons with radius propensity score matching are
presented in Tables \ref{tab:radius_means} and \ref{tab:radius_MSEs}.
The left panels of both tables display the Monte-Carlo means and MSEs
for the isotonic propensity score matching estimator, respectively.
The right panels present Monte-Carlo means and MSEs of the radius
propensity matching estimator, with radius $r=0.01,0.05,0.1,$ and
0.2.

\begin{table}[H]
	\caption{\protect\label{tab:radius_means}Monte-Carlo means (the true $\tau=0.5$)}
	
	\centering{}{\small{}%
		\begin{tabular}{ccr@{\extracolsep{0pt}.}lr@{\extracolsep{0pt}.}lr@{\extracolsep{0pt}.}lcc}
			\toprule 
			\multicolumn{2}{c}{with UC-isotonic} & \multicolumn{2}{c}{} & \multicolumn{6}{c}{radius matching with logit }\tabularnewline
			\midrule
			$N$ & $\hat{\mu}_{\tau}$ & \multicolumn{2}{c}{} & \multicolumn{2}{c}{$r=0.01$} & \multicolumn{2}{c}{$r=0.05$} & $r=0.1$ & $r=0.2$\tabularnewline
			\cmidrule{1-2}\cmidrule{5-10}
			100 & 0.4977 & \multicolumn{2}{c}{} & 0&5029 & 0&5069 & 0.5273 & 0.5822\tabularnewline
			1000 & 0.4934 & \multicolumn{2}{c}{} & 0&5010 & 0&5086 & 0.5275 & 0.5826\tabularnewline
			2000 & 0.4946 & \multicolumn{2}{c}{} & 0&4996 & 0&5072 & 0.5260 & 0.5808\tabularnewline
			5000 & 0.4963 & \multicolumn{2}{c}{} & 0&5007 & 0&5081 & 0.5270 & 0.5820\tabularnewline
			10000 & 0.4974 & \multicolumn{2}{c}{} & 0&5003 & 0&5079 & 0.5268 & 0.5816\tabularnewline
			\bottomrule
	\end{tabular}}{\small\par}
\end{table}
Table \ref{tab:radius_means} shows that the bias increases with the
matching radius. Notably, the bias does not shrink as the sample size
grows. For example, with $r=0.2$, the bias remains approximately
0.8 for both $N=100$ and $N=10000$. Table \ref{tab:radius_MSEs}
presents that while the MSE of the radius matching estimator decreases
for a small radius, it escalates for a larger radius. In fact, if
a sufficiently broad range of sample sizes is available, a U-shaped
pattern should emerge for each radius choice (see the case of $r=0.05$),
suggesting that identifying an optimal radius is crucial – a challenge
not encountered by users of the isotonic propensity score matching
estimator.

\begin{table}[H]
	\caption{\protect\label{tab:radius_MSEs}Monte-Carlo MSEs}
	
	\centering{}{\small{}%
		\begin{tabular}{ccr@{\extracolsep{0pt}.}lr@{\extracolsep{0pt}.}lr@{\extracolsep{0pt}.}lcc}
			\toprule 
			\multicolumn{2}{c}{with UC-isotonic} & \multicolumn{2}{c}{} & \multicolumn{6}{c}{radius matching with logit }\tabularnewline
			\midrule
			$N$ & $\text{MSE}$ & \multicolumn{2}{c}{} & \multicolumn{2}{c}{$r=0.01$} & \multicolumn{2}{c}{$r=0.05$} & $r=0.1$ & $r=0.2$\tabularnewline
			\cmidrule{1-2}\cmidrule{5-10}
			100 & 5.2723 & \multicolumn{2}{c}{} & 6&7361 & 5&6793 & 5.2637 & 5.4280\tabularnewline
			1000 & 5.2589 & \multicolumn{2}{c}{} & 5&2690 & 4&9969 & 5.5333 & 11.3588\tabularnewline
			2000 & 5.2158 & \multicolumn{2}{c}{} & 5&2690 & 4&8654 & 6.0080 & 17.4621\tabularnewline
			5000 & 4.9418 & \multicolumn{2}{c}{} & 5&0123 & 5&2328 & 8.4279 & 38.1188\tabularnewline
			10000 & 4.9785 & \multicolumn{2}{c}{} & 5&0199 & 5&5713 & 11.9924 & 71.1289\tabularnewline
			\bottomrule
	\end{tabular}}{\small\par}
\end{table}

\subsection{Choosing different thresholds\protect\label{supp:thresh_sim}}

In this subsection, we examine how the performance of the isotonic
propensity score matching estimator is sensitive to different thresholds
for averaging the treatment variables. We conduct 5000 simulations
for each of the five cases discussed in Section \ref{supp:thresh_theorey},
and compare their Monte-Carlo means and MSEs. For the cases (a) to
(e), we set $\alpha=\frac{1}{3},\frac{5}{12},\frac{7}{24},\frac{2}{3}$,
and $\frac{1}{8}$, corresponding respectively to the first and the
last $\lfloor N^{2/3}\rfloor$, $\lfloor N^{7/12}\rfloor$, $\lfloor N^{17/24}\rfloor$,
$\lfloor N^{1/3}\rfloor$, and $\lfloor N^{7/8}\rfloor$ observations.

Firstly, we generate data using the model \eqref{sim:uni_setup},
the same setup employed in previous subsections and in Section \ref{sec:Monte-Carlo}
of the main paper. 

\begin{table}[H]
	\caption{\protect\label{tab:iso_thresh_means}Monte-Carlo means (the true $\tau=0.5$)}
	
	\centering{}{\small{}%
		\begin{tabular}{ccccccc}
			\toprule 
			\multicolumn{7}{c}{$\hat{\mu}_{\tau}$}\tabularnewline
			\midrule
			$N$ &  & (a):$\lfloor N^{2/3}\rfloor$ & (b):$\lfloor N^{7/12}\rfloor$ & (c):$\lfloor N^{17/24}\rfloor$ & (d):$\lfloor N^{1/3}\rfloor$ & (e):$\lfloor N^{7/8}\rfloor$\tabularnewline
			\cmidrule{1-1}\cmidrule{3-7}
			\multirow{1}{*}{100} &  & 0.4977 & 0.4919  & 0.5064  & 0.4937  & 0.5409 \tabularnewline
			1000 &  & 0.4934 & 0.4922  & 0.4952  & 0.4932  & 0.5580 \tabularnewline
			2000 &  & 0.4946 & 0.4940  & 0.4958  & 0.4946  & 0.5452 \tabularnewline
			5000 &  & 0.4963 & 0.4959  & 0.4969  & 0.4962  & 0.5327 \tabularnewline
			10000 &  & 0.4974 & 0.4972  & 0.4978  & 0.4973  & 0.5257 \tabularnewline
			\bottomrule
	\end{tabular}}{\small\par}
\end{table}

As expected, Table \ref{tab:iso_thresh_means} indicates that estimates
using larger averaging thresholds tend to exhibit greater bias. Nevertheless,
the bias diminishes across all boundary averaging schemes as the sample
size increases, demonstrating the proposed isotonic matching estimator
is generally asymptotically unbiased.

\begin{table}[H]
	\caption{\protect\label{tab:iso_thresh_MSEs}Monte-Carlo MSEs}
	
	\centering{}{\small{}%
		\begin{tabular}{ccccccc}
			\toprule 
			\multicolumn{7}{c}{$N\cdot\text{MSE}$}\tabularnewline
			\midrule
			$N$ &  & (a):$\lfloor N^{2/3}\rfloor$ & (b):$\lfloor N^{7/12}\rfloor$ & (c):$\lfloor N^{17/24}\rfloor$ & (d):$\lfloor N^{1/3}\rfloor$ & (e):$\lfloor N^{7/8}\rfloor$\tabularnewline
			\cmidrule{1-1}\cmidrule{3-7}
			100 &  & 5.2723 & 5.3847  & 5.1592  & 5.3470  & 5.3505 \tabularnewline
			1000 &  & 5.2589 & 5.3038  & 5.1869  & 5.3359  & 8.2515 \tabularnewline
			2000 &  & 5.2158 & 5.2563  & 5.1678  & 5.2825  & 8.9933 \tabularnewline
			5000 &  & 4.9418 & 4.9707  & 4.9199  & 4.9884  & 10.0267 \tabularnewline
			10000 &  & 4.9785 & 4.9986  & 4.9594  & 5.0168  & 11.4050 \tabularnewline
			\bottomrule
	\end{tabular}}{\small\par}
\end{table}

Table \ref{tab:iso_thresh_MSEs} shows that the observed pattern of
estimation performance is generally in accordance with our theoretical
analysis presented in Section \ref{supp:thresh_theorey}. The averaging
schemes applied to columns (a), (b), and (c), which meet the UC-$N^{-1/4}$
condition, give small MSEs that converge to the SEB of this problem
(approximately 4.96). Conversely, column (e)'s averaging scheme, which
does not satisfy the UC-$N^{-1/4}$ condition, exhibits an increasing
$N\cdot\text{MSE}$. This indicates a convergence rate that is slower
than $N^{-1/2}$. 

The only exception is column (d), which employs a conservative averaging
scheme that does not meet the UC-$N^{-1/4}$ condition. Despite this,
there is no clear indication that estimates in column (d) perform
worse than those in columns (a), (b), and (c). It's important to note
that the primary goal of these averaging schemes is to mitigate bias
arising from estimated propensity scores near 0 and 1. Such bias would
be disproportionately magnified in the second stage of matching. However,
in our setup \eqref{sim:uni_setup}, the smallest and the largest
values of the true propensity score are 0.15 and 0.85, respectively,
which do not approach 0 and 1 closely enough. To penalize the conservativeness
of the scheme applied in column (d), in the following, we adjust the
data generating process for $X$ within the setup \eqref{sim:uni_setup}
to $X=0.01+0.98Z$, leading to the smallest and largest propensity
scores becoming 0.01 and 0.99, respectively. All other parameters
of the setup \eqref{sim:uni_setup} remain unchanged. Given this updated
data generating process, the SEB calculated according to Hahn (1998)
is:
\begin{align*}
	\Omega_{\text{SEB}} & =\text{Var}(\mathbb{E}[Y(1)-Y(0)|X])+\mathbb{E}[\text{Var}(Y(1)|X)/p_{0}(X)]+\mathbb{E}[\text{Var}(Y(0)|X)/(1-p_{0}(X))]\\
	& =\text{Var}(0.5)+\mathbb{E}[1/p_{0}(X)]+\mathbb{E}[1/(1-p_{0}(X))]\\
	& =0+\int_{0.01}^{0.99}\frac{1}{x}\frac{1}{0.98}dx+\int_{0.01}^{0.99}\frac{1}{1-x}\frac{1}{0.98}dx\approx9.38.
\end{align*}
Table \ref{tab:iso_thresh_means_nearly01} shows a pattern of Monte-Carlo
means similar to those of Table \ref{tab:iso_thresh_means}. Although
biases are generally larger, they exhibit clear trends of convergence
to zero as sample sizes increase.

\begin{table}[H]
	\caption{\protect\label{tab:iso_thresh_means_nearly01}Monte-Carlo means (the
		true $\tau=0.5$)}
	
	\centering{}{\small{}%
		\begin{tabular}{ccccccc}
			\toprule 
			\multicolumn{7}{c}{with UC-isotonic}\tabularnewline
			\midrule
			$N$ &  & (a):$\lfloor N^{2/3}\rfloor$ & (b):$\lfloor N^{7/12}\rfloor$ & (c):$\lfloor N^{17/24}\rfloor$ & (d):$\lfloor N^{1/3}\rfloor$ & (e):$\lfloor N^{7/8}\rfloor$\tabularnewline
			\cmidrule{1-1}\cmidrule{3-7}
			100 &  & 0.5691  & 0.5761  & 0.5750  & 0.5862  & 0.6411 \tabularnewline
			1000 &  & 0.5124  & 0.5217  & 0.5130  & 0.5299  & 0.6421 \tabularnewline
			2000 &  & 0.5064  & 0.5127  & 0.5070  & 0.5195  & 0.6154 \tabularnewline
			5000 &  & 0.5025  & 0.5062  & 0.5031  & 0.5110  & 0.5887 \tabularnewline
			10000 &  & 0.5011  & 0.5034  & 0.5017  & 0.5071  & 0.5726 \tabularnewline
			\bottomrule
	\end{tabular}}{\small\par}
\end{table}

\begin{table}[H]
	\caption{\protect\label{tab:iso_thresh_MSEs-nearly01}Monte-Carlo MSEs}
	
	\centering{}{\small{}%
		\begin{tabular}{ccccccc}
			\toprule 
			\multicolumn{7}{c}{with UC-isotonic}\tabularnewline
			\midrule
			$N$ &  & (a):$\lfloor N^{2/3}\rfloor$ & (b):$\lfloor N^{7/12}\rfloor$ & (c):$\lfloor N^{17/24}\rfloor$ & (d):$\lfloor N^{1/3}\rfloor$ & (e):$\lfloor N^{7/8}\rfloor$\tabularnewline
			\cmidrule{1-1}\cmidrule{3-7}
			100 &  & 7.0773  & 7.0096  & 6.9356  & 6.7309  & 7.5199 \tabularnewline
			1000 &  & 8.2009  & 9.0793  & 7.7508  & 8.9874  & 25.9746 \tabularnewline
			2000 &  & 8.8502  & 9.7641  & 8.2369  & 9.7776  & 32.7194 \tabularnewline
			5000 &  & 8.5570  & 9.4519  & 8.2101  & 9.6932  & 45.3723 \tabularnewline
			10000 &  & 9.0496  & 9.7802  & 8.6784  & 10.1159  & 58.9884 \tabularnewline
			\bottomrule
	\end{tabular}}{\small\par}
\end{table}

Table \ref{tab:iso_thresh_MSEs-nearly01} shows the Monte-Carlo MSEs
of the modified setup \eqref{sim:uni_setup}. Now, the pattern aligns
well with our theoretical analysis presented in Section \ref{supp:thresh_theorey}.
Columns (a), (b), and (c) continue to show small levels of MSEs, and
they exhibit trends of convergence to the SEB. We also note that the
scheme in column (c) performs particularly well in finite samples.
However, with the smallest and largest propensity scores being near
0 and 1, column (d) pays a price for its conservative averaging scheme.
Its MSEs exceed the SEB for both sample sizes of 5000 and 10000, showing
a clear trend of further increases. This suggests a convergence rate
slower than $N^{-1/2}$. Overall, the evidence from the simulations
supports our choice of the first and the last $N^{2/3}$ as thresholds
for averaging treatment variables.

\newpage{}

\end{document}